\documentclass[twocolumn]{aastex7}

\newcommand{\Msun}{M$_{\odot}$}
\newcommand{\Mbh}{$M_{\rm BH}$}

\newcommand{\Rsoi}{$R_{\rm SOI}$}
\newcommand{\Mstar}{$M_{\star}$}
\newcommand{\Mbulge}{$M_{\rm Bulge}$}
\newcommand{\Lbulge}{$L_{\rm Bulge}$}
\newcommand{\Lsun}{L$_\odot$}

\newcommand{\ml}{\emph{M/L}}

\newcommand{\kms}{km~s$^{-1}$}

\newcommand{\go}{G$_{\rm outer}$}

\newcommand{\cotwo}{$^{12} $CO(2$-$1)}

\newcommand{\siggas}{$\sigma_{\rm gas}$}
\newcommand{\coone}{$^{12} $CO(1$-$0)}

\hypersetup{colorlinks, linkcolor=blue, citecolor=cyan, urlcolor=magenta}

\usepackage{multirow}
\usepackage{footnote}
\usepackage{newtxtext,newtxmath}
\usepackage{graphicx}	
\usepackage{amsmath}	
\usepackage{lineno}
\linenumbers

\graphicspath{{./}{figures/}}

\received{April 29 2025}
\revised{\today}

\submitjournal{ApJ}

\shorttitle{Molecular-based SMBH measurement of NGC 7052}
\shortauthors{H.\ N.\ Ngo et al.}
\begin{document}

\title{\large{\bf Revisiting the supermassive black hole mass of NGC 7052 using high spatial resolution molecular gas observed with ALMA}}

\correspondingauthor{Dieu D. Nguyen}\email{dieun@umich.edu}
\author[0009-0006-5852-4538]{Hai N. Ngo}
\affiliation{Faculty of Physics -- Engineering Physics, University of Science, Vietnam National University, Ho Chi Minh City, Vietnam}
\email{hai10hoalk@gmail.com}

\author[0000-0002-5678-1008]{Dieu D.\ Nguyen}
\email{dieun@umich.edu}
\affiliation{Department of Astronomy, University of Michigan, 1085 South University Avenue, Ann Arbor, MI 48109, USA}

\author[0009-0004-3689-8577]{Tinh Q.\ T.\ Le}
\email{lethongquoctinh01@gmail.com}
\affiliation{Department of Physics, International University, Vietnam National University, Ho Chi Minh City, Vietnam}
\affiliation{International Centre for Interdisciplinary Science and Education, 07 Science Avenue, Ghenh Rang, 55121 Quy Nhon, Vietnam}

\author[0009-0007-3200-8751]{Khue N.\ H.\ Ho}
\email{khuehohuyngoc@gmail.com}
\affiliation{Department of Physics, International University, Vietnam National University, Ho Chi Minh City, Vietnam}

\author[0009-0005-8845-9725]{Tien H.\ T.\ Ho}
\email{htien2808@gmail.com}
\affiliation{Faculty of Physics -- Engineering Physics, University of Science, Vietnam National University, Ho Chi Minh City, Vietnam}

\author[0000-0001-5802-6041]{Elena Gallo}
\email{egallo@umich.edu}
\affiliation{Department of Astronomy, University of Michigan, 1085 South University Avenue, Ann Arbor, MI 48109, USA}

\author[0000-0003-1991-370X]{Kristina Nyland} 
\email{kristina.e.nyland.civ@us.navy.mil}
\affiliation{National Research Council, resident at the U.S. Naval Research Laboratory, 4555 Overlook Ave. SW, Washington, DC 20375, USA}

\author[0000-0001-6186-8792]{Masatoshi Imanishi}
\email{masa.imanishi@nao.ac.jp}
\affiliation{National Astronomical Observatory of Japan, National Institute of Natural Sciences, 2-21-1 Osawa, Mitaka, Tokyo 181-8588, Japan}
\affiliation{Department of Astronomical Science, Graduate University for Advanced Studies, 2-21-1 Osawa, Mitaka, Tokyo 181-8588, Japan}

\author[0000-0002-6939-0372]{Kouichiro Nakanishi}
\email{nakanisi.k@nao.ac.jp}
\affiliation{National Astronomical Observatory of Japan, National Institute of Natural Sciences, 2-21-1 Osawa, Mitaka, Tokyo 181-8588, Japan}
\affiliation{Department of Astronomical Science, Graduate University for Advanced Studies, 2-21-1 Osawa, Mitaka, Tokyo 181-8588, Japan}

\author[0000-0001-9649-2449]{Que T. Le}
\email{ltque@hcmiu.edu.vn}
\affiliation{Department of Physics, International University, Vietnam National University, Ho Chi Minh City, Vietnam}

\author[0000-0001-9879-7780]{Fabio Pacucci}
\email{fabio.pacucci@cfa.harvard.edu}
\affiliation{Center for Astrophysics—Harvard $\&$ Smithsonian, 60 Garden St., Cambridge, MA 02138, USA}
\affiliation{Black Hole Initiative, Harvard University, 20 Garden St., Cambridge, MA 02138, USA}

\author[0000-0001-7867-4572]{Eden Girma}
\email{egirma@princeton.edu}
\affiliation{Department of Astrophysical Sciences, Princeton University, Peyton Hall, Princeton, NJ 08544, USA}

\begin{abstract}
We present our dynamical mass constraints on the central supermassive black hole (SMBH) in the early-type galaxy NGC 7052 using high spatial-resolution observations of \cotwo\ emission from the Atacama Large Millimeter/submillimeter Array (ALMA). The data were obtained during ALMA Cycle 7 and have a synthesized beam size of 0\farcs29 $\times$ 0\farcs22 (97 $\times$ 73 pc$^2$). The dynamical model yielded an SMBH mass of $\approx$$(2.50 \pm 0.37 \, [{\rm statistical}] \pm 0.8 \, [{\rm systematic}]) \times10^9$ \Msun\ and a stellar-$I$ band mass-to-light ratio of $\approx$$4.08 \pm 0.23\, [{\rm statistical}] \pm 0.4 \, [{\rm systematic}]$ \Msun/\Lsun$_{,I}$ ($3\sigma$ confidence intervals). Although our new ALMA observation has three times lower spatial resolution than previous ALMA data, it still resolves the SMBH’s sphere of influence with a spatial resolution that is 1.5 times smaller than this sphere radius. While our \Mbh\ estimate is fully consistent with the previous determination, the $I$-band mass-to-light ratio is lower by 10\%. This difference arises from our improved galaxy mass model, which incorporates both the molecular gas distribution and the extended stellar mass in the outer regions of the galaxy, the components that were previously neglected.
\end{abstract}




\section{Introduction}\label{sec:intro}

Supermassive black hole (SMBH, \Mbh\;$\gtrsim10^6$~\Msun) can be found at the center of every massive galaxy with total stellar mass, \Mstar\ $\gtrsim10^{10}$ \Msun\ \citep{Kormendy13, Saglia16}. Their demographics studies have demonstrated the correlations between \Mbh\ and the luminosity \citep[\Lbulge;][]{dressler1988stellar, Magorrian98}, velocity dispersion \citep[$\sigma_\star$;][]{ferrarese2000fundamental}, stellar mass \citep[\Mbulge;][]{kormendy2001supermassive} of the bulge component of their host galaxy, or the total stellar mass of the entire galaxy \citep{Greene12, Sahu19a}. However, observational evidence also suggests that these relationships are incomplete in the regimes of both low \citep{Nguyen14, Nguyen17, Nguyen18, Nguyen19, Nguyen17conf, Nguyen19conf, Greene20} and high \citep[e.g.,][]{Lauer07, Cappellari16, Krajnovic18a, Nguyen23} mass of galaxy and black hole (BH). 

Given the number of dynamical \Mbh\ measurements has increased significantly thank to a variety of tracers, each tracer offers distinct advantages and challenges for specific galaxy types. Particularly, the stellar dynamical technique \citep[e.g.,][]{Ahn18, Nguyen18, Nguyen2025, Voggel18, Thater22, Thater23}, commonly applied to early-type galaxies (ETGs), relies heavily on absorption lines in integrated stellar spectra but is sensitive to dust extinction \citep{Alatalo13}. Although the ionized gas dynamical method \citep[e.g.,][]{Walsh13} is based on circular motion, which is often influenced by noncircular motions (e.g., inflows and outflows), it translates into significantly inconsistent estimates \Mbh\ compared to those derived from stellar dynamical technique. Recent studies have shown that this inconsistency tentatively results in gas-based \Mbh\ values that are biased low by at least a factor of two relative to stellar dynamical measurements \citep{Haring2006, Thater19_conf}. On the other hand, the maser dynamical technique is considered the “gold standard” for measuring \Mbh\ \citep[e.g.,][]{Miyoshi95}, as it probes deep within the accretion disk, close to the black hole’s sphere of influence (SOI\footnote{The space-time spherical region surrounding a BH where its gravitational influence dominates over that of other masses. This region is also defined as the vicinity of the BH in which the enclosed stellar mass is equal to \Mbh.}), using high-resolution observations from Very Long Baseline Interferometry \citep[VLBI;][]{Gao2017}. However, this method is limited to Seyfert type-2 active galactic nuclei (AGN), which are rare and constitute only about 5\% of all AGN, as found in the Cosmological Project \citep{Braatz1996}. Additionally, the reverberation mapping \cite[RM; e.g.,][]{Bentz2023a, Bentz2023b} technique estimates \Mbh\ by measuring the time lag between variations in the continuum emission from the accretion disk and the response of broad-line region (BLR) emission lines. This method is applicable to both nearby and distant AGN but is restricted to Seyfert type-1 AGN, where BLR emission lines are observable.

However, high-sensitivity and high-spatial-resolution observations of cold molecular \citep[e.g.,][]{Davis13Nature, Nguyen20, Nguyen22} and atomic \citep{Nguyen21} gas tracers provided by the Atacama Large Millimeter/submillimeter Array (ALMA) have proven to be a game changer in accurately estimating BH masses dynamically. These tracers are less affected by turbulent motions \citep{Davis20} and allow us to probe deeply into the SOI \citep{Barth16b, Barth16a, North19, Boizelle19, Boizelle21, Zhang25}, opening in a new era for precise \Mbh\ measurements. 

Generally, the flat-thin disk model is commonly used to describe the motion of a CND under the assumption that cold molecular gas moves in circular orbits, producing a central upturn velocity curve along the line of sight (LOS), which is then used to estimate \Mbh\ \citep{Davis14}. We employ this technique in this work to revisit the dynamical-\Mbh\ estimate in NGC 7052 based on our own 0\farcs29 $\times$ 0\farcs22 synthesis beamsize ALMA observation with the $^{12}$CO(2-1) emission and our improvement on the stellar-mass model \citep[i.e., compare to the mass model constrained by][see Section \ref{sec:stellar-mass}]{Smith21}. 

In Section \ref{sec:data-reduction}, we present our ALMA observation, data reduction, image processing, and LOS kinematic measurements of the \cotwo\ molecular gas. In Section \ref{sec:stellar-mass}, we refine the stellar and molecular gas mass model for NGC 7052 using Hubble Space Telescope (HST) and our ALMA observations, respectively. Section \ref{sec:dynamical-models} describes our molecular gas dynamical modeling technique for measuring the \Mbh\ and discusses sources of uncertainty, which are followed by a summary in Section \ref{sec:conclusion}.

We adopt a $\Lambda$CDM cosmology with a Hubble constant of H$_0 = 70$ \kms\ Mpc$^{-1}$, a dark energy density parameter of $\Omega_{\Lambda,0} = 0.7$, and a matter density parameter of $\Omega_{\rm m,0} = 0.3$. Given a distance of $D=69.3$ Mpc to NGC~7052 \citep{Ma14}, the physical scale for converting between arcseconds (\arcsec) and parsecs (pc) is 336 pc/\arcsec.

\begin{figure*}
\centering
\includegraphics[width=0.45\textwidth]{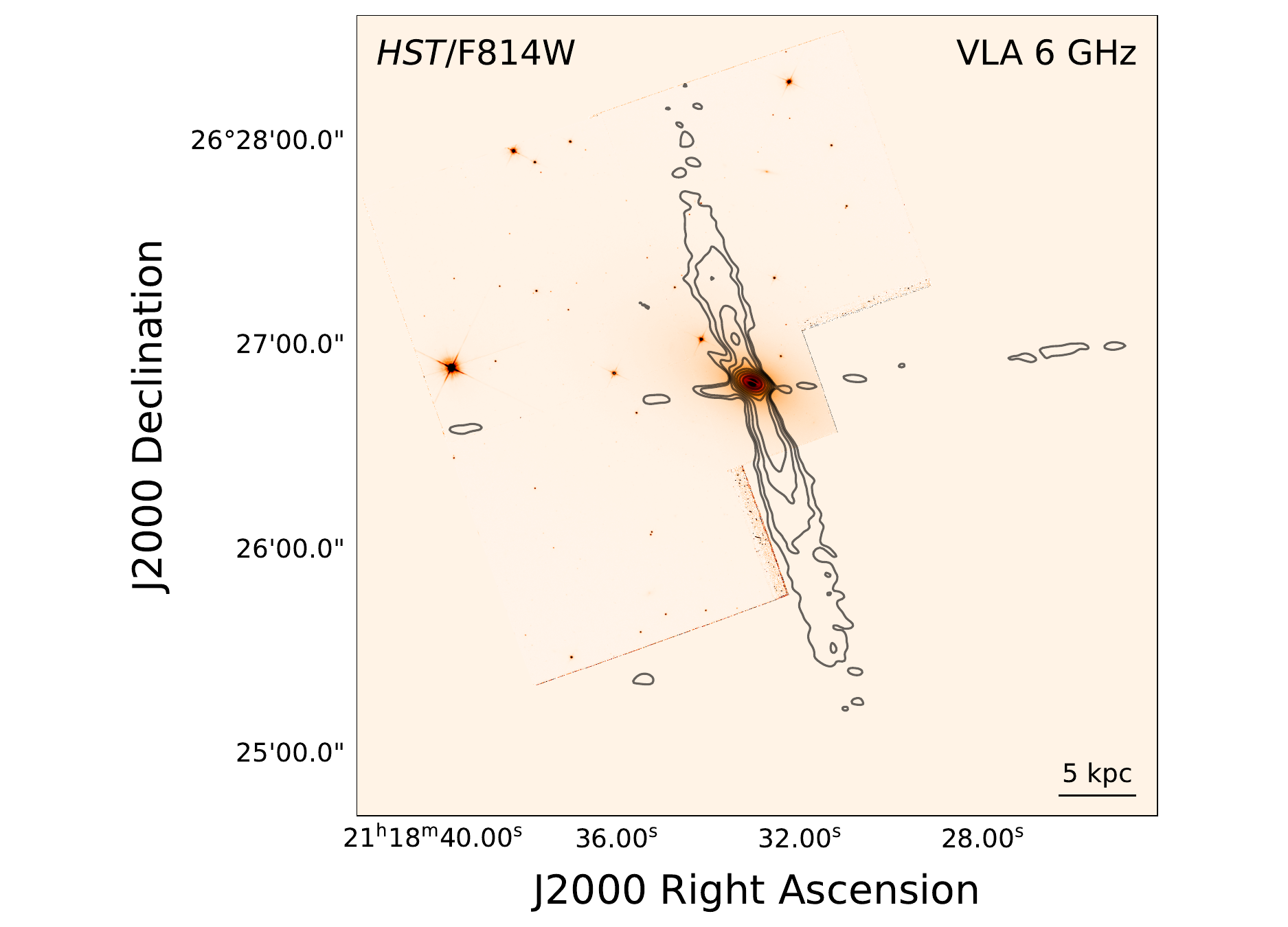}
\includegraphics[width=0.45\textwidth]{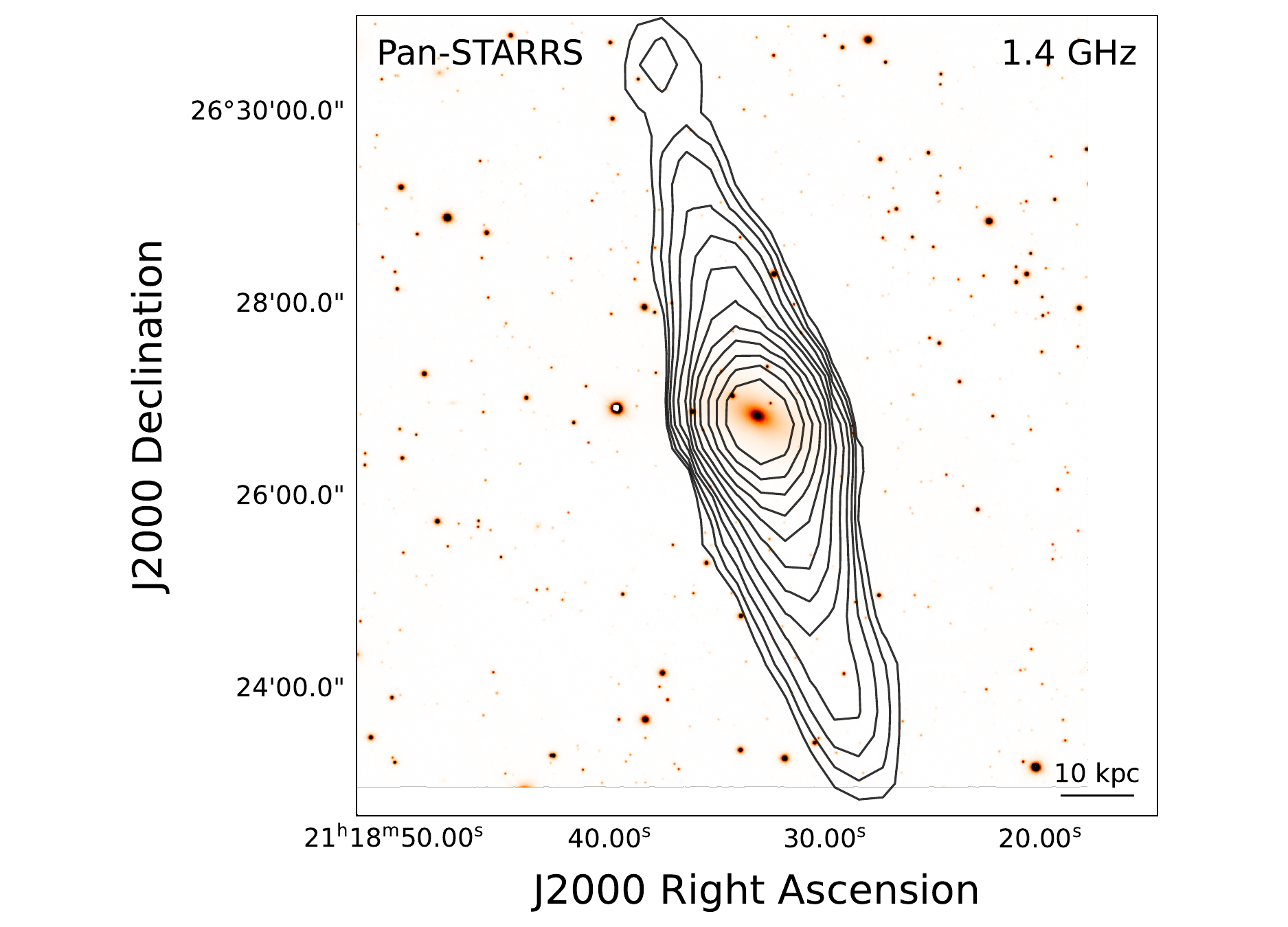}
\caption{{\bf Left:} Overlay of the HST/WFPC2 WFC F814W image with the high-resolution (beam size of 5$\arcsec$) C-band (6 GHz) VLA image clearly illustrates the jet at a position angle of $\sim$18.5$^\circ$, which corresponds to $\sim$45$^\circ$ relative to the dust plane. {\bf Right:} Overlay of the Pan-STARRS $r$-band image with the 1.4 GHz NVSS contours (beam size of 45$\arcsec$) highlighting the extent of the large-scale radio lobes.}
\label{fig:NVSS}
\end{figure*}

\subsection{The galaxy NGC~7052}\label{sec:ngc7052}

NGC~7052 (R.A., Decl. = $21^{\rm h}18^{\rm m}33\fs0380$,  ${+}26\degr26\arcmin49\farcs030$) is a massive galaxy with a stellar mass of \Mstar\ $\approx 5.6 \times 10^{11}$ \Msun\ \citep{Terrazas17, Pandya2017, Veale18, Gu22} and an effective radius of $R_e \approx 14\farcs7$ \citep{Ma14}. Its elongated, boxy outer structure suggests a history of significant merger events \citep{Nieto1991, Gonzalez-Serrano1992}. NGC~7052 is an isolated elliptical radio galaxy with a prominent core and well-defined radio jets \citep{Parma86, Morganti87}. It has been classified as a Fanaroff-Riley type I (FR I) galaxy, associated with the radio source B2 2116+26 \citep{Capetti2000, Capetti02, Donato04}.   The high-resolution (beam size of 5$\arcsec$) C-band (6 GHz) Very Large Array (VLA) image reveals that NGC~7052 hosts a jet with a position angle of 15$\degr$ relative to the dust lane, as seen in the HST image (left panel of Figure~\ref{fig:NVSS}). In addition, the NRAO VLA Sky Survey (NVSS) 1.4 GHz continuum observations (beam size of 45$\arcsec$), overlaid on an $r$-band Pan-STARRS image, show that the jet extends to at least 4 arcmin on both sides above and below the dust lane (right panel of Figure~\ref{fig:NVSS}).

Furthermore, infrared photometry from 2MASS $K_s$-band observations indicates a star formation rate of SFR $\approx 1.5$ \Msun\ yr$^{-1}$ \citep{Terrazas17} and a total luminosity of $L_K \approx 4 \times 10^{11}$ \Lsun\ \citep{Memola2009, Goulding16}. Additionally, \textit{Chandra} X-ray observations reveal a luminous AGN with an X-ray luminosity of $L_X \approx 2.2 \times 10^{41}$ erg s$^{-1}$ (0.5–2.0 keV) and a hot interstellar medium (ISM) with a temperature of $kT \approx 0.5$ keV \citep{Memola2009}. The estimated gas mass of the ISM is $2.2 \times 10^{9}$ \Msun, extending to a radius of 16 kpc ($\approx$48$\arcsec$).

The galaxy has a stellar velocity dispersion at the effective radius of $\sigma_e = 266 \pm 13$ \kms and at the central region of $\sigma_c = 298$ \kms, derived from the Mitchell/VIRUS-P IFS data \citep{Gultekin09, Veale18}. However, velocity dispersion of the ionized gas increases from $\approx$70 km s$^{-1}$ at the outer regions to $\approx$$300$ km s$^{-1}$ in its nucleus \citep{Bosch1995}.

\begin{figure}
    \centering
    \includegraphics[width=0.9\columnwidth]{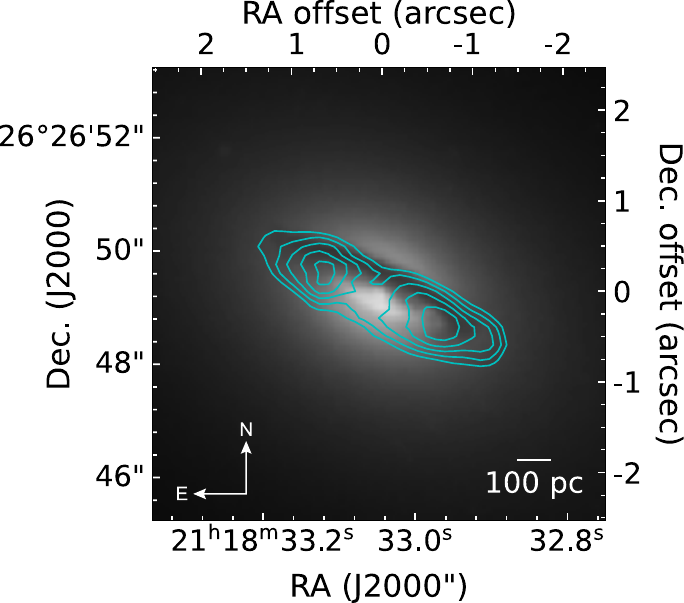}
    \caption{The ALMA \cotwo\ integrated intensity (contours), which are overlaid on the HST/WFPC2 WFC F814W image to highlight the coincidence of the gas distribution with the dust plane.}
    \label{fig:HST-image}
\end{figure}

Optical observations from the Hubble Space Telescope (HST) reveal the presence of a circumnuclear dust disk within the nucleus of NGC~7052 (Figure \ref{fig:HST-image}). In addition, observations from the Canada-France-Hawaii Telescope (CFHT) constrain the disk thickness to $h \approx 157$ pc (or $0\farcs5$) \citep{Juan96}, with a total dust mass of $M_{\rm dust} \approx 10^4$~\Msun\ \citep{Nieto1990}. This disk is nearly edge-on, with an inclination of $i = 74.8^\circ$ \citep{Smith21}, and is closely aligned with the galaxy's projected major axis. Further investigations using HST H$\alpha$ + \ion{N}{1} narrowband imaging have shown that the disk is also spatially coincident with ionized gas \citep{Bosch1995}. Analysis of the ionized gas kinematics led to an estimate of \Mbh\ $\approx 3.9 \times 10^8$ \Msun, after correcting for the adopted distance in the MASSIVE survey \citep{Marel1998}. More recent molecular gas dynamical modelling using high-resolution ALMA \cotwo\ observations (synthesized beam size of $0\farcs13\times0\farcs10$) yielded a refined \Mbh\ $=(2.5\pm0.3)\times10^9$ \Msun\ \citep{Smith21}.

Accurate \Mbh\ estimate via molecular gas kinematics requires that the angular resolution (or beam size) of the observations should be smaller than the projected radius of the SOI, which is given by $R_{\rm SOI} = GM_{\rm BH}/\sigma^2_e$  \citep[e.g.,][]{Nguyen20, Boizelle21}. Here, $G$ is the gravitational constant, $\sigma_e$ is the stellar velocity dispersion within the half-light radius. Using $\sigma_e = 266\pm13$~\kms~ reported by \citet{Gultekin09} and the central SMBH mass \Mbh\ =$(2.5\pm0.3)\times10^9$ \Msun\ by \citet{Smith21}, we estimate $R_{\rm SOI}=152\pm24$ pc $\approx$ 0\farcs45$\pm$0\farcs07. This radius is  $\approx$1.5 times larger than the synthesized beam size of our ALMA data used in this study. It is also worth noting that the \cotwo\ observations by \citet{Smith21} had nearly three times higher spatial resolution than our data, which would theoretically provide the most precise \Mbh\ measurement for NGC 7052 so far. Their final ALMA data cube was combined from one 7-m (observed in 2017) and three 12-m (observed in 2018 and 2019) configuration arrays.  However, their estimates of \Mbh\ and the mass-to-light ratio (\ml) were affected by the centrally resolved hole in the \cotwo-CND emission and the limited field of views (FoVs) of both their \cotwo-CND and their stellar mass model, as well as by the omission of point spread function (PSF) deconvolution when recovering the intrinsic central light in the HST image, the issues we will discuss further in Section \ref{sec:stellar-mass}.

\section{ALMA observation and reduction}\label{sec:data-reduction}

The \cotwo\ emission line was observed with ALMA during Cycle 7 using 39 12-m antennas in the C-6 configuration, as part of project 2019.1.00036.S (PI: Dieu Nguyen). The total on-source integration time was 2509 seconds on June 27, 2021. The baselines ranged from 15 m to 2.5 km, providing a primary beam with a full width at half maximum (FWHM) of $\approx$25\arcsec, a synthesized beam size of $\theta_{\rm FWHM}=0\farcs29 \times 0\farcs22$, a maximum recoverable scale of MRS $\approx4\farcs7$ (see ALMA Technical Handbook\footnote{\url{https://almascience.eso.org/proposing/technical-handbook}}).  The observations were conducted in Band 6 using four frequency division mode (FDM) spectral windows (SPWs), each spanning 2 GHz in width.  Each SPW is divided into 1920 channels along the frequency dimension with a channel width of 976.562 kHz ($\approx$1.3 \kms). One of these SPWs was centered on the \cotwo\ emission line at a rest frequency of $v_{\rm rest} = 230.538$ GHz, while the remaining three SPWs were used for continuum detection. During the reduction process, we employed two bright quasars, J2253+1608 and J2115+2933, to correct for bandpass and phase, respectively.

We calibrated the data using the \textsc{Common Astronomy Software Application} (\textsc{CASA}\footnote{\url{https://casa.nrao.edu/}}) package \citep{McMullin07}, version 6.5.2.26, provided by the ALMA Science Archive\footnote{\url{https://almascience.eso.org/processing/science-pipeline}}. The final data product was generated using a H{\"o}gbom deconvolver \citep{Hogbom74} and consists of a 128 $\times$ 128 pixel$^2$ image with a pixel scale of $0\farcs079$, ensuring proper sampling of the synthesized beam while reduce file size. 

\begin{table}
\centering
	\caption{Parameters of the continuum image and source.}
	\label{tab_continuum}
	\begin{tabular}{lc} 
		\hline\hline
		Image property & Value\\
		\hline
		Image size (pix$^2$) & $128\times128$\\
		Image size (\arcsec$^2$) & $10.1\times10.1$\\
		Image size (kpc$^2$) & $3.4\times3.4$\\
		Pixel scale (\arcsec$\,$pix$^{-1}$) & 0.079\\
		Pixel scale (pc$\,$pix$^{-1}$) & 26.5\\
		Sensitivity (Jy$\,$beam$^{-1}$) & 79\\
		Synthesised beam (\arcsec$^2$) & $0.29\times0.22$\\
		Synthesised beam (pc$^2$) & $97\times73$\\
		\hline
		Source property & Value \\
		\hline
		Right ascension & $21^{\rm h}18^{\rm m}33\fs0380$\\
		Declination & ${+}26\degr26\arcmin49\farcs030$\\
		Integrated flux (mJy) & $24.9\pm1.5$\\
		Deconvolved size (\arcsec$^2$) & $(0.37\pm0.01)\times(0.24\pm0.01)$\\
		Deconvolved size (pc$^2$) & $(124\pm 3)\times(81\pm9)$\\
		\hline
	\end{tabular}
\end{table}

\begin{table}
\centering
    \caption{\cotwo\ data cube properties}
    \label{tab_co}
    \begin{tabular}{lr}
      \hline  \hline
        CO image property & Value \\ \hline
        Spatial extent (pix$^2$) & 128 $\times$ 128 \\ 
        Spatial extent (\arcsec$^2$) & 10.1 $\times$ 10.1 \\ 
        Spatial extent (kpc$^2$) & 3.4 $\times$ 3.4 \\ 
        Pixel scale (\arcsec\,pix$^{-1}$) & 0.079 \\ 
        Pixel scale (pc\,pix$^{-1}$) & 26.5 \\ 
        Velocity range (km s$^{-1}$) & 4060 -- 5160 \\ 
        Channel width (km s$^{-1}$) & 10 \\ 
        Number of constraints & 76,014 \\
        Mean synthesised beam (\arcsec$^2$) & 0.31 $\times$ 0.23 \\ 
        Mean synthesised beam (pc$^2$) & 104 $\times$ 77 \\ 
        Sensitivity (mJy beam$^{-1}$ per 10 \kms) & 0.5 \\
        \hline
    \end{tabular}
\end{table}

\begin{figure*}
\centering
    \includegraphics[width=0.85\textwidth]{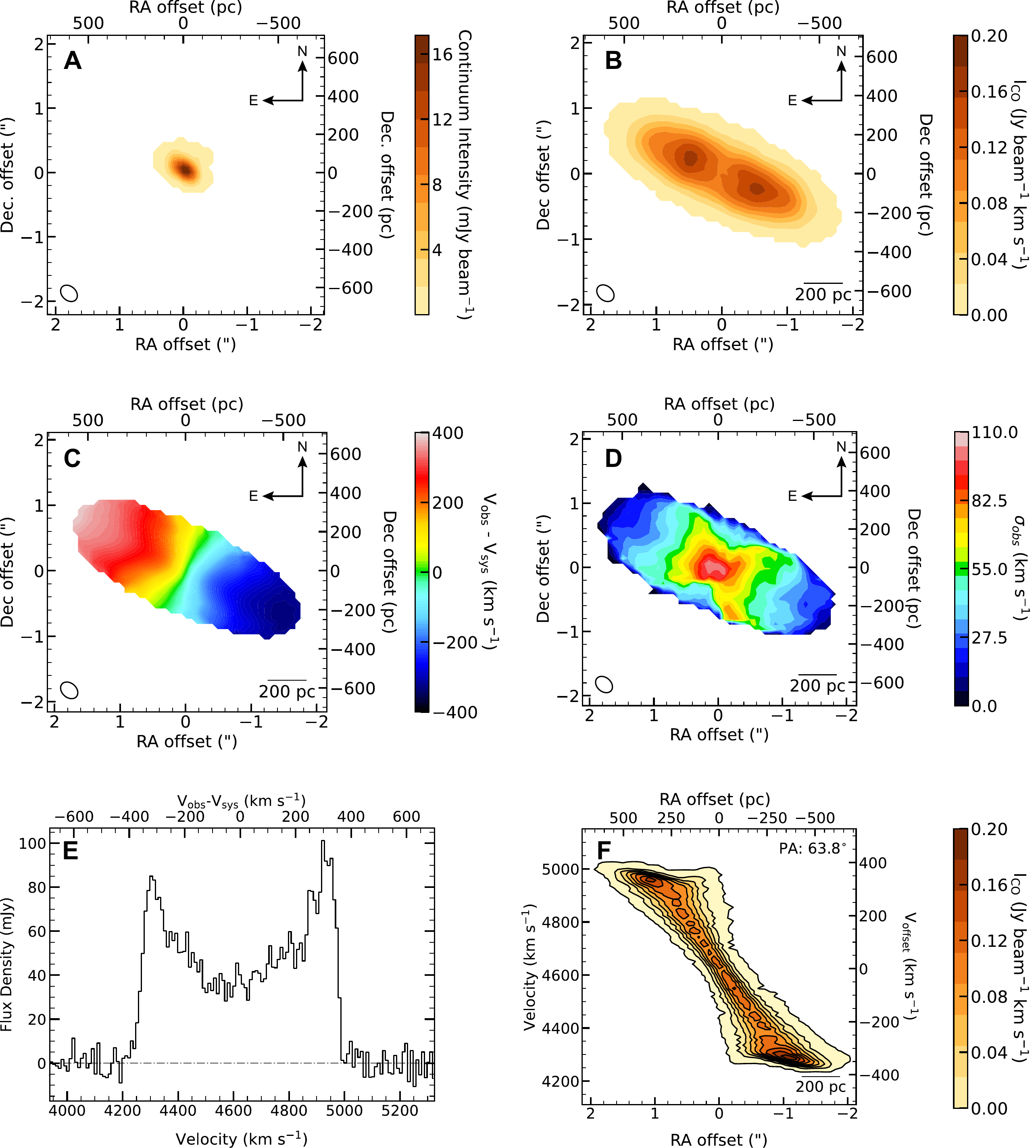}
    \caption{The 1.3 mm continuum emission (\textbf{Panel A}) and the \cotwo~emission moment maps of NGC 7052 derived from our ALMA data: integrated intensity (\textbf{Panel B}),  intensity-weighted mean LOS velocity (\textbf{Panel C}), intensity-weighted LOS velocity dispersion (\textbf{Panel D}). The synthesised beam of the observation was illustrated at the lower-left corner in each map as the black ellipse.  \textbf{Panel E}: the integrated spectrum extracted within a square box of $4\farcs4 \times 4\farcs4$ (or $1.5 \times 1.5$ kpc) with the horizontal dot-dashed line represent the zero flux level. \textbf{Panel F}: the PVD extracted along the major-axis  with an adopted systemic velocity $v_{\rm sys}=4610$ \kms\ and a position angle $\Gamma=63.8\degr$.}
    \label{fig:moment-maps}
\end{figure*}

\subsection{Continuum emission}\label{sec:cont}

We used the {\tt tclean} task in multifrequency synthesis mode \citep{Rau11} to generate the continuum image, utilizing the continuum SPWs and the line-free channels of the targeted SPW. We adopted Briggs weighting with a robust parameter of 0.5 to balance signal-to-noise ratio (S/N) and spatial resolution. The final continuum image reveals an unresolved source with a root-mean-square (RMS) noise of $\sigma_{\rm cont} = 79$ $\mu$Jy beam$^{-1}$ and a synthesized beam size of $\theta_{\rm FWHM, \rm cont} = 0\farcs29 \times 0\farcs22$ at a position angle (PA) of $\Gamma=41.4$\degr.

Panel A of Figure \ref{fig:moment-maps} shows the continuum image reveals a single source near the galaxy’s kinematic center (best-fitting SMBH position; see Section \ref{sec:bayesian-infer}), with an integrated flux density of $29.1 \pm 1.5$ mJy. This measurement is consistent (within the typical $\approx$10\% ALMA flux calibration uncertainty) with the value reported by \citet{Smith21}. The deconvolved size of the source, obtained by fitting a two-dimensional (2D) Gaussian with the \textsc{CASA} task {\tt imfit}, indicates that it is spatially unresolved (i.e., a point source). The properties of the continuum image and the detected continuum source are listed in Table \ref{tab_continuum}.

\subsection{\cotwo\ emission imaging}\label{sec:line}

After applying the continuum self-calibration to the line SPW, the \cotwo\ emission line was isolated in the $uv-$plane using the \textsc{CASA} {\tt uvcontsub}, which forms a continuum model from linear fits to line-free channels in frequency and then subtracts this model from the visibilities. We then created the \cotwo\ data cube using the {\tt tclean} task with Briggs weighting parameter of 0.5 and adopted a channel width of 10 km s$^{-1}$, which is optimal \citep{Davis14} and typically ultilized when modeling the kinematics of SMBHs \citep{Davis17, Davis20, Nguyen20}. The velocity dimension was computed in the restframe frequency of the \cotwo\ emission line (i.e., 230.538 GHz). The continuum-subtracted dirty cube was identified and cleaned interactively in regions of emission with a threshold of 1.5 times the RMS noise ($\sigma_{\rm RMS}$; measured from line-free channels). The properties of the final, self-calibrated and cleaned \cotwo\ data cube, which has a synthesized beamsize of $\theta_{\rm FWHM} = 0\farcs31 \times 0\farcs23$, are detailed in Table \ref{tab_co}.

\subsection{\cotwo\ emission moment maps}\label{sec:momentmaps}

The \cotwo\ emission extends from $\approx$4200 to 5000 \kms, with a systemic velocity of $v_{\rm sys}\approx4610$ \kms. We visualized our emission data using moment maps, including the zeroth moment (integrated intensity, panel B), first moment (intensity-weighted mean velocity, panel C), and second moment (intensity-weighted velocity dispersion, panel D), as shown in Figure \ref{fig:moment-maps}. These maps were generated directly from the \cotwo\ data cube using the masked moment method \citep{Dame11}.

First, we created a smoothed version of the original data cube by producing a copy of the original data cube, then applying a Gaussian spatial convolution with a dispersion of $\sigma = 1.5 \times \theta_{\rm FWHM}$ to that copy, followed by spectral smoothing using a Hanning window four times the channel width \citep{Smith21, Dominiak24}. We then applied a noise threshold of $0.5\sigma_{\rm RMS}$ in the unsmoothed cube (equivalent to $8\sigma_{\rm RMS}$ in the smoothed cube) to create a mask. This approach allowed us to suppress noise in the moment maps while ensuring the recovery of most of the flux in an optimization manner. All pixels in the smoothed cube (or the mask) that exceeded the threshold were selected, and the moment maps were subsequently generated using only these pixels from the unsmoothed cube. 

\begin{figure*}
\centering
\includegraphics[width=0.8\textwidth]{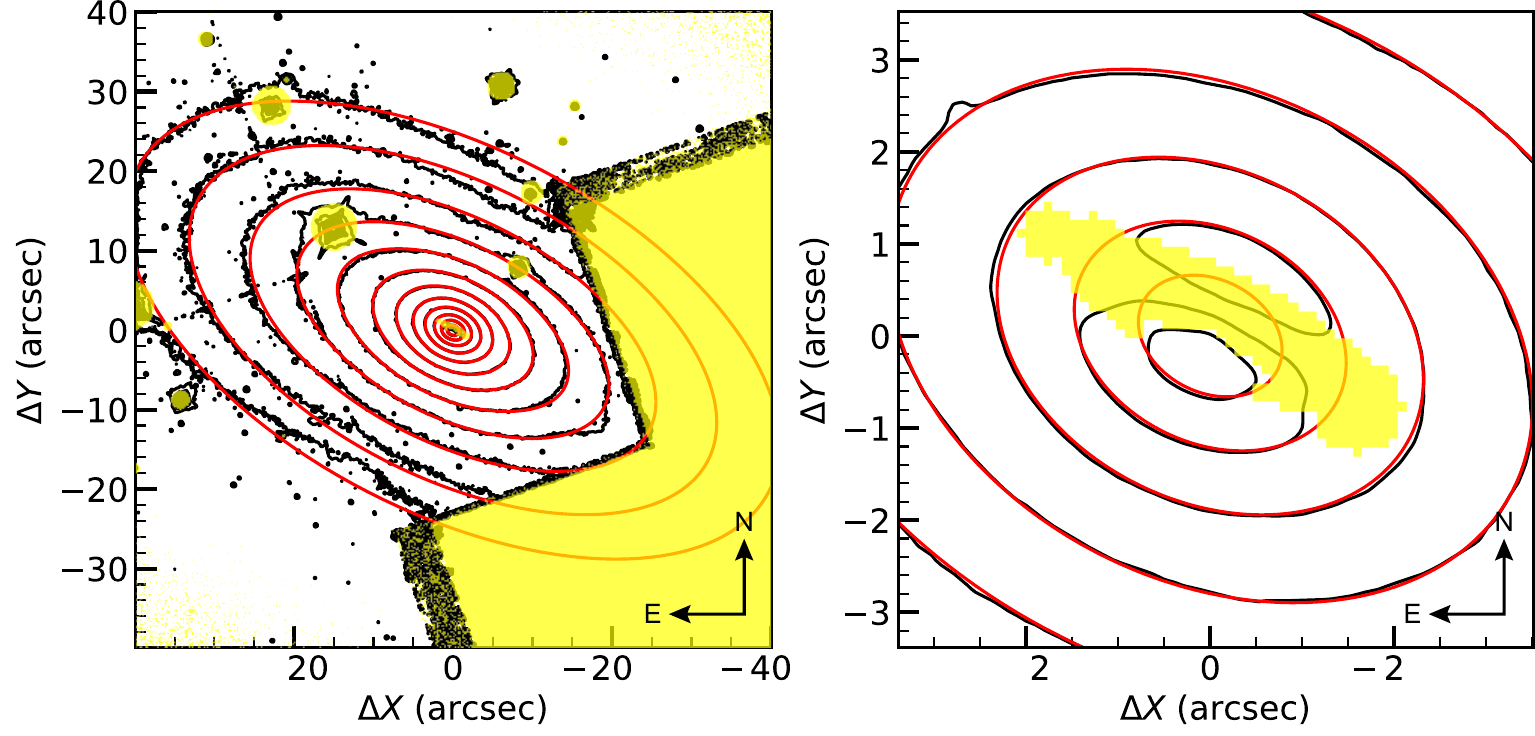}
\caption{The comparison of HST/WFPC2 WFC F814W images with its MGE model is shown in 2D surface brightness density within the FoV of $80\arcsec \times 80 \arcsec$ (\textbf{left}) and zoom into $7\arcsec \times 7\arcsec$ at the centre (\textbf{right}). Black contours represent the data, while red contours represent the model, highlighting their agreements at corresponding radii and contour levels. Yellow regions indicate the masked areas containing foreground stars, bad pixels, and the central dust disc.}
\label{fig:mge-fit}
\end{figure*}

The zeroth-moment map of integrated intensity reveals that the \cotwo\ disc extends up to $\approx$4$\arcsec$ along the major axis and 1.5$\arcsec$ along the minor axis, with a smooth variation and two intensity peaks located around $0\farcs5$ from the center. These peaks are likely a result of tidal acceleration induced by an external gravitational potential \citep{Smith21}. Additionally, \citet{Smith21} analyzed the ALMA data with nearly three times higher spatial resolution than ours and identified a small central hole in the \cotwo\ distribution, which could impact the dynamical modeling \citep{Dominiak24}. However, in our lower-resolution data, this feature is completely blurred, potentially providing a more reliable constraint on the SMBH mass because the synthesized beam size of our observations is $\approx$1.5 times smaller than the SMBH’s \Rsoi\ in NGC 7052, ensuring adequate spatial resolution for our analysis. Furthermore, the zeroth-moment map highlights the coincidence of the \cotwo\ emission with the dust disk (Figure \ref{fig:HST-image}), showing the alignment between molecular gas and the dust plane.

We estimated the total molecular gas mass using the ``CO-to-H$_2$ conversion factor'' of $X_{\rm CO} = 2 \times 10^{20}{\rm cm^{-2}(K\,km~s^{-1})^{-1}}$ \citep[or $\alpha_{\rm CO} = 4.3$ \Msun\ (K\,\kms\ pc$^{-1}$)$^{-1}$;][]{Bolatto2013}:
\begin{equation} 
\small
	\begin{split} 
M_{\rm gas} &= 1.05 \times 10^4\left( \frac{X_{\rm CO}}{2 \times 10^{20} \frac{\rm cm^{-2}}{\rm K\ km\ s^{-1}}} \right) \left( \frac{1}{1+z} \right) \left( \frac{S_{\rm CO}\Delta v}{\rm Jy\ km\ s^{-1}} \right) \left( \frac{D_L}{\rm Mpc} \right)^2 
	\end{split} 
\end{equation}
where $S_{\rm CO} \Delta v = 45$ Jy \kms\ is the integrated flux density derived from our data. In the local universe, with the redshift of NGC 7052 being $z \approx 0.015584$ \citep{Trager2000AJ119.1645}, we assumed a luminosity distance of $D_L = D\approx 69.3$ Mpc \citep{Ma14}. We also adopt a flux density ratio of unity between $^{12}$CO($2-1$) and $^{12}$CO($1-0$) \citep{Smith21}. Based on these assumptions, the total molecular gas mass is estimated to be $M_{\rm gas} \approx 2.2 \times 10^9$ \Msun. This result is higher than the measurement from \citet{Smith21}, which reported $M_{\rm gas} \approx 1.8 \times 10^9$ \Msun, but is consistent with the previously estimated mass of $2.3 \times 10^9$ \Msun\ derived from \coone\ emission using the Nobeyama 45-m single-dish telescope \citep{Wang1992}, likely providing the most reliable measurement of the gas distribution. These discrepancies arise from the higher spatial resolution of \citet{Smith21} and our ALMA observations compared to the lower-resolution measurement from the single-dish telescope.

The first-moment map of the intensity-weighted mean velocity reveals a regularly rotating, unwarped thin disc, with velocities reaching up to $\pm$400 \kms. Meanwhile, the second-moment map of the intensity-weighted velocity dispersion indicates a gas disc with moderate turbulence, where the velocity dispersion ranges from $20 \lesssim \sigma_{\rm LOS} \lesssim 70$ \kms\ outside the central boxy region of $0\farcs50\,{\rm (major\,axis)}\times0\farcs25\,{\rm (minor\,axis)}$. Within this central region, the velocity dispersion sharply increases to $\approx$100 \kms. This central peak in velocity dispersion is likely not intrinsic but rather a result of beam smearing and projection effects from a highly inclined disc ($i \gtrsim 50\degr$). The beam smearing effect is evident in the form of an ``X''-shaped structure at the center \citep{Davis17}, a phenomenon that typically arises when there is a steep intensity gradient across the beam \citep{Barth16b, Keppler19}. However, later in Section \ref{sec:results} the dynamical model fits give an intrinsic velocity dispersion of $16-17$ \kms, suggesting that is not just the central region being beam smearing matters, but over most of the disk the observed linewidths are dominated by beam smearing.

\subsection{Integrated spectrum \& Position-Velocity diagram}\label{sec:pvd}

Panel E of Figure \ref{fig:moment-maps} presents the integrated \cotwo\ spectrum of NGC 7052, extracted from a square aperture of $4\farcs4 \times 4\farcs4$ (1.5 $\times$ 1.5 kpc$^2$) to cover all line emission. The spectrum exhibits the characteristic ``double-horn'' profile, typical of a spatially resolved and rotating disc. The slight asymmetry, with the right horn has more flux than the left one, suggests a minor irregularity in the gas distribution, which is also apparent in the zeroth-moment map. This asymmetry could arise from a slight deficiency of gas in the redshifted component of the \cotwo\ CND (due to a specific gas morphology, e.g., a nuclear spiral).

Panel F of Figure \ref{fig:moment-maps} demonstrates the kinematic major-axis position–velocity diagram (PVD) of NGC 7052, extracted along a position angle (PA) of $\Gamma \approx 63.8\degr$, the best-fitting PA determined in Section \ref{sec:results}. The PVD was constructed by summing the flux within a 2-pixel-wide pseudo-slit (0$\farcs$158). When generating the PVD, we used a spatial Gaussian filter with a FWHM equal to that of the synthesized beam, rather than a larger uniform filter applied for the moment maps creation (Section \ref{sec:momentmaps}), to avoid masking out the central region. We then selected all pixels in the smoothed cube with intensities above $0.5\sigma_{\rm RMS}$ of the unsmoothed data cube. The PVD reveals a central rise in the LOS velocities within the innermost $\approx$$0\farcs5$ in radius, a characteristic kinematic signature of an SMBH. Our \cotwo\ CND kinematics are fully consistent with those reported in \citet{Smith21}, despite our ALMA observations having a spatial resolution lower by a factor of three. However, our observations recovered more flux and traced a more extended CND with $\approx$$2\arcsec$ on either side of the kinematic center, compare to that of extend only $\approx$$1\farcs5$ of \citet{Smith21}, which will help to separate the better constraints on \Mbh\ and \ml\ parameters in our dynamical models. 

\section{Improving the galaxy mass model}\label{sec:stellar-mass}

\subsection{\citet{Smith21} stellar mass model}\label{sec:smith21}

\citet{Smith21} utilized HST/Wide Field Planetary Camera 2 (WFPC2) / Planetary Camera (PC) F814W imaging (pixel scale of 0\farcs0455) taken on June 23, 1995 (Project ID: 5848, PI: van der Marel) to constrain the photometric model of NGC 7052. The dataset consists of three exposures totaling 1470 seconds. They derived the galaxy’s stellar light distribution using the Multi-Gaussian Expansion (\textsc{MGE}\footnote{v6.0.4: \url{https://pypi.org/project/mgefit/}}) algorithm \citep{Emsellem94}, implemented via the \textsc{Python} version of the {\tt mge\_fit\_sectors\_regularized} procedure \citep{Cappellari02}. During the fitting process, they masked the central pixels affected by dust in the northwestern region, which is considered as the foreground, and adopted a photometric zero-point of 20.84 mag \citep{Holtzman95} and an $I$-band Solar absolute magnitude of 4.12 mag \citep{Willmer18}, both in the Vega system. The final stellar mass model was obtained by multiplying the MGE representation by a constant \ml$_{\rm F814W}$, as presented in their section 4.1 and table 4. However, they did not deconvolve the HST image with its PSF to recover the intrinsic light. This led to an overestimation of the stellar light in a few central pixels, which significantly impacts the \Mbh\ estimates.

Given the MGE model from \citet{Smith21} and their best-fit \ml$_{\rm F814W} = 4.55$ \Msun/\Lsun, the total stellar mass of NGC 7052 is \Mstar\ $\approx 2.0 \times 10^{11}$ \Msun, which is significantly lower than the photometric estimate of \Mstar\ $\approx 5.6 \times 10^{11}$ \Msun\ reported by \citet{Veale18}. This discrepancy arises because the MGE model in \citet{Smith21} was constrained using only the Planetary Camera (PC) chip of the HST/WFPC2 image, which has a limited FoV of $40\arcsec \times 40\arcsec$. Although the gravitational influence of the SMBH is negligible at these large scales (and thus has little impact on \Mbh\ determination), the omission of outer stellar mass could lead to incorrect estimations of the \ml$_{\rm F814W}$ and the inclination of the molecular gas CND in the dynamical model \citep{Nguyen20}.

\begin{table}
\centering
\caption{Our improved HST/WFPC2 WFC F814W MGE model}    
\begin{tabular}{cccc} 
\hline\hline
$j$ &$\log \Sigma_{\star,j}$ (\Lsun\ ${\rm pc^{-2}})$ &$\log\sigma_j$ ($\arcsec$) &$q'_j=b_j/a_j$\\
(1) & (2) & (3) & (4)\\ 
\hline
1 & 3.75 & $-$0.34 & 0.75 \\
2 & 3.74 & $-$0.02 & 0.75 \\
3 & 3.54 &  0.33 & 0.75 \\
4 & 3.04 &  0.60 & 0.65 \\
5 & 2.79 &  0.68 & 0.75 \\
6 & 2.83 &  1.03 & 0.50 \\
7 & 2.40 &  1.45 & 0.51 \\
\hline
\end{tabular}
\parbox[t]{0.47\textwidth}{\small \textit{Notes:} (1) the Gaussian component, (2) the luminosity surface density, (3) the Gaussian dispersion along the major axis, and (4) the axial ratio.} 
\label{table:mge-table}
\end{table}

\begin{figure}
\centering
\includegraphics[width=0.95\columnwidth]{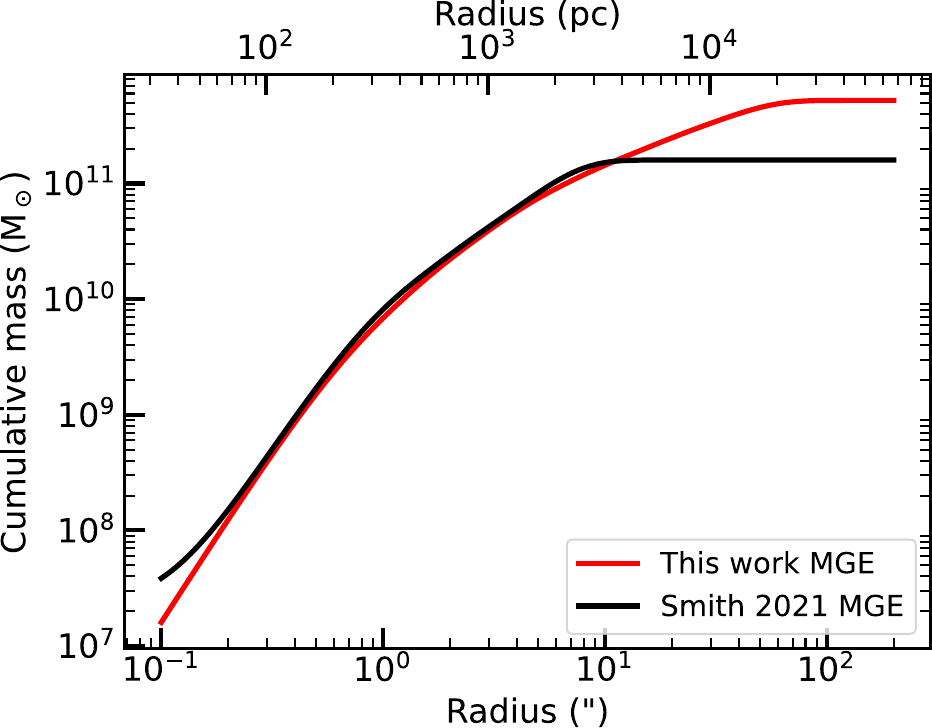}
\caption{Cumulative mass comparison between our MGE model constrained from the WFC  image and the model constrained by \citet{Smith21} with the PC image. Both constraints used the same HST observations.}
\label{fig:mge_compare}
\end{figure}

\begin{figure*}
    \centering
    \includegraphics[width=0.95\textwidth]{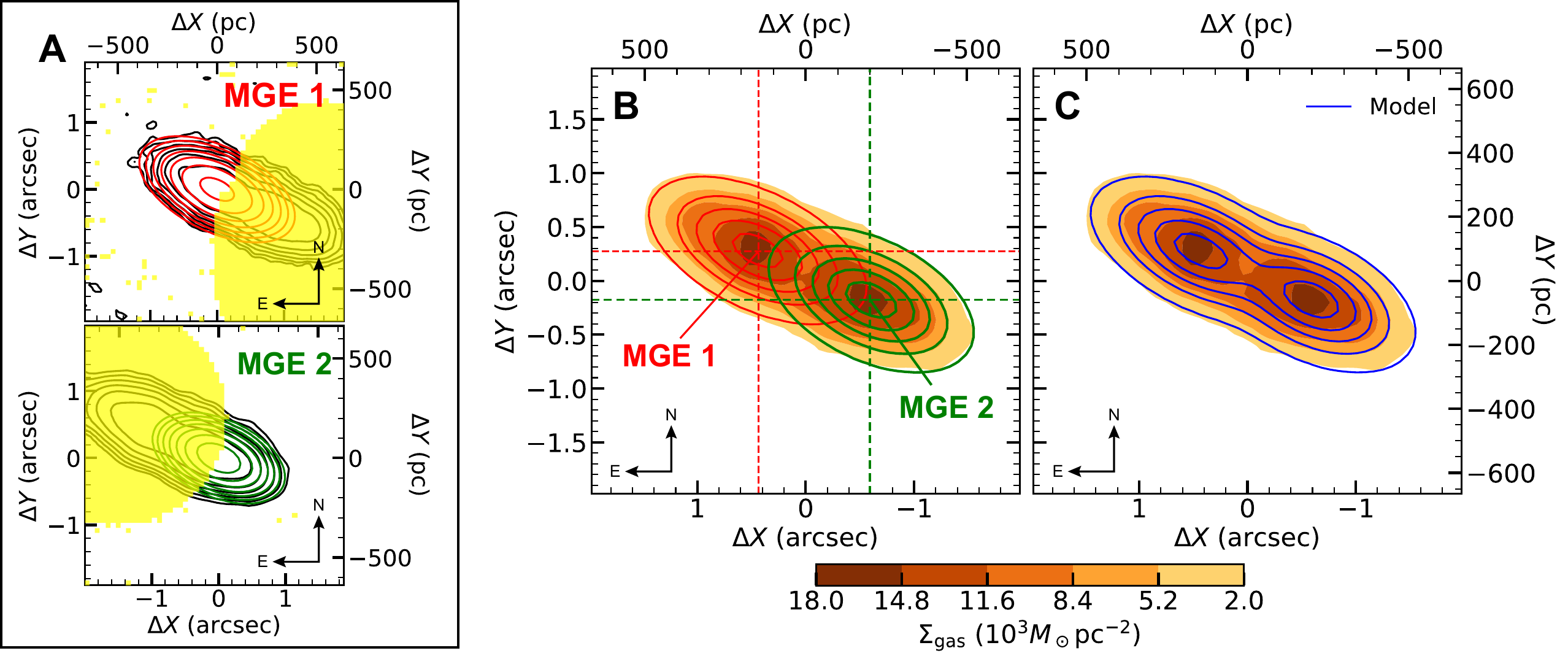}
    \caption{{\bf Panel A:} Our \textsc{MGE} fitting approach for the nuclear \cotwo\ gas emission, where we fit the two emission peaks separately. In each \textsc{MGE} fit, the yellow regions indicate masked areas, including one emission peak and the overlapping region, which are excluded from the fit.  {\bf Panel B:} A 2D superposition of the two \textsc{MGE} fits: \textsc{MGE} 1 (red) and \textsc{MGE} 2 (green) over the ALMA \cotwo\ observation. The dashed lines in the same colors mark the peak positions, which also represent the centers of \textsc{MGE} 1 and \textsc{MGE} 2. {\bf Panel C:}. Same as Figure \ref{fig:mge-fit}, showing a comparison between the ALMA surface mass density and the reconstructed model from \textsc{MGE} 1 and 2 at the same contour levels.}
    \label{fig:CO_mge}
\end{figure*}

\subsection{Our improved stellar-mass model}\label{sec:oursmass}

In this work, we rederived the photometric model of NGC 7052 using the HST/WFPC2 F814W image from the same project, principal investigator, and observation date. However, we utilized the Wide Field Camera (WFC) instead of the PC, providing a larger FoV of $80\arcsec \times 80\arcsec$ to capture the full extent of the galaxy, including both its central regions and outer edges. The image was retrieved from the Hubble Legacy Archive (HLA\footnote{\url{https://hla.stsci.edu/}}).

We estimated the sky background on the image by taking the median values from various squared boxes of $20 \times 20$ pixels$^2$ located away from light sources and in the regions beyond $40\arcsec$ in radius from the galaxy center. We then subtracted entire the image with median value to get a sky-subtracted image. 

To ensure accurate modeling, we first generated the HST/WFPC2 WFC PSF for the F814W filter using the \textsc{TinyTim} package\footnote{\url{https://github.com/spacetelescope/tinytim/releases/tag/7.5}} \citep{Krist11}. This software creates a model PSF based on the telescope instrument, detector chip, chip position, and filter used in the observations. To match the processing of the real HST observations, we generated three PSFs corresponding to three existing exposures, each positioned on a subsampled grid with sub-pixel offsets. These were designed using the same four-point box dither pattern as the original HST/WFPC2 WFC exposures. Next, we accounted for the effect of charge diffusion, where electrons leak into neighboring pixels on the CCD, by convolving each model PSF with the appropriate charge diffusion kernel. Finally, the three PSFs were combined and resampled onto a final grid with a pixel size of $0\farcs079$ using {\tt Drizzlepac}/\textsc{AstroDrizzle}\footnote{\url{https://www.stsci.edu/scientific-community/software/drizzlepac}} \citep{Avila12}.

Next, we created a mask to exclude the central dust lane, the bad/hot pixels, and the foreground stars using a point-source catalogue generated by \textsc{SExtractor}\footnote{\url{https://www.astromatic.net/software/sextractor/}}  \citep{Bertin96}. 

Our improved photometric model of NGC 7052 was derived from the HST/WFPC2 F814W sky-subtracted image, following the same procedure as \citet{Smith21}. However, in our approach, we provided the MGE fit with the HST masking image and performed a deconvolution using the PSF to recover the intrinsic light distribution of the galaxy in the form of an MGE model. Specifically, we first decomposed the PSF into an MGE, which was then used as input for the subsequent MGE fit of the F814W masked and sky-subtracted image, allowing us to derive a sum of 2D Gaussians that were convolved with the PSF MGE. This MGE can be analytically deprojected into a three-dimensional (3D) axisymmetric light distribution by assuming a free inclination ($i$). We summarized this spatially deconvolved MGE in Figure \ref{table:mge-table} and compared it with the F814W image in Figure \ref{fig:mge-fit}.

We converted this spatially deconvolved light-MGE into the galaxy mass model for NGC 7052 by assuming a constant \ml\ (taken from its best-fit constrained in Section \ref{sec:results}) and ignoring the contribution of dark matter in the central region \citep{Cappellari13a} due to the compact size of the \cotwo\ gas dics. 

For a sanity check, we compared our derived stellar mass model with the one from \citet{Smith21} in Figure \ref{fig:mge_compare}. Our model predicts slightly less stellar mass at the center (four central pixels) but includes more mass in the outer regions, although it is totally consistent with the mass estimate from \citet{Smith21} up to the radius of 10\arcsec. This is because we accounted for the PSF effect on intrinsic light and used a wider-field image to capture all the stellar light from NGC 7052. The missing mass in the extended region is unlikely to affect the \Mbh. However, the slight decrease in central light/mass and adding more mass in the extended region could lead to a higher \Mbh\ and a lower \ml\ measurement (see Section \ref{sec:results}) compare to the estimate from \citet{Smith21}, respectively.

\subsection{Interstellar medium mass model}\label{sec:ism}

Given the compact yet significant molecular gas mass of $M_{\rm gas} \approx 2.2 \times 10^9$
\Msun\ of the \cotwo-CND at the center of NGC 7052 (which is comparable to the \Mbh), its contribution cannot be ignored when modeling the galaxy total mass to estimate \Mbh\ dynamically. We, therefore, used the MGE formalism to decompose this molecular mass distribution into individual Gaussian components. This process involved converting the zeroth-moment map of the \cotwo\ integrated intensity (panel B of Figure \ref{fig:moment-maps}) into a molecular gas mass map, which was then input into the MGE algorithm. However, due to a slight attenuation of the \cotwo\ flux at the center and the presence of two emission peaks elongated along the galaxy's major axis (i.e., northeastern--southwestern orientation as seen in Figure \ref{fig:HST-image} and panel B of Figure \ref{fig:moment-maps}), we performed two separate MGE fits for the molecular gas mass and later combined the results.

The first MGE fit, referred to as MGE 1, models only half of the northeastern peak, masking the other half and completely excluding the southwestern peak. Similarly, the second MGE fit, called MGE 2, models only half of the southwestern peak while masking the other half and entirely excluding the northeastern peak. For each MGE fit, we determined the Gaussian center using the {\tt find\_galaxy} routine within the \textsc{MGE} framework. Once the Gaussian centers for both emission peaks were identified, we performed the \textsc{MGE} decompositions following the procedure described in Section \ref{sec:smith21}. However, we skip deconvolution the molecular gas mass map with the observational beam size, as this was already accounted for when generating the \cotwo\ zeroth-moment map. Each \textsc{MGE} fit for the emission peaks resulted in a single Gaussian component with specific parameters, which are listed in Table \ref{table:mge-ism} and were kept fixed in the total mass model of NGC 7052 (i.e., with no free parameters) as showed in the panels A and B of Figure \ref{fig:CO_mge}.

\begin{figure}
    \centering
    \includegraphics[width=0.47\textwidth]{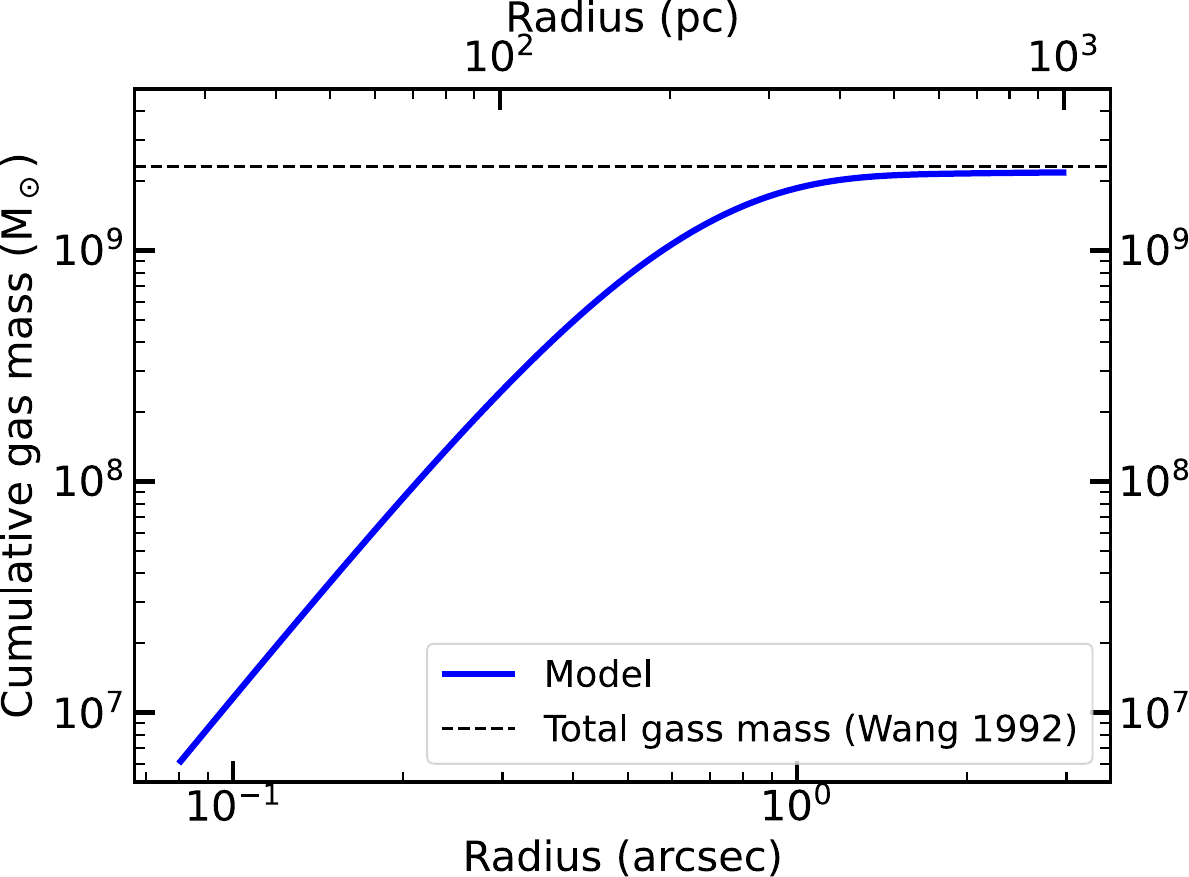}
    \caption{The cumulative molecular gas mass reconstructed from our gas MGE models is validated against the estimate from the \cotwo\ emission observed with the Nobeyama 45-m single-dish telescope \citep{Wang1992}.}
    \label{fig:total_gas_mass}
\end{figure}

We reconstructed the 2D molecular gas mass map using these two MGE 1 and 2 based on the following equation:
\begin{equation*}
\small
    \Sigma_{\rm gas}(x,y) = \sum_{j=\overline{1,2}} \frac{\Sigma_{{\rm ISM,}j}}{2\pi\sigma_{j}^2q_j^\prime} \exp\left(-\left[\frac{(x-x_{\rm peak})^2}{2\sigma_j^2} + \frac{(y-y_{\rm peak})^2}{2q_j^{\prime2}\sigma_j^2}\right]\right),
\end{equation*}
which can be compared directly to the data as shown in the panel C of Figure \ref{fig:CO_mge} at the same contour levels for both data and the reconstructed model. It appears that our reconstructed molecular gas mass model describes the data well. Figure \ref{fig:total_gas_mass} shows the accumulative gas mass calculated using our MGE 1 and 2 models, which predicts the enclosed gas mass within 2\arcsec\ that is consistent with what found by \citet{Wang1992}.

Given the compactness of the continuum emission shown in Panel A of Figure \ref{fig:moment-maps}, which is much smaller than the size of the \cotwo-CND, and the negligible dust mass inferred from the HST optical image (discussed in Section \ref{sec:ngc7052}), we  ignored  the dust mass distribution in our galaxy mass model.

\begin{table}
\centering
\caption{Gas MGE model}  
\footnotesize
\begin{tabular}{cccc} 
\hline\hline
$j$&$\log \Sigma_{\rm ISM,}$$_j$ (\Msun\ ${\rm pc^{-2}})$&$\log\sigma_j(\arcsec)$&$q_j=b_j/a_j$\\
(1) & (2) & (3) & (4)\\ 
\hline
1& $(x_{\rm peak}, y_{\rm peak}) $ & $= (+0\farcs 434, +0\farcs 277)$ &\\
& 4.225 & $-$0.375 & 0.55 \\
\hline
2& $(x_{\rm peak}, y_{\rm peak})$ & $ = (-0\farcs 593, -0\farcs174$) &\\
& 4.195 & $-$0.295 & 0.50 \\
\hline
\end{tabular}
\parbox[t]{0.47\textwidth}{\small \textit{Notes:} Same as Table \ref{table:mge-table}.} 
\label{table:mge-ism}
\end{table}

\subsection{Galaxy mass model}\label{sec:galaxy-mass}

Our galaxy mass model for NGC 7052 will be the combination of the central point mass representing the SMBH, our improve stellar mass, and the interstellar medium mass (i.e., the total gas mass). This galaxy mass model will be used to compute the circular velocity curve resulting from the gravitational potentials of those components.

\section{Dynamical Modelling}\label{sec:dynamical-models}

\subsection{\textsc{KinMS} tool} \label{sec:kinms}

To measure the \Mbh\ of NGC~7052, we analyzed our ALMA gas kinematics using the publicly available \textsc{Python} version of the \textsc{KINematic Molecular Simulation} tool \citep[\textsc{KinMS}\footnote{\url{https://github.com/TimothyADavis/KinMSpy}};][]{Davis13Nature}. This tool has been extensively used in the WISDOM \citep[e.g.,][]{Davis17, Onishi17, Ruffa23} and MBHBM$_\star$ projects \citep{Nguyen20, Nguyen22}. Below, we summarize the methodology and discuss the specifics of the NGC~7052 modelling.

\textsc{KinMS} model generates a mock data cube by simulating the gas distribution (Section \ref{sec:gas-dynamics}) and kinematics while accounting for observational effects such as beam smearing, spatial and velocity binning, and LOS projection. This simulated data cube is then directly compared to the observed data cube to determine the best-fitting values and uncertainties of the model's parameters using a Markov Chain Monte Carlo (MCMC) $\chi^2$ minimization routine and a set of priors in a Bayesian inference framework (Section \ref{sec:bayesian-infer}). When constructing the \textsc{KinMS} model, we assumed that the \cotwo\ gas circulates around the galaxy center in circular orbits, influenced by the combined gravitational potentials of the SMBH, stars, and gas/dust distributions. Among these mass components, the SMBH is treated as a central point mass. The key improvements in this study compared to \citet{Smith21} are: (1) the use of our newly derived stellar mass model (Section \ref{sec:oursmass}) and (2) the inclusion of the total molecular gas mass, which is comparable to the \Mbh\ of NGC~7052 and cannot be neglected, as was assumed in \citet{Smith21}. Thus, we calculated the gas velocity as a function of radius using the {\tt mge\_circular\_velocity} routine from the \textsc{Jeans Anisotropic Modeling} \citep[\textsc{JAM}\footnote{v7.2.4: \url{https://pypi.org/project/jampy/}};][]{Cappellari08} framework.

Our adopted \textsc{KinMS} model (Section \ref{sec:SkySampler}) matches the observations by fitting a set of free parameters. The first two are the kinematic center coordinates ($x_{\rm cen}$ and $y_{\rm cen}$), which define the SMBH location relative to the data cube’s phase center or the peak of the continuum emission identified in Section \ref{sec:cont}. This assumption is typically valid, as the offset between the kinematic and morphological centers is often much smaller than the synthesized beam size. The third parameter is the systemic velocity of the gas disc ($v_{\rm sys}$) or, equivalently, the velocity offset ($v_{\rm off}$) if $v_{\rm sys}$ has already been subtracted. The fourth parameter is the integrated intensity scaling factor ($f$) of the gas distribution described either by the \textsc{SkySampler}\footnote{\url{ https://github.com/Mark-D-Smith/KinMS- skySampler}} tool \citep{Smith19} in Section \ref{sec:SkySampler} or by an analytically axisymmetric function in Section \ref{sec:axisymmetric_function}. In addition to these, the \textsc{KinMS} model includes three parameters related to the CND morphology: the inclination angle ($i$), the position angle ($\Gamma$), and the turbulent velocity dispersion of the gas disc ($\sigma_{\rm sys}$). The final two parameters are the \Mbh\ and the \ml.  Thus, our adopted \textsc{KinMS} models consist of nine free parameters, as listed in  Table \ref{tab:mcmc-results}.

\begin{figure}
    \centering
    \includegraphics[width=0.95\columnwidth]{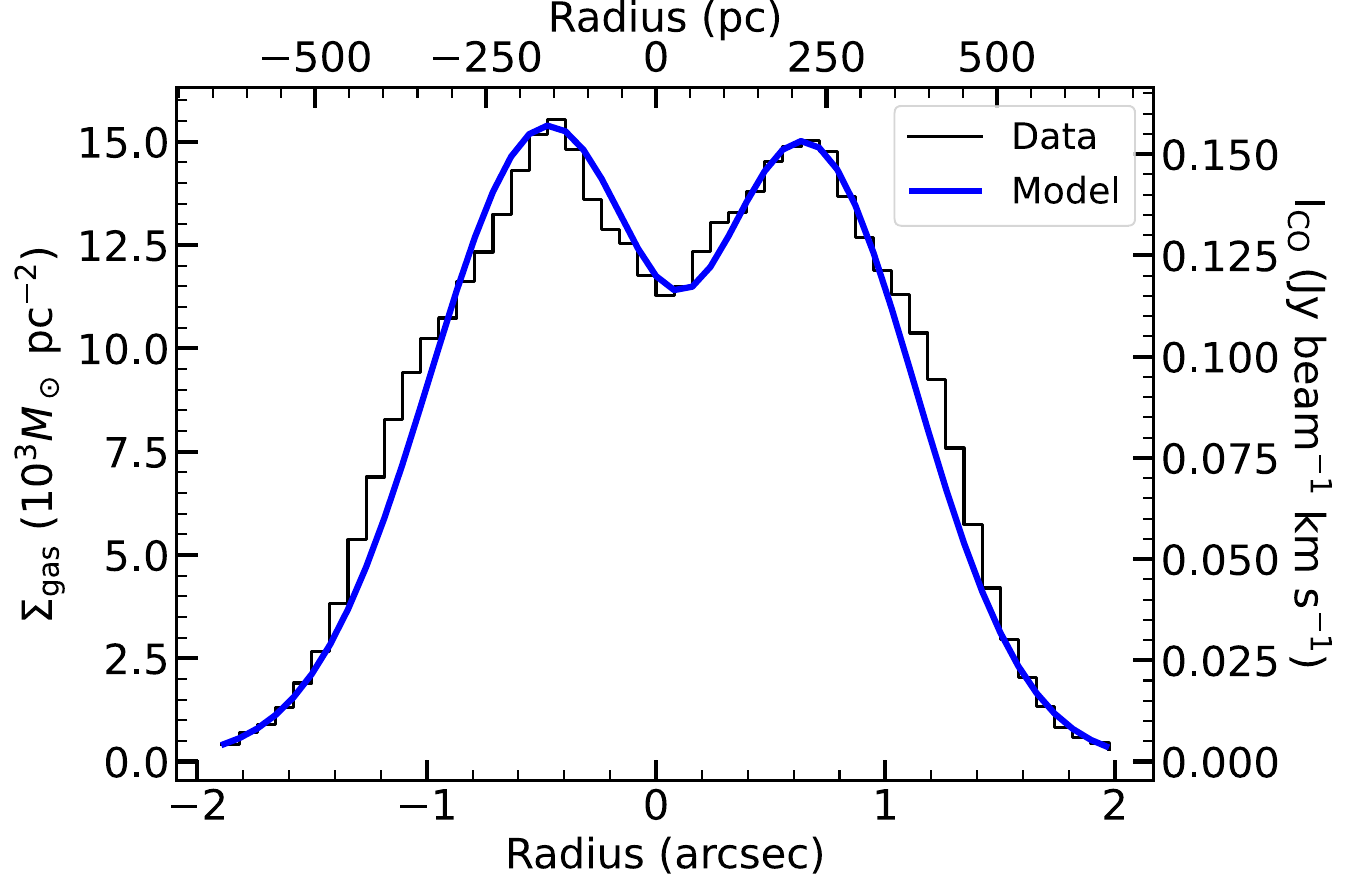}
    \caption{The morphological distribution of \cotwo\ gas is shown along a cut through the major axis, passing through the center and the two emission peaks of the integrated intensity map of NGC 7052. Our ALMA data is plotted in black, while our analytically axisymmetric model of two center-offset Gaussian functions for the gas surface brightness, is shown in green.}
    \label{axisymmetric_function}
\end{figure}

\begin{figure*}
 \centering
    \includegraphics[width=0.95\textwidth]{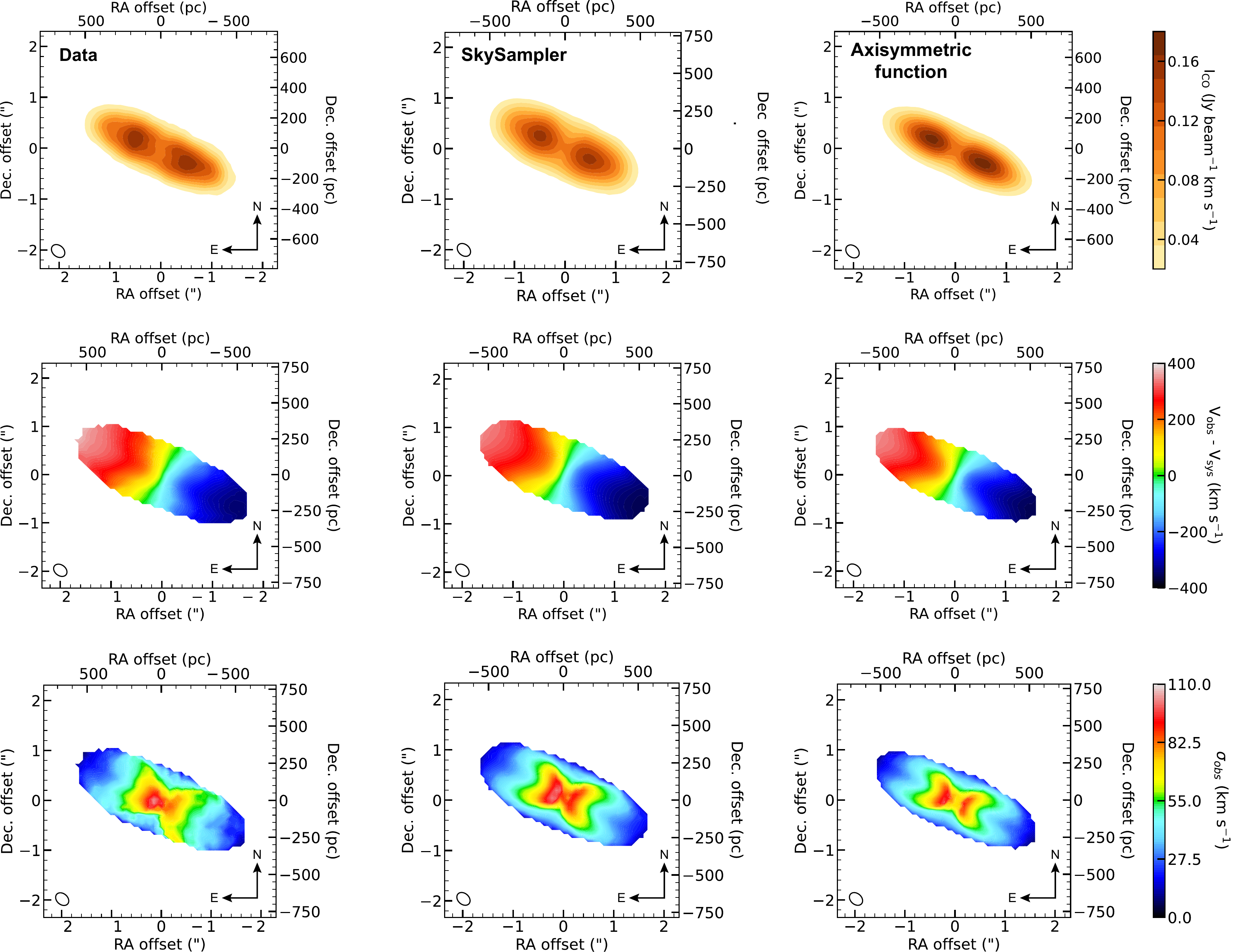}
    \caption{The comparison of \cotwo\ moment maps among our ALMA data (\textbf{left}), the best-fitting \textsc{KinMS} model applied the \textsc{SkySampler} tool (\textbf{middle}) and the best-fitting \textsc{KinMS} model assumed a sum of two center-offset Gaussians  (\textbf{right}) to spatially describe the \cotwo\ distribution, shows a strong agreement between the two. For each moment map (data versus model), we used the same colorbar. Other demonstrations are all as similar as Figure \ref{fig:mge_compare}.}
    \label{fig:data-model-momentmaps}
\end{figure*}

\subsection{Gas distribution} \label{sec:gas-dynamics}

As our gas modeling approach fits the full 3D ALMA cube, it requires a description of the gas distribution, which is then scaled by the integrated intensity scaling factor ($f$; see Section \ref{sec:kinms}) to match the observed data cube. In this study, we model the gas distribution using either the CLEAN components derived from the data cube with the \textsc{SkySampler} tool (see Section \ref{sec:SkySampler}) or a smooth, analytically defined axisymmetric function (see Section \ref{sec:axisymmetric_function}).

\subsubsection{\textsc{SkySampler}} \label{sec:SkySampler}

Given the  \cotwo\ emission of NGC~7052 exhibits two distinct peaks along its major axis (Panel B of Figure \ref{fig:moment-maps}), we employed the \textsc{SkySampler} approach \citep{Smith19} to derive the CLEAN gas component directly from the ALMA data. 

Since \textsc{SkySampler} constructs molecular gas clouds based on the CLEAN components of the data cube, this method effectively fits only the kinematics of the molecular gas while excluding assumptions about its spatial distribution. The model is thus constrained by a single free parameter, the total flux scaling factor ($f$), which rescales the entire cube. The clouds generated by \textsc{SkySampler} are assigned only relative intensities, so $f$ ensures that the model accounts for the full observed gas distribution. Additionally, since the CLEAN components do not include residuals from the deconvolution process, their total flux is slightly lower than that of the original cube. The parameter $f$ compensates for this discrepancy, allowing the model to recover the missing flux. Ideally, $f$ should correspond to the integrated flux within the fitted region of the data cube.

We uniformly sampled the CLEAN components with $10^6$ gas particles, ensuring that they precisely replicate the observed CO surface brightness distribution when convolved with the synthesized beam, using the {\tt SampleClouds} routine. The particles were then deprojected from the sky plane to the intrinsic galaxy plane using the {\tt transformClouds} algorithm, assuming a position angle of $\Gamma = 64\degr$ and an inclination of $i = 75\degr$. 

Although Panel D of Figure \ref{fig:moment-maps} shows spatial variations in the gas velocity dispersion, these variations primarily result from beam smearing and projection effects in a highly inclined disc. Therefore, we assumed a constant \siggas\ in our \textsc{KinMS} model. Additionally, we adopted a thin-disc approximation for the gas distribution, setting the disc scale height to zero in our \textsc{KinMS} model.

\subsubsection{Analytically axisymmetric function} \label{sec:axisymmetric_function}

An alternative approach to modeling the gas distribution is to use a smooth, analytically axisymmetric function, such as a Gaussian function \citep[e.g.,][]{Nguyen20} or an exponential disk \citep[e.g.,][]{Smith19, Dominiak24}. Given the morphology of the \cotwo-CND, which is well described by the sum of two center-offset Gaussian functions (as discussed in Section \ref{sec:ism}), we also use two simple Gaussians without deprojection (fixed $q_j=1$) as shown in Figure \ref{axisymmetric_function}. In this approach, only the amplitude parameter ($f$) is allowed to vary, maintaining its same interpretation as discussed in Section \ref{sec:SkySampler}. 

\begin{figure*}
\centering
    \includegraphics[width=0.85\textwidth]{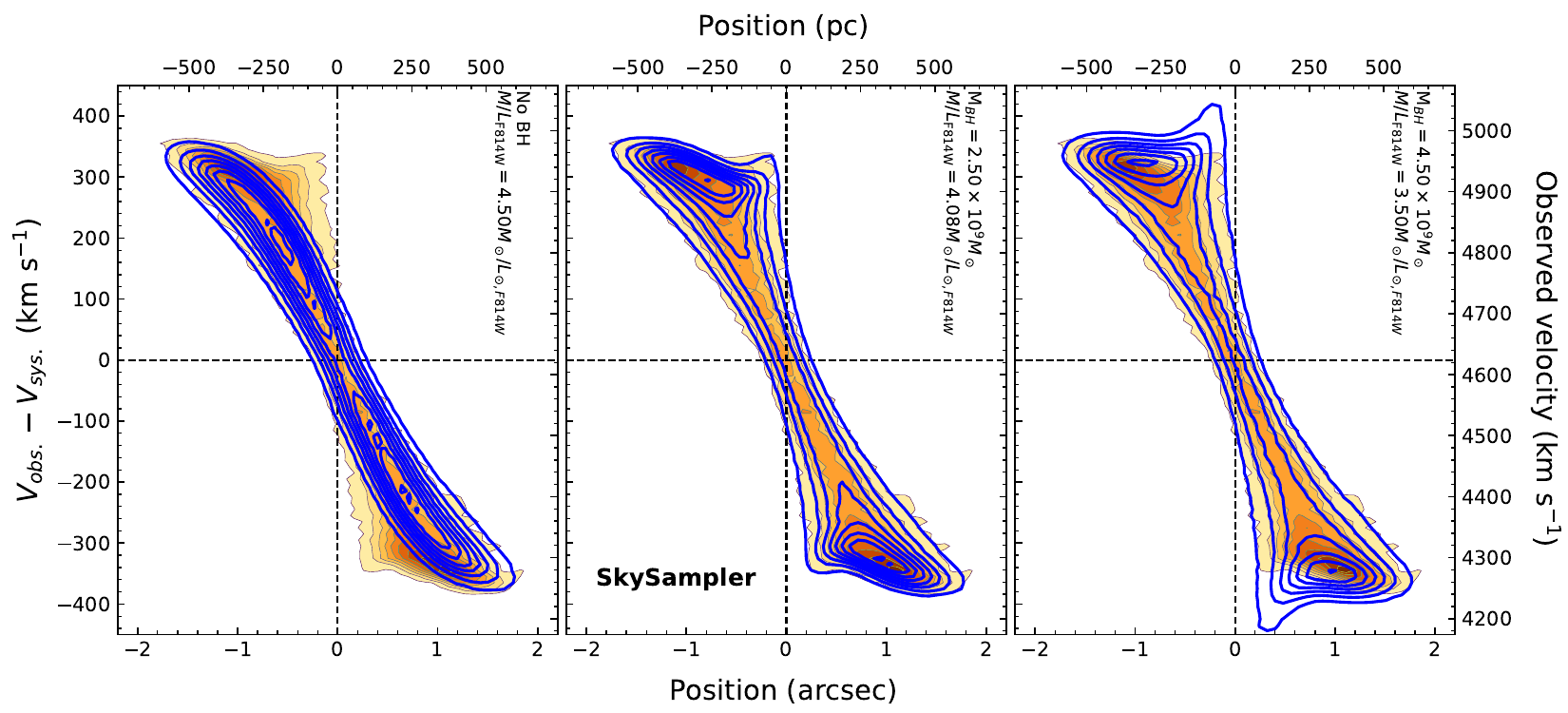}
    \hspace{1mm}\includegraphics[width=0.85\textwidth]{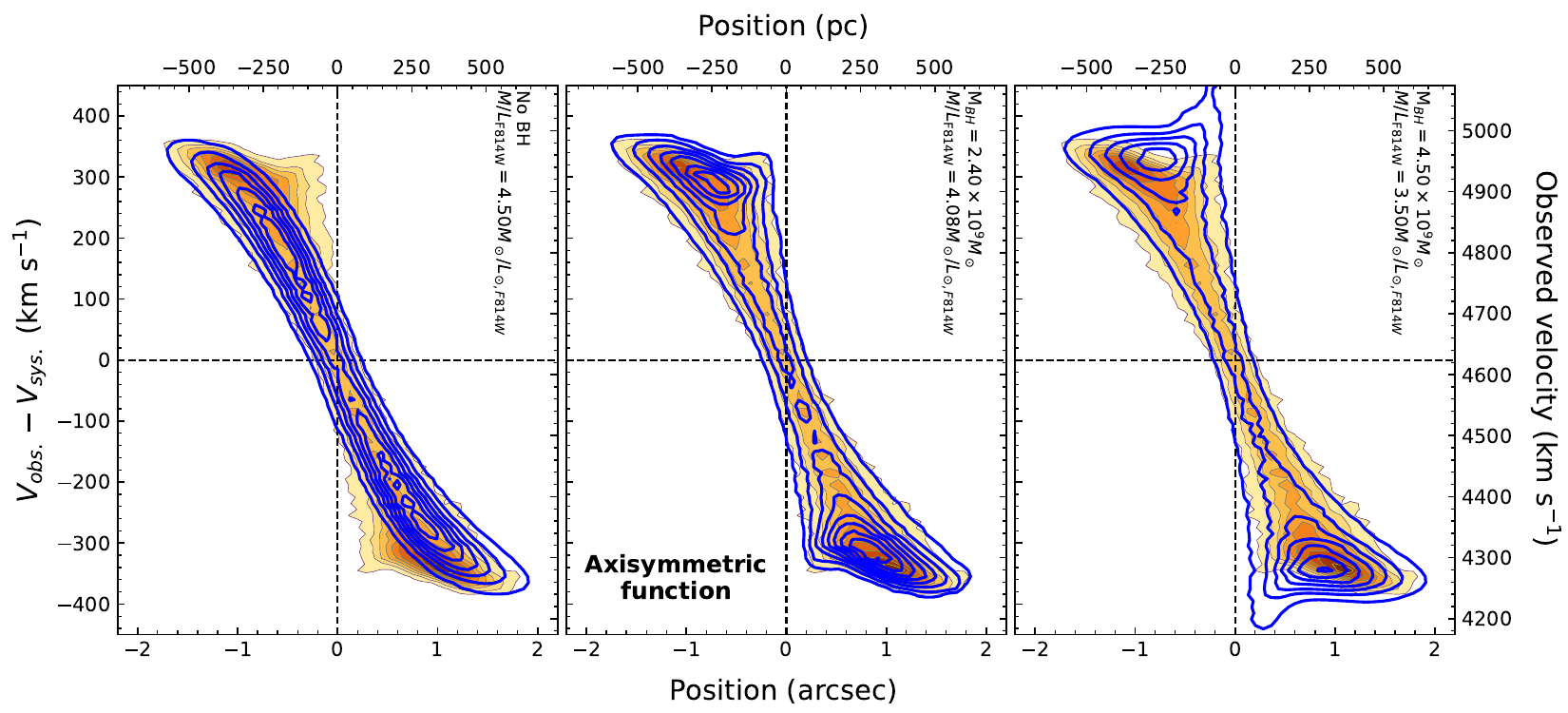}
    \caption{{\bf Upper-row panels:} The PVDs compare our ALMA observations of the \cotwo\ emission (orange-filled contours) with different \textsc{KinMS} models (blue contours), which assumed the gas distribution with the \textsc{SkySampler} tool. These models are extracted along the galaxy's major axis at a position angle of $\Gamma = 63.9^\circ$ and correspond to three different central SMBH masses: a model without a black hole (\textbf{left}), the best-fitting SMBH from this study (\textbf{center}), and an overly large SMBH (\textbf{right}). The SMBH mass and \ml$_{\rm F814W}$ for each case are indicated in the top-right corner of each panel. The black dashed lines mark the dynamical center, defined as the peak of the 1.3 mm continuum emission (Section \ref{sec:cont}). The intersection of these lines represents the systemic velocity ($v_{\rm sys}$) of the galaxy, shown on the velocity scale to the right.   {\bf Lower-row panels:}  The same \textsc{KinMS} models, which assumed the gas surface brightness distribution with an analytically axisymmetric function as the sum of two center-offset Gaussians.}
    \label{fig:compare-pvd}
\end{figure*}

\begin{figure*}
    \centering
    \includegraphics[width=0.85\columnwidth]{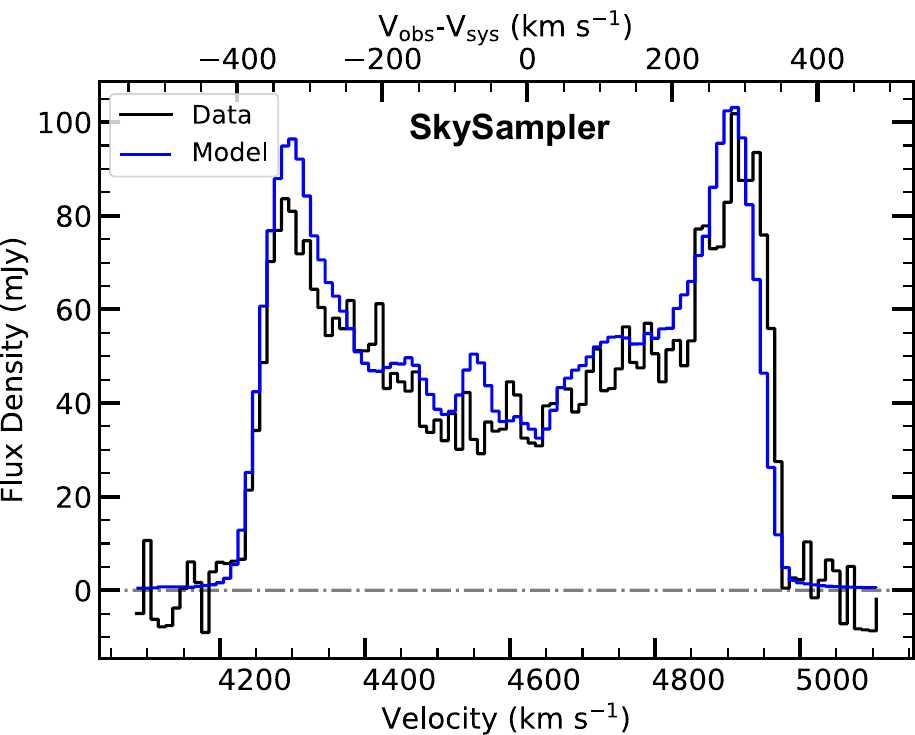}
     \hspace{8mm}\includegraphics[width=0.85\columnwidth]{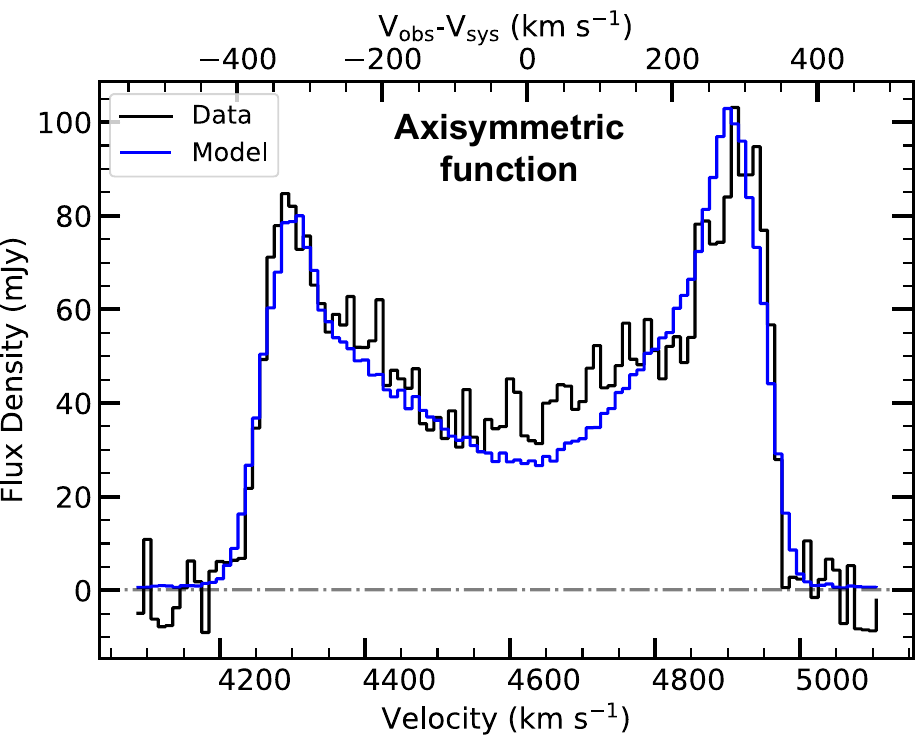}
    \caption{Overlaid of our ALMA \cotwo\ integrated spectrum (black), which is shown in panel E of Figure \ref{fig:moment-maps}, with our best-fitting \textsc{KinMS} models (red).}
    \label{fig:overlaid-spec} 
\end{figure*}

\subsection{Bayesian inference and priors} \label{sec:bayesian-infer}

The adaptive Metropolis algorithm \citep{Haario01}, implemented within a Bayesian framework using the \textsc{ADAMET}\footnote{v2.0.9: \url{https://pypi.org/project/adamet/}} package \citep{Cappellari13a}, was employed in the \textsc{KinMS} model for this analysis to constrain the best-fitting parameters and estimate their associated uncertainties from ALMA observations. The MCMC chains consisted of $10^5$ iterations, with the initial 20\% discarded as a burn-in phase. The remaining 80\% of the iterations were used to construct the full probability distribution function (PDF). The best-fit parameters were identified as those corresponding to the highest likelihood within the PDF, while statistical uncertainties were determined at the 1$\sigma$ (16–84\%) and 3$\sigma$ (0.14–99.86\%) confidence levels (CL). Given that the \Mbh\ parameter spans several orders of magnitude, we sampled it on a logarithmic scale to ensure efficient parameter exploration, while all other parameters were sampled uniformly. We verified convergence and complete sampling of the parameter space by carefully defining the parameter search ranges and initial guesses, as detailed in Table \ref{tab:mcmc-results}.

In a Bayesian method, the priors are proportional to the logarithm of the likelihood $\ln{(\rm data|model)} \propto 0.5\chi^2$, where $\chi^2$ is given by:
\begin{equation*}
    \chi^2 \equiv \sum_i{\frac{\rm (data_i - model_i)^2}{\sigma^2_i}} = \frac{1}{\sigma_{\rm RMS}^2}\sum_i{\rm (data_i - model_i)^2},
    \label{eq:chi2}
\end{equation*}
where $\sigma_{\rm RMS}$ is defined by the mask in Section \ref{sec:momentmaps} and were assumed as a constant $\sigma$ for all pixels. When computing $\chi^2$, we rescaled the uncertainties of the data cube by a factor of $(2N)^{0.25}$, where $N=76,014$ is the number of pixels with detected emission. This approach results in more realistic fit uncertainties by accounting for the potentially underestimated systematic uncertainties returned by Bayesian methods, which often dominate large datasets such as ALMA. This issue arises because the background noise of adjacent pixels is strongly correlated with the synthesized beam size due to the nature of interferometric techniques, a phenomenon known as ``noise covariance'' \citep{Barth16a, Davis17, Onishi17, North19, Nguyen20}. The idea was originally proposed by \citet{vandenBosch09}, later adapted by \citet{Mitzkus17}, and has since been widely implemented in various WISDOM \citep{North19, Smith19} and MBHBM$_\star$ \citep{Nguyen20, Nguyen22} papers.

\begin{figure*}
    \centering
    \includegraphics[width=0.95\textwidth]{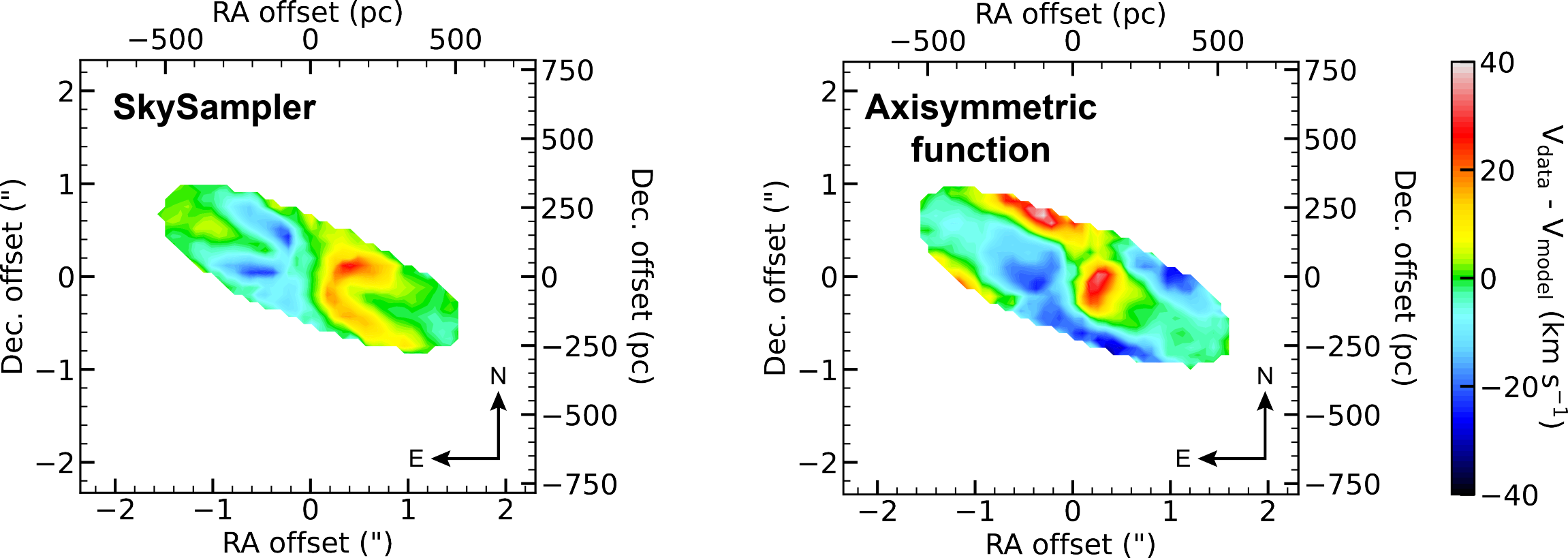}
    \caption{The first-moment residual map ({\tt data-model}) was derived by subtracting the intensity-weighted mean velocity field of the best-fit \textsc{KinMS} models from the observed data. The differences are $\lesssim$15 \kms\ (or $\lesssim$4\%) for the best-fitting \textsc{KinMS} model with \textsc{SkySampler} and $\lesssim$40 \kms\ (or $\lesssim$10\%) for the best-fitting \textsc{KinMS} model with an axisymmetric function, indicating good agreement between the data and the assumed models, and showing the absence of non-circular motions within the \cotwo-CND of NGC 7052.}
    \label{fig:residual}  
\end{figure*}

\begin{figure}
    \centering
    \includegraphics[width=0.95\columnwidth]{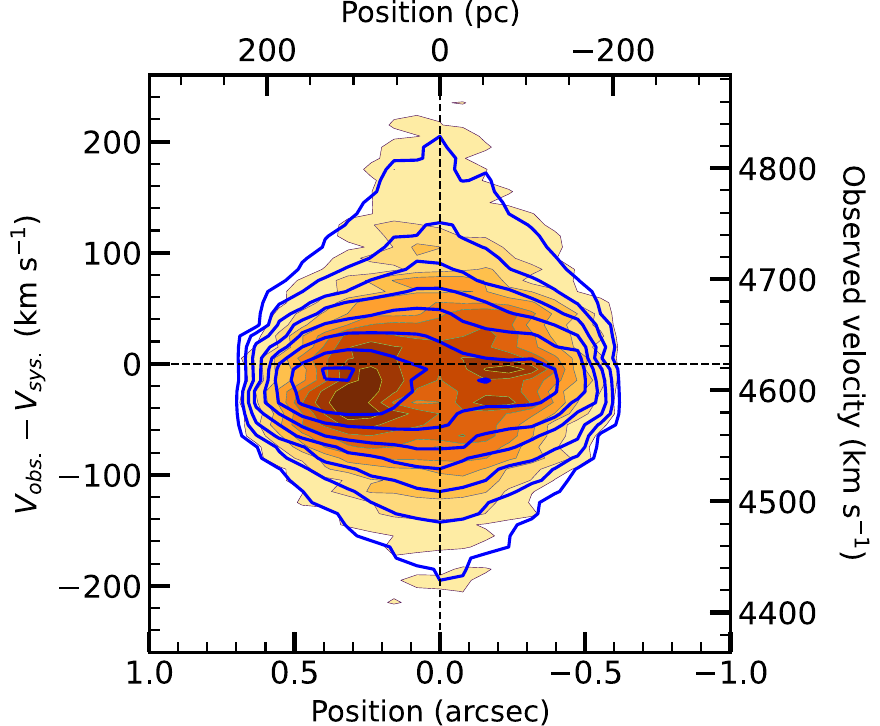}
    \caption{The PVD extracted along the CND's major-axis (elongated along the orientation of position angle $\Gamma=63.8\degr$ + 90\degr) with a systemic velocity $v_{\rm sys}=4610$ \kms.  The best-fitting \textsc{KinMS} model using the \textsc{SkySampler} tool to describe the gas distribution is overlaid on the top as the blued contours.}
    \label{fig:PVD_minoraxis}
\end{figure}

\begin{table}
\caption{Best-fitting \textsc{KinMS} parameters and their uncertainties}
\footnotesize
\centering
\scriptsize

\begin{tabular}{lccccc}
\hline \hline
Model             & Search     & Best-fit & 1$\sigma$ & 3$\sigma$ \\ 
parameters    & range       &   values & (16--84\%)  & (0.14--99.86\%) \\ 
(1)                  & (2)            & (3)       & (4)     & (5)\\ 
\hline\hline
\multicolumn{5}{c}{\textsc{SkySampler}}\\ 
\hline
Mass model  & ~ & ~ & ~ & ~ \\ \hline 
$\log({M_{\rm BH}})$       & 8    $\rightarrow$ 11  & 9.40 & +0.02,$-$0.02 & +0.06,$-$0.07 \\ 
$M/L_{\rm F814W}$             & 0    $\rightarrow$ 10  & 4.08 & +0.07,$-$0.08 & +0.23,$-$0.23 \\ 
\\
Molecular gas & ~ & ~ & ~ & ~ \\ \hline
$f$ (Jy \kms)                   & 1    $\rightarrow$ 200  & 41.81 & +0.95,$-$0.96 & +2.92,$-$2.83 \\ 
$i$ (\degr)                     & 42    $\rightarrow$ 89.9  & 73.49 & +0.42,$-$0.44 & +1.21,$-$1.38 \\ 
$\Gamma$ (\degr)               & 0    $\rightarrow$ 360  & 63.90 & +0.50,$-$0.51 & +1.44,$-$1.51 \\ 
\siggas (\kms)                 & 0    $\rightarrow$ 100  & 14.11 & +1.64,$-$1.54 & +5.12,$-$4.37 \\ 
\\
Nuisance & ~ & ~ & ~ & ~ \\ \hline
$x_c$ (\arcsec)                 & $-$0.9    $\rightarrow$ 0.9  & $-$0.010 & +0.00,$-$0.01 & +0.01,$-$0.01 \\ 
$y_c$ (\arcsec)                 & $-$0.9    $\rightarrow$ 0.9  & $-$0.017 & +0.01,$-$0.01 & +0.02,$-$0.02 \\ 
$v_{\rm off}$ (\kms)         & $-$75    $\rightarrow$ 75  & $-$13.234 & +1.52,$-$1.52 & +4.49,$-$4.49 \\ 
\hline\hline
\multicolumn{5}{c}{Analytically axisymmetric function}\\ 
\hline
Mass model:  & ~ & ~ & ~ & ~ \\ 
$\log(M_{\rm BH}$/\Msun)  & 8    $\rightarrow$ 11   & 9.37 & +0.03,$-$0.03 & +0.08,$-$0.11 \\ 
$M/L_{\rm F814W}$ (\Msun/\Lsun)      & 0    $\rightarrow$ 10   & 4.09 & +0.12,$-$0.11 & +0.35,$-$0.34 \\ 
\\
Mass model  & ~ & ~ & ~ & ~ \\ \hline 
$\log({M_{\rm BH}})$       & 8    $\rightarrow$ 11  & 9.38 & +0.02,$-$0.02 & +0.06,$-$0.07 \\ 
$M/L_{\rm F814W}$             & 0    $\rightarrow$ 10  & 4.08 & +0.08,$-$0.08 & +0.24,$-$0.23 \\ 
\\
Molecular gas & ~ & ~ & ~ & ~ \\ \hline
$f$ (Jy \kms)                   & 1    $\rightarrow$ 200  & 36.78 & +0.91,$-$0.90 & +2.72,$-$2.58 \\ 
$i$ (\degr)                     & 42    $\rightarrow$ 89.9  & 76.14 & +0.50,$-$0.51 & +1.45,$-$1.54 \\ 
$\Gamma$ (\degr)               & 0    $\rightarrow$ 360  & 64.57 & +0.43,$-$0.43 & +1.27,$-$1.25 \\ 
\siggas (\kms)                 & 0    $\rightarrow$ 100  & 17.32 & +1.66,$-$1.73 & +5.37,$-$4.74 \\ 
\\
\hline
\end{tabular}
\parbox[t]{0.47\textwidth}{\small \textit{Notes:} When model the gas distribution with the sum of two center-offset simple Gaussian with \textsc{KinMS}, we fixed nuisance parameters at their best-fit values in the previous case constraining the gas distribution with the \textsc{SkySampler} tool.}
\label{tab:mcmc-results}
\end{table}

\begin{figure*}
    \centering
    \includegraphics[width=0.95\textwidth]{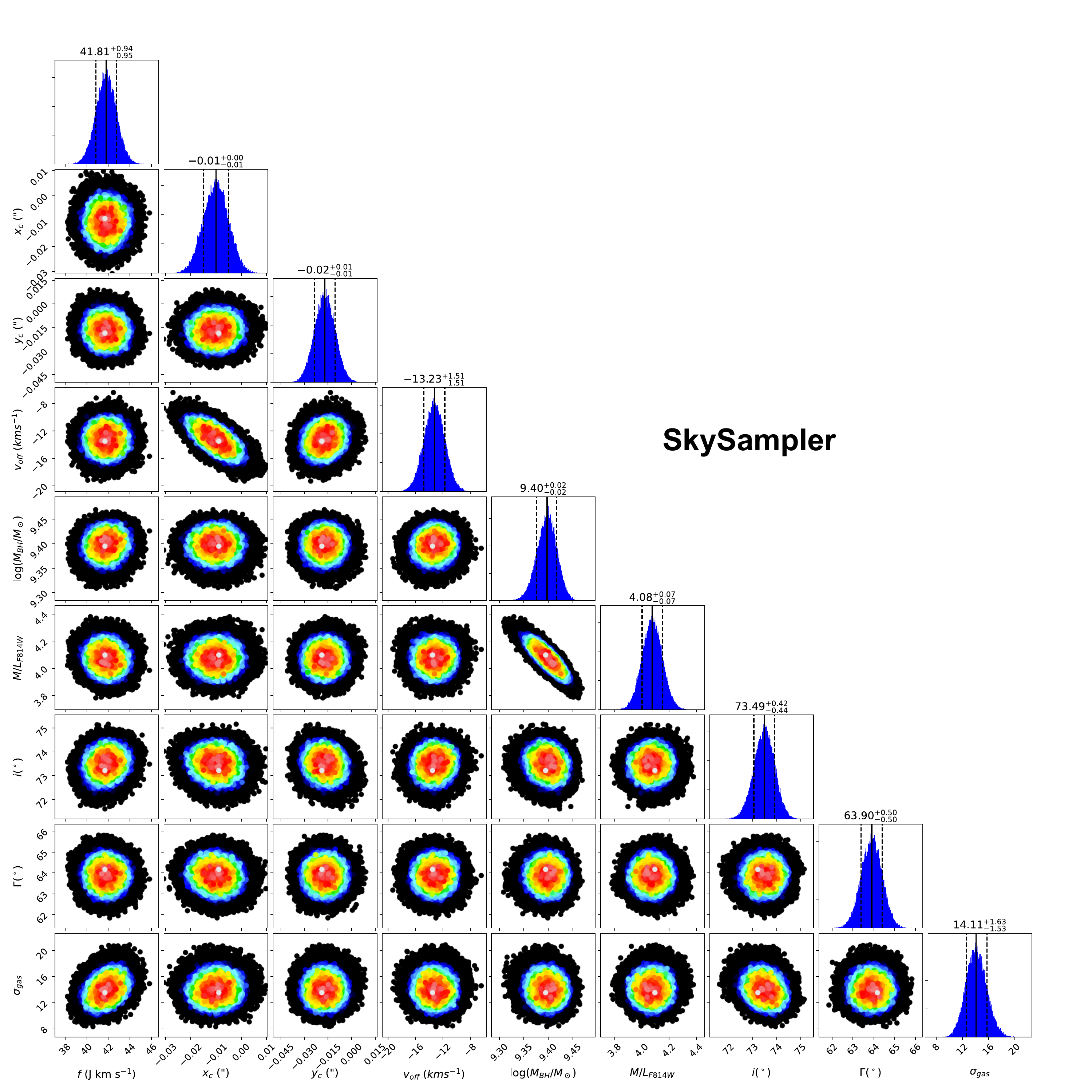}
    \caption{The corner plot shows the posterior distributions obtained after removing the initial 20\% of the post-burn-in phase from a total of $10^5$ MCMC iterations using the \textsc{KinMS} model, which assumed the gas distribution with the \textsc{SkySampler} tool. The top 1D histograms display the marginalized posterior distributions for each parameter, along with their 1$\sigma$ uncertainties (see text for details). The lower panels present 2D scatter plots of parameter pairs, where colors indicate CLs, ranging from 1$\sigma$ (white) to 3$\sigma$ (blue), with black representing CLs below 3$\sigma$. Detailed results are listed in Table \ref{tab:mcmc-results}.} 
    \label{fig:triangle}
\end{figure*}

\begin{figure*}
    \centering
    \includegraphics[width=0.85\textwidth]{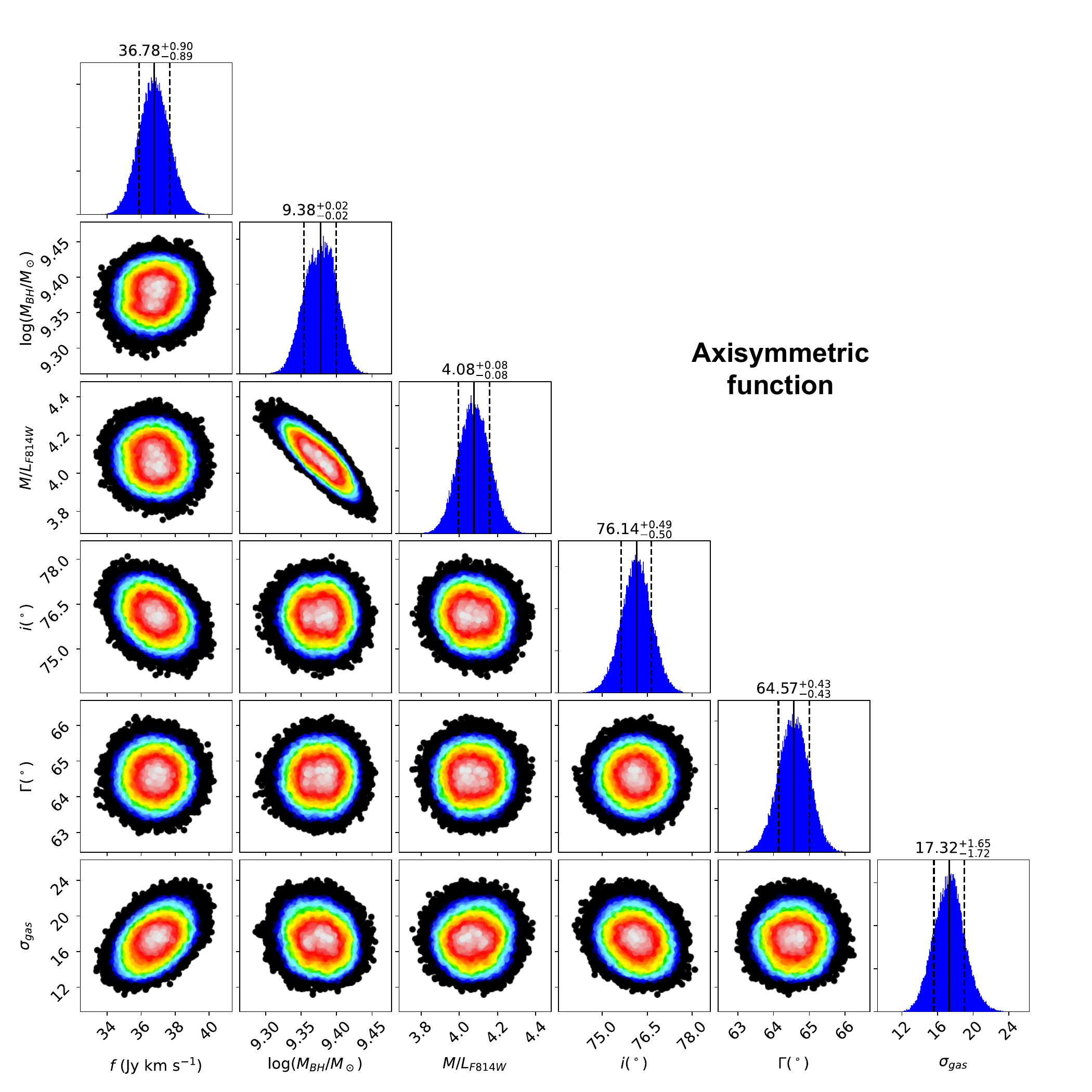}
    \caption{Same as Figure \ref{fig:triangle} but the \textsc{KinMS} model was assumed the gas surface brightness as the sum of two center-offset Gaussians.}  
     \label{fig:triangle1}
\end{figure*}

\subsection{Results}\label{sec:results}

The observed molecular gas kinematics clearly indicate the presence of a central SMBH, as the rotation speed increases toward the center for radii smaller than $0\farcs5$. As listed in Table \ref{tab:mcmc-results}, the best-fitting \textsc{KinMS} model using the \textsc{SkySampler} tool gives a \Mbh\ $=(2.50\pm 0.37)\times10^9$ \Msun\ and a $M/L_{\rm F814W}=4.08 \pm 0.23$ (\Msun/\Lsun), while that same model using an analytically axisymmetric function of two center-offset Gaussians provides a \Mbh\ $=(2.34^{+0.39}_{-0.52})\times10^9$ \Msun\ and a $M/L_{\rm F814W}=4.08^{+0.24}_{-0.23}$ (\Msun/\Lsun). The former best-fitting model has a minimum chi-squared of $\chi^2_{\rm min} = 65,903$, corresponding to a reduced chi-squared of $\chi^2_{\rm red, min} = 0.867$ (i.e., $\chi^2_{\rm min}$ per degree of freedom), while the latter model has $\chi^2_{\rm min} = 57,273$ and $\chi^2_{\rm red, min} = 0.754$. Additionally, our intermediate angular resolution ALMA data does not resolve the central hole in the \cotwo~CND (though it is sufficient to resolve the SMBH’s SOI). This helps minimize mismatches between the data and the model.

All uncertainties are given at the $1\sigma$ confidence level (CL). These two best-fitting models fit the ALMA data very well at all positions of the \cotwo-CND, as seen in Figure \ref{fig:data-model-momentmaps} for the best-fit \textsc{KinMS} models using either \textsc{SkySampler} or the analytically axisymmetric function, where we compared the consistencies in all moment maps. The well agreements of these two best-fitting models with the data are also presented in the PVDs extracted along the major axis of the \cotwo-CND also illustrated in the middle panels of Figure \ref{fig:compare-pvd}.

In each approach with either \textsc{SkySampler} or sum of two center-offset Gaussians, for comparison, Figure \ref{fig:compare-pvd} also shows two other \textsc{KinMS} models: a model with no SMBH (\Mbh\ $=0$ \Msun) and $M/L_{\rm F814W}=4.5$ (\Msun/\Lsun). These models match the extended kinematics of the CND but fail to reproduce the increase in rotation speed toward the center. Another model with a larger \Mbh\ $=4.5\times10^9$ \Msun\ and $M/L_{\rm F814W}=3.5$ (\Msun/\Lsun) fit the extended kinematics but produce too much centrally rising circular motion at small radii. In all these alternative models only $M/L_{\rm F814W}$ were allowed to vary, while \Mbh\ were fixed at the above values and other parameters were also fixed to their best-fit values from Table \ref{tab:mcmc-results}. In addition, we further assessed the agreement between the observed \cotwo\ emission and the best-fit \textsc{KinMS} models by comparing their integrated spectra in Figure \ref{fig:overlaid-spec}. Our best-fit models not only match the kinematics but also reproduce the asymmetry in the integrated flux caused by the two peaks along the major axis. These comparisons confirm that the best-fit models accurately represent the observed gas kinematics.

 Furthermore, we validated our results by examining the possible presence of non-circular motions (e.g., gas inflows/outflows) within the \cotwo-CND. We checked the residuals map of the intensity-weighted mean LOS velocity ($V_{\rm residual}=V_{\rm data} - V_{\rm model}$, Figure \ref{fig:residual}). Our best-fit \textsc{KinMS} model with \textsc{SkySampler} provides $|V_{\rm residual}| \lesssim 15$ \kms\ ($\lesssim$4\%) across the CND, which is approximately equal to the channel width of our reduced ALMA cube ($\approx$10 \kms), suggesting there is no non-circular motions in the \cotwo-CND.  However, the best-fit \textsc{KinMS} model with two center-offset Gaussians yields residual velocities of $|V_{\rm residual}| \lesssim 40$ \kms\ ($\lesssim$10\%). This is because we assumed a smooth function for the gas distribution, which provides a reasonable approximation. In contrast, the \textsc{KinMS} model with \textsc{SkySampler} tool utilized the actual spatial gas distribution from the data cube, significantly reducing differences in the intensity-weighted mean LOS velocity field. Therefore, we adopt the results from the best-fit \textsc{KinMS} model with \textsc{SkySampler} as our final measurement, while using the alternative model to assess the uncertainty.
  
 Another verification was performed to confirm the absence of non-circular motions or kinematic warps (i.e., a change in position angle that twists the isovelocity contours in the velocity map along the CND minor-axis) within the \cotwo\ CND. These effects could significantly impact our dynamical modeling and the accurate measurement of \Mbh. The minor-axis PVD of NGC 7052 as shown in Figure \ref{fig:PVD_minoraxis}, extracted along the direction of $\Gamma+90\degr$, exhibits symmetry in all four ‘forbidden quadrants’ of the PVD. A slightly higher velocity in the redshifted component of the CND is likely due to a gas deficiency, possibly caused by a specific gas morphology, e.g., a nuclear spiral.
   

Within $1\sigma$ uncertainty, our \Mbh\ constraint is fully consistent (though a bit higher) with the measurement obtained by \citet{Smith21} using the ALMA observations with nearly three times higher angular resolution than our ALMA data. All other parameters also agree with their results, except for $M/L_{\rm F814W}$. Our estimated $M/L_{\rm F814W}$ value is 10\% lower than that reported by \citet{Smith21}. This difference arises because our updated stellar mass model accounts for the total mass of the entire galaxy by modeling the HST WFC image. While including the extended mass of the galaxy does not impact the \Mbh\ measurement, it provides a stronger constraint on $M/L_{\rm F814W}$.

Figure \ref{fig:triangle} (for the best-fitting \textsc{KinMS} model using \textsc{SkySampler}) and Figure \ref{fig:triangle1} (for the best-fitting \textsc{KinMS} model using an analytically axisymmetric function of two center-offset Gaussians) present the 2D posterior distributions for each pair of free parameters, with colors representing their likelihood. White corresponds to the maximum likelihood within 1$\sigma$ CL, while blue marks the likelihood within 3$\sigma$ CL. The 1D histograms show the marginalized distributions for each parameter. The thick black vertical lines indicate the best-fit values with the highest likelihood, while the dashed vertical lines on either side represent the 1$\sigma$ uncertainties. All histograms exhibit a Gaussian-like shape, demonstrating that our MCMC optimization with the \textsc{KinMS} model achieved a well convergence.

While other parameters are well constrained, the well-known anti-correlation between \Mbh\ and $M/L_{\rm F814W}$ is clearly evident, a common effect when working with spatially resolved data. Additionally, correlations exist between the nuisance parameters ($x_{\rm cen}$, $y_{\rm cen}$, and $v_{\rm off}$). These arise when the observational beam size is large, as we constrain the kinematic center to align with the peak of the spatially unresolved continuum emission (Section \ref{sec:cont}).

Our best-fitting \textsc{KinMS} models determined an inclination of $i \approx 73$\degr, which agrees well with previous estimates based on the dust disk \citep[$i\approx$70\degr;][]{Bosch1995, Juan96}. This agreement is crucial because inclination plays a significant role in the overall uncertainty of our measurements.

\begin{figure*}
    \centering
    \includegraphics[width=0.75\textwidth]{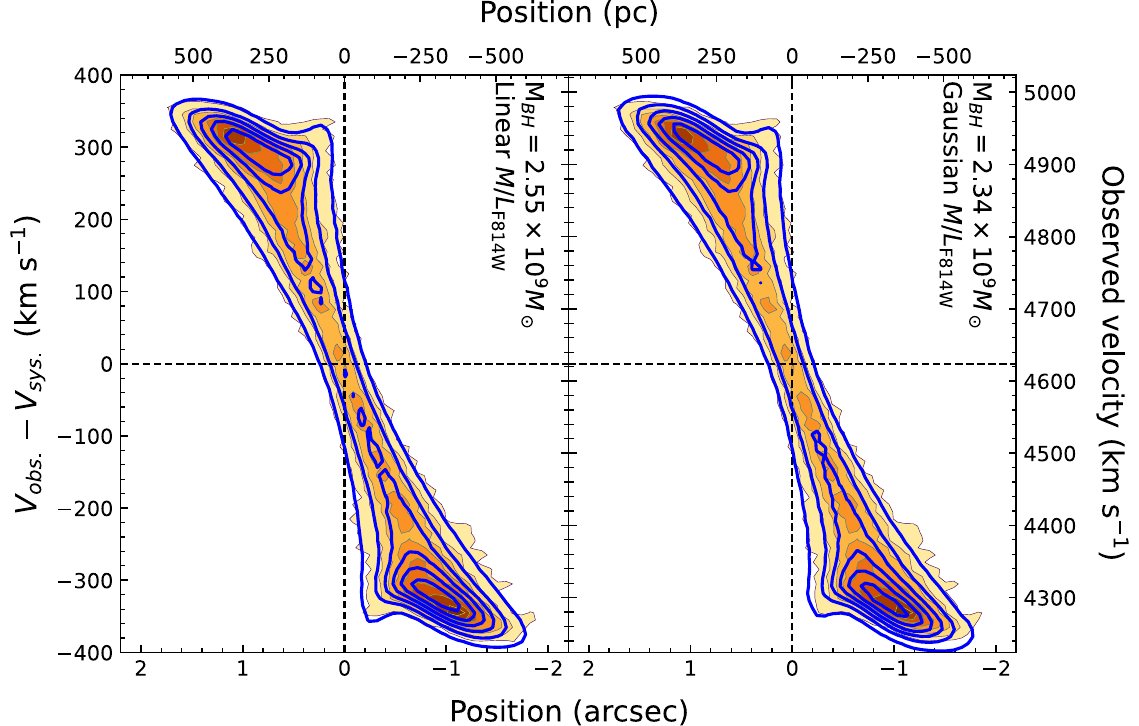}
    \caption{Position–velocity diagrams along the major axis of the best-fitting \textsc{KinMS} models using the \textsc{SkySampler} tool to describe the \cotwo\ gas surface brightness distribution, shown for models with linear (left) and Gaussian (right) $M/L_{\rm F814W}(r)$ profiles.}
    \label{fig:pvd-compare-mlvary}
\end{figure*}

\subsection{Uncertainties} \label{sec:uncertainties}

Given the different assumptions of the \cotwo-CND distribution in our \textsc{KinMS} models, using whether \textsc{SkySampler} or an analytically axisymmetric function, and based on the results in Table \ref{tab:mcmc-results}, the differences in our derived \Mbh\ and $M/L_{\rm F814W}$ are less than 2.5\%. Compared to \citet{Smith21}, our \Mbh\ measurement is either higher or lower by 4\%, while our $M/L_{\rm F814W}$ is less than 10\%.  Thus, we consider our results to be robust against these sources of uncertainty and fully consistent with the \citet{Smith21} estimate, despite their higher-resolution ALMA observations.

In the following subsections, we discuss a variety additional sources of uncertainty in dynamical modeling, including (i) the adopted distance to NGC 7052, (ii) the assumption of a thick disc (by setting the $z$-coordinate perpendicular to the disc plane), (iii) the turbulent velocity dispersion of the gas, (iv) the inclination, and (v) the AGN contamination MGE without masking. 

\subsubsection{Distances}\label{sec:uncer_dist}

The \Mbh\ estimate is systematically affected by the assumed distance to the galaxy, following the relation $M_{\rm BH} \propto D$. For NGC 7052, only two distance estimates are available in the literature. We adopt the value from the MASSIVE survey, which derives a distance of 69.3 Mpc based on redshift measurements \citep{Ma14}. An alternative estimate, based on 21-cm line kinematics observed with the Nançay radio telescope and the $JHK$ Tully-Fisher relation, yields a distance of 46.4 Mpc \citep{Theureau2007}. This discrepancy results in a systematic uncertainty of $\sim$30\% in \Mbh, which exceeds both the random and other systematic uncertainties and thus represents the dominant source of error in the black hole mass measurement.

\subsubsection{Thick disk assumption} \label{sec:uncer_thick_disc}

In our \textsc{KinMS} models, we assumed that the \cotwo\ disk is thin by setting its vertical thickness to zero. However, the \cotwo\ disk is expected to have a finite thickness along the $z$-axis, perpendicular to the disc plane. To test the impact of this assumption on the \Mbh\ estimate for NGC 7052, we introduced an additional free parameter in the \textsc{KinMS} models to represent a constant vertical thickness. This test was performed for both the model using the \textsc{SkySampler} tool and the model employing an analytic axisymmetric surface brightness profile composed of two center-offset Gaussians. In both cases, the parameters of the best-fitting model remained nearly unchanged compared to those in Table \ref{tab:mcmc-results}, that is all differences are less than 3\%, with fitted vertical thicknesses of $z = 0\farcs01\pm0\farcs01$ for the model with \textsc{SkySampler} and $z = 0\farcs02\pm0\farcs01$ for the model with two center-offset Gaussians. Both are consistent with our original assumption of a razor-thin disc ($z = 0$) for the \cotwo\ disk of NGC~7052.

\subsubsection{Turbulent velocity dispersion of the gas}\label{sec:uncer_sigma_vary}

In the analysis above, we assumed a constant turbulent velocity dispersion for the gas. However, in practice, the velocity dispersion can vary with both radius and azimuth across the disk. Moreover, an increase in velocity dispersion near the galaxy center due to beam smearing can lead to an overestimation of \Mbh. To assess the impact of these effects on the error budget of \Mbh, we allowed the velocity dispersion to vary as a function of radius. Specifically, we tested a range of radial profiles for the \cotwo\ velocity dispersion, adopting several functional forms for $\sigma_{\rm gas}(r)$: 

{\it (a) Linear gradient:} $\sigma_{\rm gas}(r) = a \times r + b$, where $a$ and $b$ are free parameters. We found $a \approx 0$, with $b = 15.8$ \kms\ for the model using the \textsc{SkySampler} tool and $b = 19.3$ \kms\ for the model with two center-offset Gaussians. The other best-fitting \textsc{KinMS} parameters are consistent with those from the default constant velocity dispersion models discussed in Section \ref{sec:results} and listed in Table \ref{tab:mcmc-results}. 

{\it (b) Exponential:} $\sigma_{\rm gas}(r) = \sigma_0 \exp(-r/r_0) + \sigma_1$, where $\sigma_0$, $\sigma_1$, and $r_0$ are free parameters. To avoid unrealistically narrow line profiles during the fitting process, we impose a lower limit of $\sigma_{\rm gas,min} = 1$ \kms\ \citep{Barth16b, Nguyen20}. The best-fitting \textsc{KinMS} model using the \textsc{SkySampler} tool yields \Mbh\ $= (2.38 \pm 0.12) \times 10^9$ \Msun\ and $M/L_{\rm F814W} = 4.22\pm 0.18$ (\Msun/\Lsun), with the exponential velocity dispersion profile characterized by $\sigma_0 = 64.76 \pm 3.25$ \kms, $\sigma_1 = 17.11 \pm 0.85$ \kms, and $r_0 = -0\farcs07 \pm 0.05$. The corresponding model using two center-offset Gaussians gives \Mbh\ $= (2.27\pm 0.15) \times 10^9$ \Msun\ and $M/L_{\rm F814W} = 4.16 \pm 0.17$ (\Msun/\Lsun), with the same exponential dispersion parameters: $\sigma_0 = 144.62 \pm 32.73$ \kms, $\sigma_1 = 7.94 \pm 1.68$ \kms, and $r_0 = -0\farcs09\pm 0.24$.

{\it (c) Gaussian:}  $\sigma_{\rm gas}(r) = \sigma_0 \exp\left[-(r - r_0)^2 / 2\mu^2\right] + \sigma_1$, where $\sigma_0$, $\sigma_1$, $\mu$, and $r_0$ are free parameters. We allow $r_0$ to vary over both positive and negative values to account for cases where the line width is offset from the center. During the fits, we also set a lower limit of $\sigma_{\rm gas,min} = 1$ \kms. The best-fitting \textsc{KinMS} model using the \textsc{SkySampler} tool yields \Mbh\ $= (2.32 \pm 0.12) \times 10^9$ \Msun\ and $M/L_{\rm F814W} = 4.20 \pm 0.14$ (\Msun/\Lsun), with a Gaussian dispersion profile characterized by $\sigma_0 = 61.90 \pm 3.02$ \kms, $\sigma_1 = 11.93 \pm 0.60$ \kms, $r_0 = -0\farcs08 \pm 0.05$, and $\mu = 0\farcs39 \pm 0.05$. Similarly, the best-fitting \textsc{KinMS} model using two center-offset Gaussians returns \Mbh\ $= (2.53 \pm0.16) \times 10^9$ \Msun\ and $M/L_{\rm F814W} = 4.03 \pm 0.17$ (\Msun/\Lsun), with  Gaussian dispersion parameters of $\sigma_0 = 61.15 \pm 7.30$ \kms, $\sigma_1 = 15.59 \pm 1.70$ \kms, $r_0 = -0.08 \pm 0.03$, and $\mu = 0\farcs42 \pm 0.10$.

These results indicate that the assumption of a constant $\sigma_{\rm gas}$ provides an adequate description of the \cotwo\ disc’s kinematics for the purpose of dynamical modeling of \Mbh. Overall, our choice of radial functional forms for the gas velocity dispersion has some impact on the \Mbh\ measurements. Given the minimal effect when assuming a linear gradient in $\sigma_{\rm gas}(r)$, the resulting uncertainties in the \Mbh\ constraints are less than 14\% and 22\% for the exponential and Gaussian $\sigma_{\rm gas}(r)$ profiles, respectively.

\subsubsection{Inclination} \label{sec:uncer_mge_deprojection}

The MGE deprojection with an assumed inclination for constructing the 3D intrinsic stellar mass model can be a significant source of uncertainty, particularly when the galaxy is viewed close to face-on (i.e., inclination $\lesssim 40^\circ$). \citet{Smith19} found that low inclinations lead to asymmetric posteriors and introduce substantial uncertainties in both the SMBH mass and stellar $M/L$, as demonstrated in the face-on galaxy NGC 524. In contrast, NGC 7052 has a well-constrained kinematic inclination of $i \approx 73^\circ$, which results in a unique 3D intrinsic mass model when deprojected from the MGE. Therefore, the contribution of inclination-related uncertainties to our measurements of \Mbh\ and $M/L_{\rm F814W}$ is minimal (see Figures \ref{fig:triangle} and \ref{fig:triangle1}).

\subsubsection{MGE without dust masking} \label{sec:uncer_mge_agn}

We tested the uncertainties in the \Mbh\ and $M/L_{\rm F814W}$ measurements using an HST/WFC3 F814W MGE model for NGC 7052 constructed without masking the central pixels, which are affected by dust extinction. The best-fitting \textsc{KinMS} model using the \textsc{SkySampler} tool yields \Mbh\ $= (2.49_{-0.18}^{+0.15}) \times 10^9$ \Msun\ and $M/L_{\rm F814W} = 4.09^{+0.20}_{-0.18}$ (\Msun/\Lsun), while the corresponding model using two center-offset Gaussians gives \Mbh\ $= (2.40^{+0.22}_{-0.20}) \times 10^9$ \Msun\ and $M/L_{\rm F814W} = 4.10^{+0.12}_{-0.12}$ (\Msun/\Lsun). The other molecular gas and nuisance parameters differ by less than 5\% from the corresponding default model values listed in Table \ref{tab:mcmc-results} and described in Section \ref{sec:results}. Notably, the \Mbh\ and $M/L_{\rm F814W}$ values derived from the unmasked MGE model are fully consistent with the default models, likely due to the low dust mass, which is at least five orders of magnitude smaller than the black hole mass \citep[i.e., $M_{\rm dust} \approx 10^4$~\Msun;][Section \ref{sec:ngc7052}]{Nieto1990}.

\subsubsection{Mass model with M/L$_{\rm F814W}$ variations} \label{sec:uncer_mass_vary}

In our analysis, we assumed a constant \ml\ across the galaxy. However, the \ml\ profile may vary with radius due to mass segregation \citep{Nguyen2025b}, potentially producing a central peak that mimics and adds to the effect of a massive central dark object. To test this possibility, we ran test models following the same manner described in Section \ref{sec:bayesian-infer}, and allowed for either a linearly varying $M/L_{\rm F814W} (r) = M/L_0 + \alpha \times r$ or a Gaussian profile $M/L_{\rm F814W} (r) = M/L_0 \exp{(-r^2/2\sigma_{\rm Gaussian}^2)} + M/L_1$, where $M/L_0$ is the central \ml\ value of both profiles, $\alpha$ is the slope of the linear function, $M/L_1$ is a constant, and $\sigma_{\rm Gaussian}$ is the width of the Gaussian. During these model tests, we fixed all nuisance and molecular gas parameters to their best-fit values from the default models listed in Table \ref{tab:mcmc-results}, but left the \Mbh\ as a free parameter. These new best-fitting \textsc{KinMS} models were run with the Bayesian method as described in Section \ref{sec:bayesian-infer} and their results are recorded in Table \ref{tab:mcmc-results_mlvary}.

We presented a comparison of the two best-fitting \textsc{KinMS} models that use the \textsc{SkySampler} tool to describe the \cotwo\ gas surface brightness distribution in Figure \ref{fig:pvd-compare-mlvary}, 
for both cases of radial variation in the $M/L_{\rm F814W}(r)$ profile. Interestingly, both the linear and Gaussian $M/L_{\rm F814W}(r)$ profiles fit the data well across the \cotwo\ CND, despite the lack of observational evidence for color or stellar population variations in the nucleus of NGC 7052. The best-fitting \Mbh\ values from these models differ by less than 10\% and 2\% for the linear and Gaussian profiles, respectively, compared to the default constant $M/L_{\rm F814W}$ models (Section \ref{sec:results} and Table \ref{tab:mcmc-results}), and remain fully consistent within the 1$\sigma$ uncertainties.   We also found similar results for the best-fitting \textsc{KinMS} models that use two center-offset Gaussians to describe the \cotwo\ surface brightness distribution. These results suggest that our \Mbh\ measurement using ALMA data is relatively insensitive to the detailed form of the $M/L_{\rm F814W}$ profile (e.g., variations due to dust extinction or stellar population changes across the \cotwo\ CND), because the gravitational influence of the central black hole dominates on the spatial scale of the \cotwo\ gas disk. We therefore conclude that variations in the stellar mass model due to changes in $M/L_{\rm F814W}$ contribute approximately 10\% to the overall error budget of \Mbh.

\begin{table}
\caption{Best-fitting \ml\ models' parameters and their uncertainties}
\footnotesize
\centering
\begin{tabular}{lccccc}
\hline \hline
Model         & Search  & Best-fit & 1$\sigma$  & 3$\sigma$ \\ 
parameters    & range   &   values &(16--84\%)  & (0.14--99.86\%)\\ 
(1)           & (2)     & (3)      & (4)        & (5)\\ 
\hline\hline
\multicolumn{5}{c}{\textsc{SkySampler}}\\ 
\hline
Linear \ml$_{\rm F814W}$ & \multicolumn{3}{c}{$\chi^2_{\rm red, min} \approx 0.684$}  & ~ \\ \hline 
$\log(M_{\rm BH}$/\Msun) &8$\rightarrow$11& 9.41 & $\pm$0.02 & $\pm$0.06 \\ 
$M/L_0$ (\Msun/\Lsun)    &0$\rightarrow$10& 3.76 & $\pm$0.06 & $\pm$0.24 \\ 
$\alpha$ (\Msun/\Lsun\ per \arcsec)& 0$\rightarrow$10& 0.044 & $\pm$0.01 & $\pm$0.03\\ 
\hline
Gaussian \ml$_{\rm F814W}$ & \multicolumn{3}{c}{$\chi^2_{\rm red, min} \approx 0.671$}  & ~ \\ \hline 
$\log(M_{\rm BH}$/\Msun)  &8$\rightarrow$11& 9.37 & $\pm$0.03 & $\pm$0.09 \\ 
$M/L_0$ (\Msun/\Lsun)     &0$\rightarrow$10& 2.16 & $\pm$0.16 & $\pm$0.37 \\ 
$M/L_1$ (\Msun/\Lsun)     &0$\rightarrow$10& 3.90 & $\pm$0.18 & $\pm$0.54 \\ 
$\sigma_{\rm Gaussian}$ (\arcsec)&0$\rightarrow$10& 0.04 & $\pm$1.08 & $\pm$3.24 \\ 
\\
\hline\hline
\multicolumn{5}{c}{Analytically axisymmetric function}\\ 
\hline
Linear \ml$_{\rm F814W}$  & \multicolumn{3}{c}{$\chi^2_{\rm red, min} \approx 0.610$}  & ~ \\ \hline 
$\log(M_{\rm BH}$/\Msun)  &8$\rightarrow$11& 9.38 & $\pm$0.02 & $\pm$0.06 \\ 
$M/L_0$ (\Msun/\Lsun)     &0$\rightarrow$10& 3.76 & $\pm$0.10 & $\pm$0.30 \\ 
$\alpha$ (\Msun/\Lsun\ per \arcsec)& 0$\rightarrow$10 & 0.09 & $\pm$0.03 & $\pm$0.09 \\ 
\hline
Gaussian \ml$_{\rm F814W}$ & \multicolumn{3}{c}{$\chi^2_{\rm red, min} \approx 0.605$}  & ~ \\ \hline 
$\log(M_{\rm BH}$/\Msun)  &8$\rightarrow$11& 9.36 & $\pm$0.02 & $\pm$0.06 \\ 
$M/L_0$ (\Msun/\Lsun)     &0$\rightarrow$10& 1.87 & $\pm$0.09 & $\pm$0.27 \\ 
$M/L_1$ (\Msun/\Lsun)     &0$\rightarrow$10& 2.37 & $\pm$0.09 & $\pm$0.27 \\
$\sigma_{\rm Gaussian}$ (\arcsec)&0$\rightarrow$10& 2.70 & $\pm$0.21 & $\pm$0.63 \\ 
\hline
\end{tabular}
\parbox[t]{0.47\textwidth}{\small \textit{Notes:} In these \textsc{KinMS} models, we fixed all molecular gas and nuisance parameters at their best-fit values as of their default models listed in Table \ref{tab:mcmc-results}.}
\label{tab:mcmc-results_mlvary}
\end{table}

\begin{figure}
\centering
    \includegraphics[width=\columnwidth]{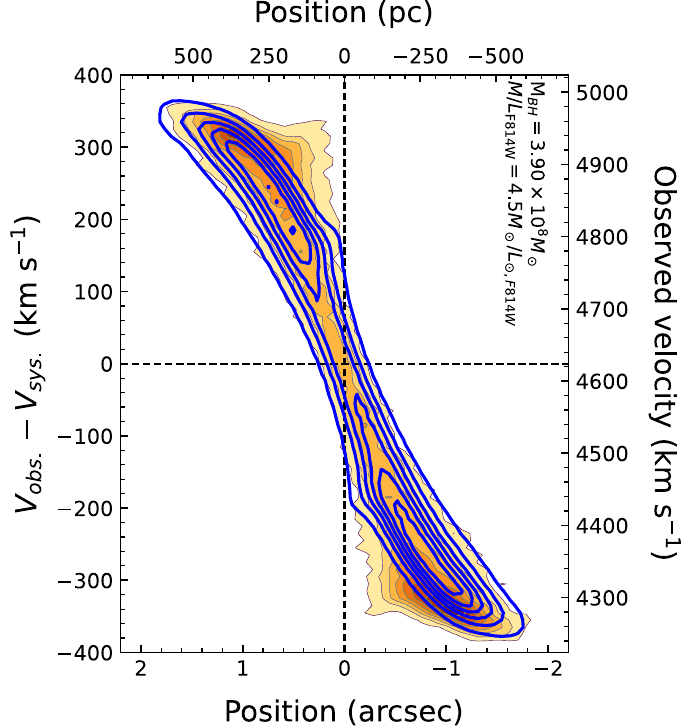}
    \caption{The position-velocity diagram along the major-axis of models consist of a SMBH mass of \Mbh$=3.9\times10^8$ \Msun\ derived from ionized gas by \citet{Marel1998} with a constant $M/L_{\rm F814W}$.}
    \label{fig:pvd-compare-bosch}
\end{figure}

\begin{figure}
    \includegraphics[width=0.98\columnwidth]{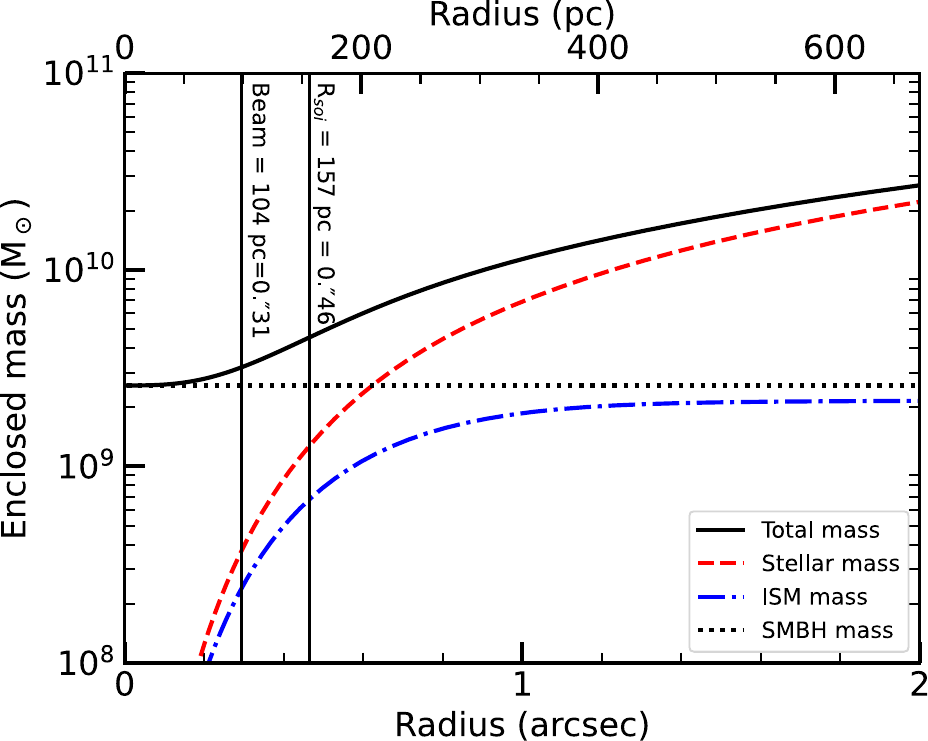}
    \caption{Enclosed mass of NGC 7052 (black solid line) as a function of radius, showing the contributions all mass components: \Mbh, stars, and ISM (i.e., gas and dust).}
    \label{fig:NGC7052_cumulative_mass}
\end{figure}

\begin{figure}
    \centering
    \includegraphics[width=0.9\columnwidth]{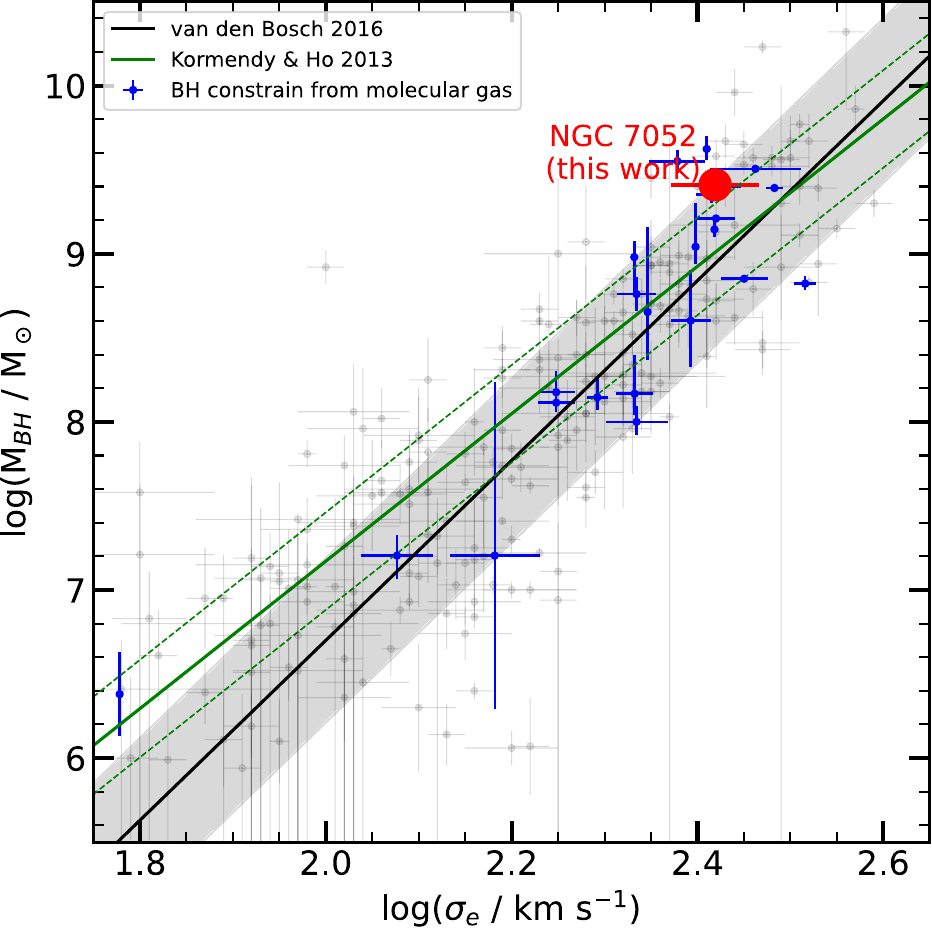}
    \caption{Our \Mbh\ estimate for NGC 7052 in the context of various \Mbh--$\sigma$ scaling relations and their intrinsic scatters.}
    \label{fig:bhmass-sigma}
\end{figure}

\subsection{The reliability of our measurements} \label{sec:previous-work}

Our results show a discrepancy compared to \citet{Marel1998}, who used ionized-gas kinematics and reported an \Mbh\ = $3.9^{+2.7}_{-1.5} \times 10^8$ \Msun. This discrepancy can be explained by differences in the tracers used (H$\alpha$ + [\ion{N}{2}] versus \cotwo\ emission) and the extent of the gas disk sampled. Ionized gas is likely affected by significant turbulence from the AGN and was observed at only six positions along the major axis, rather than across the entire gas disk. In contrast, cold molecular gas is much less impacted by turbulence. Our measurement using our ALMA observation in this work is more consistent with \citet[][with \Mbh\, = $2.5\times 10^9$ \Msun]{Smith21}. To directly compare with our findings, we adopted the \Mbh\ from \citet{Marel1998} and adjusted other parameters to achieve the best fit. The PVD along the major axis of this model is shown in Figure \ref{fig:pvd-compare-bosch}. 
Even with an increased \ml\ of 4.5 \Msun/\Lsun, the central circular velocity rise could not be reproduced with this lower \Mbh. Adopting a higher \ml\ value causes the outer regions of the \cotwo\ CND to deviate significantly from the observed data, resulting in an unphysical model.

Previous studies suggest that accurate \Mbh\ measurements require the beam size to be smaller than, or at least equal to, the SMBH's SOI. \citep[e.g.,][]{Rusli13, Davis14, Boizelle21, Nguyen20}. To assess the resolving power of our data for the SMBH's \Rsoi\ in NGC 7052, we used the ratio $\xi = 2R_{\rm SOI}/\theta_{\rm FWHM}$. Observations with $\xi < 2$ (or \Rsoi\ $\lesssim \theta_{\rm FWHM}$) can still yield \Mbh\ estimates but are more susceptible to systematic uncertainties from stellar mass contributions and disk structural properties \citep[e.g.,][see Figure \ref{fig:NGC7052_cumulative_mass}]{Nguyen21, Nguyen22}. Our data, with $\xi \approx 3.5$, provides sufficient resolution for a reliable \Mbh\ measurement. Figure \ref{fig:NGC7052_cumulative_mass} clearly shows that within the beam size of our ALMA observations, the \Mbh\ dominates over all other mass components. As a result, its kinematic influence on the inner region of the \cotwo\ CND is clearly detected and well resolved, strengthening the reliability of our \Mbh\ measurement.

\subsection{\Mbh-$\sigma$ scaling relation} \label{sec:Mbh-sigma}

Our SMBH mass estimate for NGC 7052 is consistent within the $+1\sigma$ uncertainty of the \Mbh--$\sigma$ relations compiled by \citet{Bosch2016} and \citet{Kormendy13}, as shown in Figure \ref{fig:bhmass-sigma}. The predicted \Mbh\ values from these correlations are $0.9 \times 10^9$ \Msun\ and $1.1 \times 10^9$ \Msun, respectively. While NGC 7052 appears as a slight positive outlier, it remains within the upper bounds of these correlations. This suggests that NGC 7052 is at a transition point where SMBH growth begins to shift from bulge-dominated processes to dry mergers \citep{Cappellari16, Krajnovic18a}.

\section{Conclusions}\label{sec:conclusion}

We revisited the \Mbh\ in NGC 7052 using cold gas dynamical modeling and our ALMA \cotwo\ observations from Cycle 7. The data were taken with a synthesized beam size of 0\farcs31 $\times$ 0\farcs23 (or 104 $\times$ 77 pc$^2$). Our estimates of \Mbh\ and \ml$_{\rm F814W}$ and other parameters related to the \cotwo-CND, using various approaches of spatially gas distribution, are fully consistent with the measurements from \citet{Smith21} within $3\sigma$ uncertainties: \Mbh\ = $(2.50 \pm 0.37 \, [{\rm statistical}] \pm 0.8 \, [{\rm systematic}]) \times10^9$ \Msun\ and \ml$_{\rm F814W} = 4.08 \pm 0.23 \, [{\rm statistical}] \pm 0.4 \, [{\rm systematic}]$ \Msun/\Lsun.  The results further emphasize the critical role of our newly obtained intermediate-spatial-resolution ALMA observations (e.g., \cotwo\ emission) in accurately measuring \Mbh, as long as the observational beam size is still smaller than or comparable to the SMBH's SOI, compared to warm gas tracers that are often disturbed and influenced by non-circular motions. Additionally, our intrinsic and wide-field stellar mass model plays an important role in precisely constraining \ml$_{\rm F814W}$, which is essential for effectively disentangling the stellar mass contribution from \Mbh, leading to more accurate \Mbh\ measurements. In our analysis, we accounted for the molecular gas mass distribution, which is comparable to the \Mbh\ but ignored in the previous works, and refined the stellar mass model of NGC~7052. 

\section*{Acknowledgements}

The authors would like to thank the anonymous referee for their careful reading and useful comments, that helped to improve the paper greatly. Research conducted by H.N.N. is funded by University of Science, VNU-HCM under grant number T2023-105. T.Q.T.L.'s work is partially supported by a grant from the Simons Foundation to IFIRSE, ICISE (916424, N.H.). This paper makes use of the following ALMA data: ADS/JAO.ALMA\#2019.1.00036.S. ALMA is a partnership of ESO (representing its member states), NSF (USA) and NINS (Japan), together with NRC (Canada) and NSC and ASIAA (Taiwan) and KASI (Republic of Korea), in cooperation with the Republic of Chile. The Joint ALMA Observatory is operated by ESO, AUI/NRAO, and NAOJ. The National Radio Astronomy Observatory is a facility of the National Science Foundation operated under cooperative agreement by Associated Universities, Inc.

\facility{ALMA, HST, Pan-STARRS, and VLA.}

\software{{\tt Python~3.12:} \citep{VanRossum2009}, 
{\tt Matplotlib~3.6:} \citep{Hunter2007}, 
{\tt NumPy~1.22:} \citep{Harris2020}, 
{\tt SciPy~1.3:} \citep{Virtanen2020},  
{\tt photutils~0.7:} \citep{bradley2024}, 
{\tt AstroPy~5.1} \citep{AstropyCollaboration22}, 
{\tt AdaMet 2.0} \citep{Cappellari13a}, 
{\tt JamPy~7.2} \citep{Cappellari20}, 
{\tt MgeFit~5.0} \citep{Cappellari02},
{\tt SkySampler} \citep{Smith19}.
}

\bibliographystyle{aasjournalv7}
\bibliography{references} 

\begin{thebibliography}{}
\expandafter\ifx\csname natexlab\endcsname\relax\def\natexlab#1{#1}\fi
\providecommand{\url}[1]{\href{#1}{#1}}
\providecommand{\dodoi}[1]{doi:~\href{http://doi.org/#1}{\nolinkurl{#1}}}
\providecommand{\doeprint}[1]{\href{http://ascl.net/#1}{\nolinkurl{http://ascl.net/#1}}}
\providecommand{\doarXiv}[1]{\href{https://arxiv.org/abs/#1}{\nolinkurl{https://arxiv.org/abs/#1}}}

\bibitem[{C.~P. {Ahn} {et~al.}(2018){Ahn}, {Seth}, {Cappellari},
  {Krajnovi{\'c}}, {Strader}, {Voggel}, {Walsh}, {Bahramian}, {Baumgardt},
  {Brodie}, {Chilingarian}, {Chomiuk}, {den Brok}, {Frank}, {Hilker},
  {McDermid}, {Mieske}, {Neumayer}, {Nguyen}, {Pechetti}, {Romanowsky}, \&
  {Spitler}}]{Ahn18}
{Ahn}, C.~P., {Seth}, A.~C., {Cappellari}, M., {et~al.} 2018,
  \bibinfo{title}{{The Black Hole in the Most Massive Ultracompact Dwarf Galaxy
  M59-UCD3},} \apj, 858, 102, \dodoi{10.3847/1538-4357/aabc57}

\bibitem[{K. {Alatalo} {et~al.}(2013){Alatalo}, {Davis}, {Bureau}, {Young},
  {Blitz}, {Crocker}, {Bayet}, {Bois}, {Bournaud}, {Cappellari}, {Davies}, {de
  Zeeuw}, {Duc}, {Emsellem}, {Khochfar}, {Krajnovi{\'c}}, {Kuntschner},
  {Lablanche}, {Morganti}, {McDermid}, {Naab}, {Oosterloo}, {Sarzi}, {Scott},
  {Serra}, \& {Weijmans}}]{Alatalo13}
{Alatalo}, K., {Davis}, T.~A., {Bureau}, M., {et~al.} 2013,
  \bibinfo{title}{{The ATLAS$^{3D}$ project - XVIII. CARMA CO imaging survey of
  early-type galaxies},} \mnras, 432, 1796, \dodoi{10.1093/mnras/sts299}

\bibitem[{ {Astropy Collaboration} {et~al.}(2022){Astropy Collaboration},
  {Price-Whelan}, {Lim}, {Earl}, {Starkman}, {Bradley}, {Shupe}, {Patil},
  {Corrales}, {Brasseur}, {N{\"o}the}, {Donath}, {Tollerud}, {Morris},
  {Ginsburg}, {Vaher}, {Weaver}, {Tocknell}, {Jamieson}, {van Kerkwijk},
  {Robitaille}, {Merry}, {Bachetti}, {G{\"u}nther}, {Aldcroft},
  {Alvarado-Montes}, {Archibald}, {B{\'o}di}, {Bapat}, {Barentsen},
  {Baz{\'a}n}, {Biswas}, {Boquien}, {Burke}, {Cara}, {Cara}, {Conroy},
  {Conseil}, {Craig}, {Cross}, {Cruz}, {D'Eugenio}, {Dencheva}, {Devillepoix},
  {Dietrich}, {Eigenbrot}, {Erben}, {Ferreira}, {Foreman-Mackey}, {Fox},
  {Freij}, {Garg}, {Geda}, {Glattly}, {Gondhalekar}, {Gordon}, {Grant},
  {Greenfield}, {Groener}, {Guest}, {Gurovich}, {Handberg}, {Hart},
  {Hatfield-Dodds}, {Homeier}, {Hosseinzadeh}, {Jenness}, {Jones}, {Joseph},
  {Kalmbach}, {Karamehmetoglu}, {Ka{\l}uszy{\'n}ski}, {Kelley}, {Kern},
  {Kerzendorf}, {Koch}, {Kulumani}, {Lee}, {Ly}, {Ma}, {MacBride}, {Maljaars},
  {Muna}, {Murphy}, {Norman}, {O'Steen}, {Oman}, {Pacifici}, {Pascual},
  {Pascual-Granado}, {Patil}, {Perren}, {Pickering}, {Rastogi}, {Roulston},
  {Ryan}, {Rykoff}, {Sabater}, {Sakurikar}, {Salgado}, {Sanghi}, {Saunders},
  {Savchenko}, {Schwardt}, {Seifert-Eckert}, {Shih}, {Jain}, {Shukla}, {Sick},
  {Simpson}, {Singanamalla}, {Singer}, {Singhal}, {Sinha}, {Sip{\H{o}}cz},
  {Spitler}, {Stansby}, {Streicher}, {{\v{S}}umak}, {Swinbank}, {Taranu},
  {Tewary}, {Tremblay}, {de Val-Borro}, {Van Kooten}, {Vasovi{\'c}}, {Verma},
  {de Miranda Cardoso}, {Williams}, {Wilson}, {Winkel}, {Wood-Vasey}, {Xue},
  {Yoachim}, {Zhang}, {Zonca}, \& {Astropy Project
  Contributors}}]{AstropyCollaboration22}
{Astropy Collaboration}, {Price-Whelan}, A.~M., {Lim}, P.~L., {et~al.} 2022,
  \bibinfo{title}{{The Astropy Project: Sustaining and Growing a
  Community-oriented Open-source Project and the Latest Major Release (v5.0) of
  the Core Package},} \apj, 935, 167, \dodoi{10.3847/1538-4357/ac7c74}

\bibitem[{R.~J. {Avila} {et~al.}(2012){Avila}, {Hack}, \& {STScI AstroDrizzle
  Team}}]{Avila12}
{Avila}, R.~J., {Hack}, W.~J., \& {STScI AstroDrizzle Team}. 2012, in American
  Astronomical Society Meeting Abstracts, Vol. 220, American Astronomical
  Society Meeting Abstracts \#220, 135.13

\bibitem[{A.~J. {Barth} {et~al.}(2016{\natexlab{a}}){Barth}, {Boizelle},
  {Darling}, {Baker}, {Buote}, {Ho}, \& {Walsh}}]{Barth16b}
{Barth}, A.~J., {Boizelle}, B.~D., {Darling}, J., {et~al.} 2016{\natexlab{a}},
  \bibinfo{title}{{Measurement of the Black Hole Mass in NGC 1332 from ALMA
  Observations at 0.044 arcsecond Resolution},} \apjl, 822, L28,
  \dodoi{10.3847/2041-8205/822/2/L28}

\bibitem[{A.~J. {Barth} {et~al.}(2016{\natexlab{b}}){Barth}, {Darling},
  {Baker}, {Boizelle}, {Buote}, {Ho}, \& {Walsh}}]{Barth16a}
{Barth}, A.~J., {Darling}, J., {Baker}, A.~J., {et~al.} 2016{\natexlab{b}},
  \bibinfo{title}{{Toward Precision Black Hole Masses with ALMA: NGC 1332 as a
  Case Study in Molecular Disk Dynamics},} \apj, 823, 51,
  \dodoi{10.3847/0004-637X/823/1/51}

\bibitem[{M.~C. {Bentz} {et~al.}(2023{\natexlab{a}}){Bentz}, {Markham},
  {Rosborough}, {Onken}, {Street}, {Valluri}, \& {Treu}}]{Bentz2023b}
{Bentz}, M.~C., {Markham}, M., {Rosborough}, S., {et~al.} 2023{\natexlab{a}},
  \bibinfo{title}{{Velocity-resolved Reverberation Mapping of NGC 3227},} \apj,
  959, 25, \dodoi{10.3847/1538-4357/ad08b8}

\bibitem[{M.~C. {Bentz} {et~al.}(2023{\natexlab{b}}){Bentz}, {Onken}, {Street},
  \& {Valluri}}]{Bentz2023a}
{Bentz}, M.~C., {Onken}, C.~A., {Street}, R., \& {Valluri}, M.
  2023{\natexlab{b}}, \bibinfo{title}{{Reverberation Mapping of IC 4329A},}
  \apj, 944, 29, \dodoi{10.3847/1538-4357/acab62}

\bibitem[{E. {Bertin} \& S. {Arnouts}(1996){Bertin} \& {Arnouts}}]{Bertin96}
{Bertin}, E., \& {Arnouts}, S. 1996, \bibinfo{title}{{SExtractor: Software for
  source extraction.},} \aaps, 117, 393, \dodoi{10.1051/aas:1996164}

\bibitem[{B.~D. {Boizelle} {et~al.}(2019){Boizelle}, {Barth}, {Walsh}, {Buote},
  {Baker}, {Darling}, \& {Ho}}]{Boizelle19}
{Boizelle}, B.~D., {Barth}, A.~J., {Walsh}, J.~L., {et~al.} 2019,
  \bibinfo{title}{{A Precision Measurement of the Mass of the Black Hole in NGC
  3258 from High-resolution ALMA Observations of Its Circumnuclear Disk},}
  \apj, 881, 10, \dodoi{10.3847/1538-4357/ab2a0a}

\bibitem[{B.~D. {Boizelle} {et~al.}(2021){Boizelle}, {Walsh}, {Barth}, {Buote},
  {Baker}, {Darling}, {Ho}, {Cohn}, \& {Kabasares}}]{Boizelle21}
{Boizelle}, B.~D., {Walsh}, J.~L., {Barth}, A.~J., {et~al.} 2021,
  \bibinfo{title}{{Black Hole Mass Measurements of Radio Galaxies NGC 315 and
  NGC 4261 Using ALMA CO Observations},} \apj, 908, 19,
  \dodoi{10.3847/1538-4357/abd24d}

\bibitem[{A.~D. {Bolatto} {et~al.}(2013){Bolatto}, {Wolfire}, \&
  {Leroy}}]{Bolatto2013}
{Bolatto}, A.~D., {Wolfire}, M., \& {Leroy}, A.~K. 2013, \bibinfo{title}{{The
  CO-to-H$_{2}$ Conversion Factor},} \araa, 51, 207,
  \dodoi{10.1146/annurev-astro-082812-140944}

\bibitem[{J.~A. {Braatz} {et~al.}(1996){Braatz}, {Wilson}, \&
  {Henkel}}]{Braatz1996}
{Braatz}, J.~A., {Wilson}, A.~S., \& {Henkel}, C. 1996, \bibinfo{title}{{A
  Survey for H 2O Megamasers in Active Galactic Nuclei. I. Observations},}
  \apjs, 106, 51, \dodoi{10.1086/192328}

\bibitem[{L. Bradley {et~al.}(2024)Bradley, Sip{\H o}cz, Robitaille, Tollerud,
  Vin{\'{\i}}cius, Deil, Barbary, Wilson, Busko, Donath, G{\"u}nther, Cara,
  Lim, Me{\ss}linger, Conseil, Burnett, Bostroem, Droettboom, Bray, Bratholm,
  Ginsburg, Jamieson, Barentsen, Craig, Morris, Perrin, Rathi, Pascual, \&
  Georgiev}]{bradley2024}
Bradley, L., Sip{\H o}cz, B., Robitaille, T., {et~al.} 2024,
  \bibinfo{title}{astropy/photutils: 2.0.2,}, 2.0.2 Zenodo,
  \dodoi{10.5281/zenodo.13989456}

\bibitem[{A. {Capetti} {et~al.}(2002){Capetti}, {Celotti}, {Chiaberge}, {de
  Ruiter}, {Fanti}, {Morganti}, \& {Parma}}]{Capetti02}
{Capetti}, A., {Celotti}, A., {Chiaberge}, M., {et~al.} 2002,
  \bibinfo{title}{{The HST survey of the B2 sample of radio-galaxies: Optical
  nuclei and the FR I/BL Lac unified scheme},} \aap, 383, 104,
  \dodoi{10.1051/0004-6361:20011714}

\bibitem[{A. Capetti {et~al.}(2000)Capetti, Trussoni, Celotti, Feretti, \&
  Chiaberge}]{Capetti2000}
Capetti, A., Trussoni, E., Celotti, A., Feretti, L., \& Chiaberge, M. 2000,
  \bibinfo{title}{Spectral energy distributions of FR I nuclei and the FR I/BL
  Lac unifying model,} Monthly Notices of the Royal Astronomical Society, 318,
  493, \dodoi{10.1046/j.1365-8711.2000.03823.x}

\bibitem[{M. Cappellari(2002)Cappellari}]{Cappellari02}
Cappellari, M. 2002, \bibinfo{title}{Efficient multi-Gaussian expansion of
  galaxies,} \mnras, 333, 400, \dodoi{10.1046/j.1365-8711.2002.05412.x}

\bibitem[{M. Cappellari(2008)Cappellari}]{Cappellari08}
Cappellari, M. 2008, \bibinfo{title}{Measuring the inclination and
  mass-to-light ratio of axisymmetric galaxies via anisotropic Jeans models of
  stellar kinematics,} \mnras, 390, 71,
  \dodoi{10.1111/j.1365-2966.2008.13754.x}

\bibitem[{M. {Cappellari}(2016){Cappellari}}]{Cappellari16}
{Cappellari}, M. 2016, \bibinfo{title}{{Structure and Kinematics of Early-Type
  Galaxies from Integral Field Spectroscopy},} \araa, 54, 597,
  \dodoi{10.1146/annurev-astro-082214-122432}

\bibitem[{M. Cappellari(2020)Cappellari}]{Cappellari20}
Cappellari, M. 2020, \bibinfo{title}{{Efficient solution of the anisotropic
  spherically aligned axisymmetric Jeans equations of stellar hydrodynamics for
  galactic dynamics},} \mnras, 494, 4819, \dodoi{10.1093/mnras/staa959}

\bibitem[{M. {Cappellari} {et~al.}(2013){Cappellari}, {Scott}, {Alatalo},
  {Blitz}, {Bois}, {Bournaud}, {Bureau}, {Crocker}, {Davies}, {Davis}, {de
  Zeeuw}, {Duc}, {Emsellem}, {Khochfar}, {Krajnovi{\'c}}, {Kuntschner},
  {McDermid}, {Morganti}, {Naab}, {Oosterloo}, {Sarzi}, {Serra}, {Weijmans}, \&
  {Young}}]{Cappellari13a}
{Cappellari}, M., {Scott}, N., {Alatalo}, K., {et~al.} 2013,
  \bibinfo{title}{{The ATLAS$^{3D}$ project - XV. Benchmark for early-type
  galaxies scaling relations from 260 dynamical models: mass-to-light ratio,
  dark matter, Fundamental Plane and Mass Plane},} \mnras, 432, 1709,
  \dodoi{10.1093/mnras/stt562}

\bibitem[{T.~M. {Dame}(2011){Dame}}]{Dame11}
{Dame}, T.~M. 2011, \bibinfo{title}{{Optimization of Moment Masking for CO
  Spectral Line Surveys},} arXiv e-prints, arXiv:1101.1499.
\newblock \doarXiv{1101.1499}

\bibitem[{T.~A. {Davis}(2014){Davis}}]{Davis14}
{Davis}, T.~A. 2014, \bibinfo{title}{{A figure of merit for black hole mass
  measurements with molecular gas},} \mnras, 443, 911,
  \dodoi{10.1093/mnras/stu1163}

\bibitem[{T.~A. {Davis} {et~al.}(2013){Davis}, {Bureau}, {Cappellari}, {Sarzi},
  \& {Blitz}}]{Davis13Nature}
{Davis}, T.~A., {Bureau}, M., {Cappellari}, M., {Sarzi}, M., \& {Blitz}, L.
  2013, \bibinfo{title}{{A black-hole mass measurement from molecular gas
  kinematics in NGC4526},} \nat, 494, 328, \dodoi{10.1038/nature11819}

\bibitem[{T.~A. {Davis} {et~al.}(2017){Davis}, {Bureau}, {Onishi},
  {Cappellari}, {Iguchi}, \& {Sarzi}}]{Davis17}
{Davis}, T.~A., {Bureau}, M., {Onishi}, K., {et~al.} 2017,
  \bibinfo{title}{{WISDOM Project - II. Molecular gas measurement of the
  supermassive black hole mass in NGC 4697},} \mnras, 468, 4675,
  \dodoi{10.1093/mnras/stw3217}

\bibitem[{T.~A. {Davis} {et~al.}(2020){Davis}, {Nguyen}, {Seth}, {Greene},
  {Nyland}, {Barth}, {Bureau}, {Cappellari}, {den Brok}, {Iguchi}, {Lelli},
  {Liu}, {Neumayer}, {North}, {Onishi}, {Sarzi}, {Smith}, \&
  {Williams}}]{Davis20}
{Davis}, T.~A., {Nguyen}, D.~D., {Seth}, A.~C., {et~al.} 2020,
  \bibinfo{title}{{Revealing the intermediate-mass black hole at the heart of
  the dwarf galaxy NGC 404 with sub-parsec resolution ALMA observations},}
  \mnras, 496, 4061, \dodoi{10.1093/mnras/staa1567}

\bibitem[{L. {de Juan} {et~al.}(1996){de Juan}, {Colina}, \&
  {Golombek}}]{Juan96}
{de Juan}, L., {Colina}, L., \& {Golombek}, D. 1996, \bibinfo{title}{{Nuclear
  disks of gas and dust in Fanaroff-Riley type I radio host galaxies: 3C 402N
  and NGC 7052.},} \aap, 305, 776

\bibitem[{P. Dominiak {et~al.}(2024)Dominiak, Bureau, Davis, Ma, Greene, \&
  Gu}]{Dominiak24}
Dominiak, P., Bureau, M., Davis, T.~A., {et~al.} 2024, \bibinfo{title}{{The
  MASSIVE survey – XIX. Molecular gas measurements of the supermassive black
  hole masses in the elliptical galaxies NGC 1684 and NGC 0997},} Monthly
  Notices of the Royal Astronomical Society, 529, 1597,
  \dodoi{10.1093/mnras/stae314}

\bibitem[{D. {Donato} {et~al.}(2004){Donato}, {Sambruna}, \&
  {Gliozzi}}]{Donato04}
{Donato}, D., {Sambruna}, R.~M., \& {Gliozzi}, M. 2004,
  \bibinfo{title}{{Obscuration and Origin of Nuclear X-Ray Emission in FR I
  Radio Galaxies},} \apj, 617, 915, \dodoi{10.1086/425575}

\bibitem[{A. {Dressler} \& D.~O. {Richstone}(1988){Dressler} \&
  {Richstone}}]{dressler1988stellar}
{Dressler}, A., \& {Richstone}, D.~O. 1988, \bibinfo{title}{{Stellar Dynamics
  in the Nuclei of M31 and M32: Evidence for Massive Black Holes},} \apj, 324,
  701, \dodoi{10.1086/165930}

\bibitem[{E. {Emsellem} {et~al.}(1994){Emsellem}, {Monnet}, \&
  {Bacon}}]{Emsellem94}
{Emsellem}, E., {Monnet}, G., \& {Bacon}, R. 1994, \bibinfo{title}{{The
  multi-gaussian expansion method: a tool for building realistic photometric
  and kinematical models of stellar systems I. The formalism},} \aap, 285, 723

\bibitem[{L. {Ferrarese} \& D. {Merritt}(2000){Ferrarese} \&
  {Merritt}}]{ferrarese2000fundamental}
{Ferrarese}, L., \& {Merritt}, D. 2000, \bibinfo{title}{{A Fundamental Relation
  between Supermassive Black Holes and Their Host Galaxies},} \apjl, 539, L9,
  \dodoi{10.1086/312838}

\bibitem[{F. {Gao} {et~al.}(2017){Gao}, {Braatz}, {Reid}, {Condon}, {Greene},
  {Henkel}, {Impellizzeri}, {Lo}, {Kuo}, {Pesce}, {Wagner}, \&
  {Zhao}}]{Gao2017}
{Gao}, F., {Braatz}, J.~A., {Reid}, M.~J., {et~al.} 2017, \bibinfo{title}{{The
  Megamaser Cosmology Project. IX. Black Hole Masses for Three Maser
  Galaxies},} \apj, 834, 52, \dodoi{10.3847/1538-4357/834/1/52}

\bibitem[{J.~I. {Gonzalez-Serrano} \& I.
  {Perez-Fournon}(1992){Gonzalez-Serrano} \&
  {Perez-Fournon}}]{Gonzalez-Serrano1992}
{Gonzalez-Serrano}, J.~I., \& {Perez-Fournon}, I. 1992, \bibinfo{title}{{CCD
  Surface Photometry of Three Low-Luminosity radio Galaxies Containing Radio
  Jets},} \aj, 104, 535, \dodoi{10.1086/116252}

\bibitem[{A.~D. {Goulding} {et~al.}(2016){Goulding}, {Greene}, {Ma}, {Veale},
  {Bogdan}, {Nyland}, {Blakeslee}, {McConnell}, \& {Thomas}}]{Goulding16}
{Goulding}, A.~D., {Greene}, J.~E., {Ma}, C.-P., {et~al.} 2016,
  \bibinfo{title}{{The MASSIVE Survey. IV. The X-ray Halos of the Most Massive
  Early-type Galaxies in the Nearby Universe},} \apj, 826, 167,
  \dodoi{10.3847/0004-637X/826/2/167}

\bibitem[{J.~E. {Greene}(2012){Greene}}]{Greene12}
{Greene}, J.~E. 2012, \bibinfo{title}{{Low-mass black holes as the remnants of
  primordial black hole formation},} Nature Communications, 3, 1304,
  \dodoi{10.1038/ncomms2314}

\bibitem[{J.~E. {Greene} {et~al.}(2020){Greene}, {Strader}, \& {Ho}}]{Greene20}
{Greene}, J.~E., {Strader}, J., \& {Ho}, L.~C. 2020,
  \bibinfo{title}{{Intermediate-Mass Black Holes},} \araa, 58, 257,
  \dodoi{10.1146/annurev-astro-032620-021835}

\bibitem[{M. {Gu}(2022){Gu}}]{Gu22}
{Gu}, M. 2022, in American Astronomical Society Meeting Abstracts, Vol. 240,
  American Astronomical Society Meeting \#240, 123.04

\bibitem[{K. {G{\"u}ltekin} {et~al.}(2009){G{\"u}ltekin}, {Richstone},
  {Gebhardt}, {Lauer}, {Tremaine}, {Aller}, {Bender}, {Dressler}, {Faber},
  {Filippenko}, {Green}, {Ho}, {Kormendy}, {Magorrian}, {Pinkney}, \&
  {Siopis}}]{Gultekin09}
{G{\"u}ltekin}, K., {Richstone}, D.~O., {Gebhardt}, K., {et~al.} 2009,
  \bibinfo{title}{{The M-{\ensuremath{\sigma}} and M-L Relations in Galactic
  Bulges, and Determinations of Their Intrinsic Scatter},} \apj, 698, 198,
  \dodoi{10.1088/0004-637X/698/1/198}

\bibitem[{H. Haario {et~al.}(2001)Haario, Saksman, \& Tamminen}]{Haario01}
Haario, H., Saksman, E., \& Tamminen, J. 2001, \bibinfo{title}{{An adaptive
  Metropolis algorithm},} Bernoulli, 7, 223

\bibitem[{N. {H{\"a}ring-Neumayer} {et~al.}(2006){H{\"a}ring-Neumayer},
  {Cappellari}, {Rix}, {Hartung}, {Prieto}, {Meisenheimer}, \&
  {Lenzen}}]{Haring2006}
{H{\"a}ring-Neumayer}, N., {Cappellari}, M., {Rix}, H.~W., {et~al.} 2006,
  \bibinfo{title}{{VLT Diffraction-limited Imaging and Spectroscopy in the NIR:
  Weighing the Black Hole in Centaurus A with NACO},} \apj, 643, 226,
  \dodoi{10.1086/501494}

\bibitem[{C.~R. Harris {et~al.}(2020)Harris, Millman, van~der Walt, Gommers,
  Virtanen, Cournapeau, Wieser, Taylor, Berg, Smith, Kern, Picus, Hoyer, van
  Kerkwijk, Brett, Haldane, del R{\'{\i}}o, Wiebe, Peterson,
  G{\'{e}}rard-Marchant, Sheppard, Reddy, Weckesser, Abbasi, Gohlke, \&
  Oliphant}]{Harris2020}
Harris, C.~R., Millman, K.~J., van~der Walt, S.~J., {et~al.} 2020,
  \bibinfo{title}{Array programming with {NumPy},} Nature, 585, 357,
  \dodoi{10.1038/s41586-020-2649-2}

\bibitem[{J.~A. {H{\"o}gbom}(1974){H{\"o}gbom}}]{Hogbom74}
{H{\"o}gbom}, J.~A. 1974, \bibinfo{title}{{Aperture Synthesis with a
  Non-Regular Distribution of Interferometer Baselines},} \aaps, 15, 417

\bibitem[{J.~A. {Holtzman} {et~al.}(1995){Holtzman}, {Hester}, {Casertano},
  {Trauger}, {Watson}, {Ballester}, {Burrows}, {Clarke}, {Crisp}, {Evans},
  {Gallagher}, {Griffiths}, {Hoessel}, {Matthews}, {Mould}, {Scowen},
  {Stapelfeldt}, \& {Westphal}}]{Holtzman95}
{Holtzman}, J.~A., {Hester}, J.~J., {Casertano}, S., {et~al.} 1995,
  \bibinfo{title}{{The Performance and Calibration of WFPC2 on the Hubble Space
  Telescope},} \pasp, 107, 156, \dodoi{10.1086/133533}

\bibitem[{J.~D. Hunter(2007)Hunter}]{Hunter2007}
Hunter, J.~D. 2007, \bibinfo{title}{Matplotlib: A 2D graphics environment,}
  Computing In Science \& Engineering, 9, 90, \dodoi{10.1109/MCSE.2007.55}

\bibitem[{M. {Keppler} {et~al.}(2019){Keppler}, {Teague}, {Bae}, {Benisty},
  {Henning}, {van Boekel}, {Chapillon}, {Pinilla}, {Williams}, {Bertrang},
  {Facchini}, {Flock}, {Ginski}, {Juhasz}, {Klahr}, {Liu}, {M{\"u}ller},
  {P{\'e}rez}, {Pohl}, {Rosotti}, {Samland}, \& {Semenov}}]{Keppler19}
{Keppler}, M., {Teague}, R., {Bae}, J., {et~al.} 2019, \bibinfo{title}{{Highly
  structured disk around the planet host PDS 70 revealed by high-angular
  resolution observations with ALMA},} \aap, 625, A118,
  \dodoi{10.1051/0004-6361/201935034}

\bibitem[{J. {Kormendy} \& K. {Gebhardt}(2001){Kormendy} \&
  {Gebhardt}}]{kormendy2001supermassive}
{Kormendy}, J., \& {Gebhardt}, K. 2001, in American Institute of Physics
  Conference Series, Vol. 586, 20th Texas Symposium on relativistic
  astrophysics, ed. J.~C. {Wheeler} \& H.~{Martel}, 363--381,
  \dodoi{10.1063/1.1419581}

\bibitem[{J. Kormendy \& L.~C. Ho(2013)Kormendy \& Ho}]{Kormendy13}
Kormendy, J., \& Ho, L.~C. 2013, \bibinfo{title}{{Coevolution (Or Not) of
  Supermassive Black Holes and Host Galaxies},} \araa, 51, 511,
  \dodoi{10.1146/annurev-astro-082708-101811}

\bibitem[{D. {Krajnovi{\'c}} {et~al.}(2018){Krajnovi{\'c}}, {Cappellari}, \&
  {McDermid}}]{Krajnovic18a}
{Krajnovi{\'c}}, D., {Cappellari}, M., \& {McDermid}, R.~M. 2018,
  \bibinfo{title}{{Two channels of supermassive black hole growth as seen on
  the galaxies mass-size plane},} \mnras, 473, 5237,
  \dodoi{10.1093/mnras/stx2704}

\bibitem[{J.~E. {Krist} {et~al.}(2011){Krist}, {Hook}, \& {Stoehr}}]{Krist11}
{Krist}, J.~E., {Hook}, R.~N., \& {Stoehr}, F. 2011, in Society of
  Photo-Optical Instrumentation Engineers (SPIE) Conference Series, Vol. 8127,
  Optical Modeling and Performance Predictions V, ed. M.~A. {Kahan}, 81270J,
  \dodoi{10.1117/12.892762}

\bibitem[{T.~R. {Lauer} {et~al.}(2007){Lauer}, {Faber}, {Richstone},
  {Gebhardt}, {Tremaine}, {Postman}, {Dressler}, {Aller}, {Filippenko},
  {Green}, {Ho}, {Kormendy}, {Magorrian}, \& {Pinkney}}]{Lauer07}
{Lauer}, T.~R., {Faber}, S.~M., {Richstone}, D., {et~al.} 2007,
  \bibinfo{title}{{The Masses of Nuclear Black Holes in Luminous Elliptical
  Galaxies and Implications for the Space Density of the Most Massive Black
  Holes},} \apj, 662, 808, \dodoi{10.1086/518223}

\bibitem[{C.-P. {Ma} {et~al.}(2014){Ma}, {Greene}, {McConnell}, {Janish},
  {Blakeslee}, {Thomas}, \& {Murphy}}]{Ma14}
{Ma}, C.-P., {Greene}, J.~E., {McConnell}, N., {et~al.} 2014,
  \bibinfo{title}{{The MASSIVE Survey. I. A Volume-limited Integral-field
  Spectroscopic Study of the Most Massive Early-type Galaxies within 108 Mpc},}
  \apj, 795, 158, \dodoi{10.1088/0004-637X/795/2/158}

\bibitem[{J. {Magorrian} {et~al.}(1998){Magorrian}, {Tremaine}, {Richstone},
  {Bender}, {Bower}, {Dressler}, {Faber}, {Gebhardt}, {Green}, {Grillmair},
  {Kormendy}, \& {Lauer}}]{Magorrian98}
{Magorrian}, J., {Tremaine}, S., {Richstone}, D., {et~al.} 1998,
  \bibinfo{title}{{The Demography of Massive Dark Objects in Galaxy Centers},}
  \aj, 115, 2285, \dodoi{10.1086/300353}

\bibitem[{J.~P. {McMullin} {et~al.}(2007){McMullin}, {Waters}, {Schiebel},
  {Young}, \& {Golap}}]{McMullin07}
{McMullin}, J.~P., {Waters}, B., {Schiebel}, D., {Young}, W., \& {Golap}, K.
  2007, in Astronomical Society of the Pacific Conference Series, Vol. 376,
  Astronomical Data Analysis Software and Systems XVI, ed. R.~A. {Shaw},
  F.~{Hill}, \& D.~J. {Bell}, 127

\bibitem[{E. {Memola} {et~al.}(2009){Memola}, {Trinchieri}, {Wolter},
  {Focardi}, \& {Kelm}}]{Memola2009}
{Memola}, E., {Trinchieri}, G., {Wolter}, A., {Focardi}, P., \& {Kelm}, B.
  2009, \bibinfo{title}{{The diverse X-ray properties of four truly isolated
  elliptical galaxies: NGC 2954, NGC 6172, NGC 7052, and NGC 7785},} \aap, 497,
  359, \dodoi{10.1051/0004-6361/200810801}

\bibitem[{M. {Mitzkus} {et~al.}(2017){Mitzkus}, {Cappellari}, \&
  {Walcher}}]{Mitzkus17}
{Mitzkus}, M., {Cappellari}, M., \& {Walcher}, C.~J. 2017,
  \bibinfo{title}{{Dominant dark matter and a counter-rotating disc: MUSE view
  of the low-luminosity S0 galaxy NGC 5102},} \mnras, 464, 4789,
  \dodoi{10.1093/mnras/stw2677}

\bibitem[{M. {Miyoshi} {et~al.}(1995){Miyoshi}, {Moran}, {Herrnstein},
  {Greenhill}, {Nakai}, {Diamond}, \& {Inoue}}]{Miyoshi95}
{Miyoshi}, M., {Moran}, J., {Herrnstein}, J., {et~al.} 1995,
  \bibinfo{title}{{Evidence for a black hole from high rotation velocities in a
  sub-parsec region of NGC4258},} \nat, 373, 127, \dodoi{10.1038/373127a0}

\bibitem[{R. {Morganti} {et~al.}(1987){Morganti}, {Fanti}, {Fanti}, {Parma}, \&
  {de Ruiter}}]{Morganti87}
{Morganti}, R., {Fanti}, C., {Fanti}, R., {Parma}, P., \& {de Ruiter}, H.~R.
  1987, \bibinfo{title}{{VLA observations of low luminosity radio galaxies. V.
  A detailed radio study of five jets.},} \aap, 183, 203

\bibitem[{D.~D. {Nguyen}(2017){Nguyen}}]{Nguyen17conf}
{Nguyen}, D.~D. 2017, \bibinfo{title}{{Improved dynamical constraints on the
  mass of the central black hole in NGC 404},} arXiv e-prints,
  arXiv:1712.02470, \dodoi{10.48550/arXiv.1712.02470}

\bibitem[{D.~D. {Nguyen}(2019){Nguyen}}]{Nguyen19conf}
{Nguyen}, D.~D. 2019, in ALMA2019: Science Results and Cross-Facility
  Synergies, 106, \dodoi{10.5281/zenodo.3585410}

\bibitem[{D.~D. {Nguyen} {et~al.}(2023){Nguyen}, {Cappellari}, \&
  {Pereira-Santaella}}]{Nguyen23}
{Nguyen}, D.~D., {Cappellari}, M., \& {Pereira-Santaella}, M. 2023,
  \bibinfo{title}{{Simulating supermassive black hole mass measurements for a
  sample of ultramassive galaxies using ELT/HARMONI high-spatial-resolution
  integral-field stellar kinematics},} \mnras, 526, 3548,
  \dodoi{10.1093/mnras/stad2860}

\bibitem[{D.~D. {Nguyen} {et~al.}(2014){Nguyen}, {Seth}, {Reines}, {den Brok},
  {Sand}, \& {McLeod}}]{Nguyen14}
{Nguyen}, D.~D., {Seth}, A.~C., {Reines}, A.~E., {et~al.} 2014,
  \bibinfo{title}{{Extended Structure and Fate of the Nucleus in Henize 2-10},}
  \apj, 794, 34, \dodoi{10.1088/0004-637X/794/1/34}

\bibitem[{D.~D. {Nguyen} {et~al.}(2017){Nguyen}, {Seth}, {den Brok},
  {Neumayer}, {Cappellari}, {Barth}, {Caldwell}, {Williams}, \&
  {Binder}}]{Nguyen17}
{Nguyen}, D.~D., {Seth}, A.~C., {den Brok}, M., {et~al.} 2017,
  \bibinfo{title}{{Improved Dynamical Constraints on the Mass of the Central
  Black Hole in NGC 404},} \apj, 836, 237, \dodoi{10.3847/1538-4357/aa5cb4}

\bibitem[{D.~D. {Nguyen} {et~al.}(2018){Nguyen}, {Seth}, {Neumayer}, {Kamann},
  {Voggel}, {Cappellari}, {Picotti}, {Nguyen}, {B{\"o}ker}, {Debattista},
  {Caldwell}, {McDermid}, {Bastian}, {Ahn}, \& {Pechetti}}]{Nguyen18}
{Nguyen}, D.~D., {Seth}, A.~C., {Neumayer}, N., {et~al.} 2018,
  \bibinfo{title}{{Nearby Early-type Galactic Nuclei at High Resolution:
  Dynamical Black Hole and Nuclear Star Cluster Mass Measurements},} \apj, 858,
  118, \dodoi{10.3847/1538-4357/aabe28}

\bibitem[{D.~D. {Nguyen} {et~al.}(2019){Nguyen}, {Seth}, {Neumayer}, {Iguchi},
  {Cappellari}, {Strader}, {Chomiuk}, {Tremou}, {Pacucci}, {Nakanishi},
  {Bahramian}, {Nguyen}, {den Brok}, {Ahn}, {Voggel}, {Kacharov}, {Tsukui},
  {Ly}, {Dumont}, \& {Pechetti}}]{Nguyen19}
{Nguyen}, D.~D., {Seth}, A.~C., {Neumayer}, N., {et~al.} 2019,
  \bibinfo{title}{{Improved Dynamical Constraints on the Masses of the Central
  Black Holes in Nearby Low-mass Early-type Galactic Nuclei and the First Black
  Hole Determination for NGC 205},} \apj, 872, 104,
  \dodoi{10.3847/1538-4357/aafe7a}

\bibitem[{D.~D. {Nguyen} {et~al.}(2020){Nguyen}, {den Brok}, {Seth}, {Davis},
  {Greene}, {Cappellari}, {Jensen}, {Thater}, {Iguchi}, {Imanishi}, {Izumi},
  {Nyland}, {Neumayer}, {Nakanishi}, {Nguyen}, {Tsukui}, {Bureau}, {Onishi},
  {Nguyen}, \& {Le}}]{Nguyen20}
{Nguyen}, D.~D., {den Brok}, M., {Seth}, A.~C., {et~al.} 2020,
  \bibinfo{title}{{The MBHBM$_{{\ensuremath{\star}}}$ Project. I. Measurement
  of the Central Black Hole Mass in Spiral Galaxy NGC 3504 Using Molecular Gas
  Kinematics},} \apj, 892, 68, \dodoi{10.3847/1538-4357/ab77aa}

\bibitem[{D.~D. {Nguyen} {et~al.}(2021){Nguyen}, {Izumi}, {Thater}, {Imanishi},
  {Kawamuro}, {Baba}, {Nakano}, {Turner}, {Kohno}, {Matsushita}, {Mart{\'\i}n},
  {Meier}, {Nguyen}, \& {Nguyen}}]{Nguyen21}
{Nguyen}, D.~D., {Izumi}, T., {Thater}, S., {et~al.} 2021,
  \bibinfo{title}{{Black hole mass measurement using ALMA observations of [CI]
  and CO emissions in the Seyfert 1 galaxy NGC 7469},} \mnras, 504, 4123,
  \dodoi{10.1093/mnras/stab1002}

\bibitem[{D.~D. {Nguyen} {et~al.}(2022){Nguyen}, {Bureau}, {Thater}, {Nyland},
  {den Brok}, {Cappellari}, {Davis}, {Greene}, {Neumayer}, {Imanishi}, {Izumi},
  {Kawamuro}, {Baba}, {Nguyen}, {Iguchi}, {Tsukui}, {Lam}, \& {Ho}}]{Nguyen22}
{Nguyen}, D.~D., {Bureau}, M., {Thater}, S., {et~al.} 2022,
  \bibinfo{title}{{The MBHBM$^{{\ensuremath{\star}}}$ Project - II. Molecular
  gas kinematics in the lenticular galaxy NGC 3593 reveal a supermassive black
  hole},} \mnras, 509, 2920, \dodoi{10.1093/mnras/stab3016}

\bibitem[{D.~D. {Nguyen} {et~al.}(2025{\natexlab{a}}){Nguyen}, {Ngo}, {Le},
  {Graham}, {Soria}, {Chilingarian}, {Thatte}, {Phuong}, {Hoang},
  {Pereira-Santaella}, {Durre}, {Pham}, {Ngoc Tram}, {Ngoc}, \&
  {L{\^e}}}]{Nguyen2025}
{Nguyen}, D.~D., {Ngo}, H.~N., {Le}, T. Q.~T., {et~al.} 2025{\natexlab{a}},
  \bibinfo{title}{{Supermassive black hole mass measurement in the spiral
  galaxy NGC 4736 using JWST/NIRSpec stellar kinematics},} \aap, 698, L9,
  \dodoi{10.1051/0004-6361/202554672}

\bibitem[{D.~D. {Nguyen} {et~al.}(2025{\natexlab{b}}){Nguyen}, {Cappellari},
  {Ngo}, {Le}, {Le}, {Ho}, {Nguyen}, {On}, {Tong}, {Thatte}, \&
  {Pereira-Santaella}}]{Nguyen2025b}
{Nguyen}, D.~D., {Cappellari}, M., {Ngo}, H.~N., {et~al.} 2025{\natexlab{b}},
  \bibinfo{title}{{Simulating Intermediate Black Hole Mass Measurements for a
  Sample of Galaxies with Nuclear Star Clusters Using ELT/HARMONI High Spatial
  Resolution Integral-field Stellar Kinematics},} \aj, 170, 124,
  \dodoi{10.3847/1538-3881/ade9ba}

\bibitem[{J.-L. {Nieto} {et~al.}(1991){Nieto}, {Bender}, \&
  {Surma}}]{Nieto1991}
{Nieto}, J.-L., {Bender}, R., \& {Surma}, P. 1991, \bibinfo{title}{{Central
  brightness profiles and isophotal shapes in elliptical galaxies.},} \aap,
  244, L37

\bibitem[{J.~L. {Nieto} {et~al.}(1990){Nieto}, {McClure}, {Fletcher}, {Arnaud},
  {Bacon}, {Bender}, {Comte}, \& {Poulain}}]{Nieto1990}
{Nieto}, J.~L., {McClure}, R., {Fletcher}, J.~M., {et~al.} 1990,
  \bibinfo{title}{{The core of the elliptical galaxy NGC 7052.},} \aap, 235,
  L17

\bibitem[{E.~V. {North} {et~al.}(2019){North}, {Davis}, {Bureau}, {Cappellari},
  {Iguchi}, {Liu}, {Onishi}, {Sarzi}, {Smith}, \& {Williams}}]{North19}
{North}, E.~V., {Davis}, T.~A., {Bureau}, M., {et~al.} 2019,
  \bibinfo{title}{{WISDOM project - V. Resolving molecular gas in Keplerian
  rotation around the supermassive black hole in NGC 0383},} \mnras, 490, 319,
  \dodoi{10.1093/mnras/stz2598}

\bibitem[{K. {Onishi} {et~al.}(2017){Onishi}, {Iguchi}, {Davis}, {Bureau},
  {Cappellari}, {Sarzi}, \& {Blitz}}]{Onishi17}
{Onishi}, K., {Iguchi}, S., {Davis}, T.~A., {et~al.} 2017,
  \bibinfo{title}{{WISDOM project - I. Black hole mass measurement using
  molecular gas kinematics in NGC 3665},} \mnras, 468, 4663,
  \dodoi{10.1093/mnras/stx631}

\bibitem[{V. {Pandya} {et~al.}(2017){Pandya}, {Greene}, {Ma}, {Veale}, {Ene},
  {Davis}, {Blakeslee}, {Goulding}, {McConnell}, {Nyland}, \&
  {Thomas}}]{Pandya2017}
{Pandya}, V., {Greene}, J.~E., {Ma}, C.-P., {et~al.} 2017, \bibinfo{title}{{The
  MASSIVE Survey. VI. The Spatial Distribution and Kinematics of Warm Ionized
  Gas in the Most Massive Local Early-type Galaxies},} \apj, 837, 40,
  \dodoi{10.3847/1538-4357/aa5ebc}

\bibitem[{P. {Parma} {et~al.}(1986){Parma}, {de Ruiter}, {Fanti}, \&
  {Fanti}}]{Parma86}
{Parma}, P., {de Ruiter}, H.~R., {Fanti}, C., \& {Fanti}, R. 1986,
  \bibinfo{title}{{VLA observations of low luminosity radio galaxies. I.
  Sources with angular size smaller than two arcminutes.},} \aaps, 64, 135

\bibitem[{U. {Rau} \& T.~J. {Cornwell}(2011){Rau} \& {Cornwell}}]{Rau11}
{Rau}, U., \& {Cornwell}, T.~J. 2011, \bibinfo{title}{{A multi-scale
  multi-frequency deconvolution algorithm for synthesis imaging in radio
  interferometry},} \aap, 532, A71, \dodoi{10.1051/0004-6361/201117104}

\bibitem[{I. Ruffa {et~al.}(2023)Ruffa, Davis, Cappellari, Bureau, Elford,
  Iguchi, Lelli, Liang, Liu, Lu, Sarzi, \& Williams}]{Ruffa23}
Ruffa, I., Davis, T.~A., Cappellari, M., {et~al.} 2023, \bibinfo{title}{{WISDOM
  project – XIV. SMBH mass in the early-type galaxies NGC 0612, NGC 1574,
  and NGC 4261 from CO dynamical modelling},} \mnras, 522, 6170,
  \dodoi{10.1093/mnras/stad1119}

\bibitem[{S.~P. Rusli {et~al.}(2013)Rusli, Erwin, Saglia, Thomas, Fabricius,
  Bender, \& Nowak}]{Rusli13}
Rusli, S.~P., Erwin, P., Saglia, R.~P., {et~al.} 2013, \bibinfo{title}{DEPLETED
  GALAXY CORES AND DYNAMICAL BLACK HOLE MASSES,} The Astronomical Journal, 146,
  160, \dodoi{10.1088/0004-6256/146/6/160}

\bibitem[{R.~P. {Saglia} {et~al.}(2016){Saglia}, {Opitsch}, {Erwin}, {Thomas},
  {Beifiori}, {Fabricius}, {Mazzalay}, {Nowak}, {Rusli}, \&
  {Bender}}]{Saglia16}
{Saglia}, R.~P., {Opitsch}, M., {Erwin}, P., {et~al.} 2016,
  \bibinfo{title}{{The SINFONI Black Hole Survey: The Black Hole Fundamental
  Plane Revisited and the Paths of (Co)evolution of Supermassive Black Holes
  and Bulges},} \apj, 818, 47, \dodoi{10.3847/0004-637X/818/1/47}

\bibitem[{N. {Sahu} {et~al.}(2019){Sahu}, {Graham}, \& {Davis}}]{Sahu19a}
{Sahu}, N., {Graham}, A.~W., \& {Davis}, B.~L. 2019, \bibinfo{title}{{Black
  Hole Mass Scaling Relations for Early-type Galaxies. I. M $_{BH}$-M $_{*,}$
  $_{sph}$ and M $_{BH}$-M $_{*,gal}$},} \apj, 876, 155,
  \dodoi{10.3847/1538-4357/ab0f32}

\bibitem[{M.~D. {Smith} {et~al.}(2019){Smith}, {Bureau}, {Davis}, {Cappellari},
  {Liu}, {North}, {Onishi}, {Iguchi}, \& {Sarzi}}]{Smith19}
{Smith}, M.~D., {Bureau}, M., {Davis}, T.~A., {et~al.} 2019,
  \bibinfo{title}{{WISDOM project - IV. A molecular gas dynamical measurement
  of the supermassive black hole mass in NGC 524},} \mnras, 485, 4359,
  \dodoi{10.1093/mnras/stz625}

\bibitem[{M.~D. {Smith} {et~al.}(2021){Smith}, {Bureau}, {Davis}, {Cappellari},
  {Liu}, {Onishi}, {Iguchi}, {North}, {Sarzi}, \& {Williams}}]{Smith21}
{Smith}, M.~D., {Bureau}, M., {Davis}, T.~A., {et~al.} 2021,
  \bibinfo{title}{{WISDOM project - VII. Molecular gas measurement of the
  supermassive black hole mass in the elliptical galaxy NGC 7052},} \mnras,
  503, 5984, \dodoi{10.1093/mnras/stab791}

\bibitem[{B.~A. {Terrazas} {et~al.}(2017){Terrazas}, {Bell}, {Woo}, \&
  {Henriques}}]{Terrazas17}
{Terrazas}, B.~A., {Bell}, E.~F., {Woo}, J., \& {Henriques}, B. M.~B. 2017,
  \bibinfo{title}{{Supermassive Black Holes as the Regulators of Star Formation
  in Central Galaxies},} \apj, 844, 170, \dodoi{10.3847/1538-4357/aa7d07}

\bibitem[{S. {Thater}(2019){Thater}}]{Thater19_conf}
{Thater}, S. 2019, in ALMA2019: Science Results and Cross-Facility Synergies,
  129, \dodoi{10.5281/zenodo.3585459}

\bibitem[{S. {Thater} {et~al.}(2023){Thater}, {Lyubenova}, {Fahrion},
  {Mart{\'\i}n-Navarro}, {Jethwa}, {Nguyen}, \& {van de Ven}}]{Thater23}
{Thater}, S., {Lyubenova}, M., {Fahrion}, K., {et~al.} 2023,
  \bibinfo{title}{{Effect of the initial mass function on the dynamical SMBH
  mass estimate in the nucleated early-type galaxy FCC 47},} \aap, 675, A18,
  \dodoi{10.1051/0004-6361/202245362}

\bibitem[{S. {Thater} {et~al.}(2022){Thater}, {Krajnovi{\'c}}, {Weilbacher},
  {Nguyen}, {Bureau}, {Cappellari}, {Davis}, {Iguchi}, {McDermid}, {Onishi},
  {Sarzi}, \& {van de Ven}}]{Thater22}
{Thater}, S., {Krajnovi{\'c}}, D., {Weilbacher}, P.~M., {et~al.} 2022,
  \bibinfo{title}{{Cross-checking SMBH mass estimates in NGC 6958 - I. Stellar
  dynamics from adaptive optics-assisted MUSE observations},} \mnras, 509,
  5416, \dodoi{10.1093/mnras/stab3210}

\bibitem[{G. {Theureau} {et~al.}(2007){Theureau}, {Hanski}, {Coudreau},
  {Hallet}, \& {Martin}}]{Theureau2007}
{Theureau}, G., {Hanski}, M.~O., {Coudreau}, N., {Hallet}, N., \& {Martin},
  J.~M. 2007, \bibinfo{title}{{Kinematics of the Local Universe. XIII. 21-cm
  line measurements of 452 galaxies with the Nan{\c{c}}ay radiotelescope, JHK
  Tully-Fisher relation, and preliminary maps of the peculiar velocity field},}
  \aap, 465, 71, \dodoi{10.1051/0004-6361:20066187}

\bibitem[{S.~C. {Trager} {et~al.}(2000){Trager}, {Faber}, {Worthey}, \&
  {Gonz{\'a}lez}}]{Trager2000AJ119.1645}
{Trager}, S.~C., {Faber}, S.~M., {Worthey}, G., \& {Gonz{\'a}lez}, J.~J. 2000,
  \bibinfo{title}{{The Stellar Population Histories of Local Early-Type
  Galaxies. I. Population Parameters},} \aj, 119, 1645, \dodoi{10.1086/301299}

\bibitem[{F.~C. {van den Bosch} \& R.~P. {van der Marel}(1995){van den Bosch}
  \& {van der Marel}}]{Bosch1995}
{van den Bosch}, F.~C., \& {van der Marel}, R.~P. 1995,
  \bibinfo{title}{{Dynamics of the nuclear gas and dust disc in the E4 radio
  galaxy NGC 7052},} \mnras, 274, 884, \dodoi{10.1093/mnras/274.3.884}

\bibitem[{R.~C.~E. {van den Bosch}(2016){van den Bosch}}]{Bosch2016}
{van den Bosch}, R. C.~E. 2016, \bibinfo{title}{{Unification of the fundamental
  plane and Super Massive Black Hole Masses},} \apj, 831, 134,
  \dodoi{10.3847/0004-637X/831/2/134}

\bibitem[{R.~C.~E. {van den Bosch} \& G. {van de Ven}(2009){van den Bosch} \&
  {van de Ven}}]{vandenBosch09}
{van den Bosch}, R. C.~E., \& {van de Ven}, G. 2009,
  \bibinfo{title}{{Recovering the intrinsic shape of early-type galaxies},}
  \mnras, 398, 1117, \dodoi{10.1111/j.1365-2966.2009.15177.x}

\bibitem[{R.~P. {van der Marel} \& F.~C. {van den Bosch}(1998){van der Marel}
  \& {van den Bosch}}]{Marel1998}
{van der Marel}, R.~P., \& {van den Bosch}, F.~C. 1998,
  \bibinfo{title}{{Evidence for a 3 X 10\^8 M\_{\ensuremath{\odot}} Black Hole
  in NGC 7052 from Hubble Space Telescope Observations of the Nuclear Gas
  Disk},} \aj, 116, 2220, \dodoi{10.1086/300593}

\bibitem[{G. Van~Rossum \& F.~L. Drake(2009)Van~Rossum \&
  Drake}]{VanRossum2009}
Van~Rossum, G., \& Drake, F.~L. 2009, Python 3 Reference Manual (Scotts Valley,
  CA: CreateSpace)

\bibitem[{M. Veale {et~al.}(2017)Veale, Ma, Greene, Thomas, Blakeslee, Walsh,
  \& Ito}]{Veale18}
Veale, M., Ma, C.-P., Greene, J.~E., {et~al.} 2017, \bibinfo{title}{The MASSIVE
  survey – VIII. Stellar velocity dispersion profiles and environmental
  dependence of early-type galaxies,} Monthly Notices of the Royal Astronomical
  Society, 473, 5446, \dodoi{10.1093/mnras/stx2717}

\bibitem[{P. Virtanen {et~al.}(2020)Virtanen, Gommers, Oliphant, Haberland,
  Reddy, Cournapeau, Burovski, Peterson, Weckesser, Bright, van~der Walt,
  Brett, Wilson, Millman, Mayorov, Nelson, Jones, Kern, Larson, Carey, Polat,
  Feng, Moore, VanderPlas, Laxalde, Perktold, Cimrman, Henriksen, Quintero,
  Harris, Archibald, Ribeiro, Pedregosa, van Mulbregt, Vijaykumar, Bardelli,
  Rothberg, Hilboll, Kloeckner, Scopatz, Lee, Rokem, Woods, Fulton, Masson,
  Häggström, Fitzgerald, Nicholson, Hagen, Pasechnik, Olivetti, Martin,
  Wieser, Silva, Lenders, Wilhelm, Young, Price, Ingold, Allen, Lee, Audren,
  Probst, Dietrich, Silterra, Webber, Slavi{\v{c}}, Nothman, Buchner, Kulick,
  Schönberger, de~Miranda~Cardoso, Reimer, Harrington, Rodr{\'{\i}}guez,
  Nunez-Iglesias, Kuczynski, Tritz, Thoma, Newville, Kümmerer, Bolingbroke,
  Tartre, Pak, Smith, Nowaczyk, Shebanov, Pavlyk, Brodtkorb, Lee, McGibbon,
  Feldbauer, Lewis, Tygier, Sievert, Vigna, Peterson, More, Pudlik, Oshima,
  Pingel, Robitaille, Spura, Jones, Cera, Leslie, Zito, Krauss, Upadhyay,
  Halchenko, \& V{\'{a}}zquez-Baeza}]{Virtanen2020}
Virtanen, P., Gommers, R., Oliphant, T.~E., {et~al.} 2020,
  \bibinfo{title}{{SciPy} 1.0: fundamental algorithms for scientific computing
  in Python,} Nature Methods, 17, 261, \dodoi{10.1038/s41592-019-0686-2}

\bibitem[{K.~T. {Voggel} {et~al.}(2018){Voggel}, {Seth}, {Neumayer}, {Mieske},
  {Chilingarian}, {Ahn}, {Baumgardt}, {Hilker}, {Nguyen}, {Romanowsky},
  {Walsh}, {den Brok}, \& {Strader}}]{Voggel18}
{Voggel}, K.~T., {Seth}, A.~C., {Neumayer}, N., {et~al.} 2018,
  \bibinfo{title}{{Upper Limits on the Presence of Central Massive Black Holes
  in Two Ultra-compact Dwarf Galaxies in Centaurus A},} \apj, 858, 20,
  \dodoi{10.3847/1538-4357/aabae5}

\bibitem[{J.~L. {Walsh} {et~al.}(2013){Walsh}, {Barth}, {Ho}, \&
  {Sarzi}}]{Walsh13}
{Walsh}, J.~L., {Barth}, A.~J., {Ho}, L.~C., \& {Sarzi}, M. 2013,
  \bibinfo{title}{{The M87 Black Hole Mass from Gas-dynamical Models of Space
  Telescope Imaging Spectrograph Observations},} \apj, 770, 86,
  \dodoi{10.1088/0004-637X/770/2/86}

\bibitem[{Z. {Wang} {et~al.}(1992){Wang}, {Kenney}, \& {Ishizuki}}]{Wang1992}
{Wang}, Z., {Kenney}, J. D.~P., \& {Ishizuki}, S. 1992,
  \bibinfo{title}{{Molecular Gas in Elliptical Galaxies With Dust Lanes},} \aj,
  104, 2097, \dodoi{10.1086/116385}

\bibitem[{C.~N.~A. {Willmer}(2018){Willmer}}]{Willmer18}
{Willmer}, C. N.~A. 2018, \bibinfo{title}{{The Absolute Magnitude of the Sun in
  Several Filters},} \apjs, 236, 47, \dodoi{10.3847/1538-4365/aabfdf}

\bibitem[{H. {Zhang} {et~al.}(2025){Zhang}, {Bureau}, {Ruffa}, {Cappellari},
  {Davis}, {Dominiak}, {Elford}, {Iguchi}, {Lelli}, {Sarzi}, \&
  {Williams}}]{Zhang25}
{Zhang}, H., {Bureau}, M., {Ruffa}, I., {et~al.} 2025, \bibinfo{title}{{WISDOM
  Project - XXII. A 5 per cent precision CO-dynamical supermassive black hole
  mass measurement in the galaxy NGC 383},} \mnras, 537, 520,
  \dodoi{10.1093/mnras/staf055}

\end{thebibliography}


\label{lastpage}
\end{document}